\documentclass[11pt]{article} 
\usepackage{epsfig}
\def\theequation{\thesection.\arabic{equation}}
\begin{document}
\makeatletter
\setlength\@fptop{0\p@ }
\setlength\@fpsep{12\p@ }
\setlength\@fpbot{0\p@ plus 1fil }
\makeatother

\def\textfraction{.01}
\def\floatpagefraction{.8}

\intextsep 20pt plus 2pt minus 2pt
\setcounter{topnumber}{4}
\def\topfraction{.9}
\setcounter{bottomnumber}{2}
\def\bottomfraction{.8}
\setcounter{totalnumber}{6}

%
\makeatletter
\newinsert \@kludgeins
\global\dimen\@kludgeins \maxdimen
\global\count\@kludgeins 1000
\gdef \enlargethispage {%
   \@ifstar
     {%
      \@enlargepage{\hbox{\kern\p@}}}%
     {%
      \@enlargepage\@empty}%
}
\gdef\@enlargepage#1#2{%
   \@tempskipa#2\relax
   \ifdim \@tempskipa>.5\maxdimen
     \@latexerr{Suggested\space extra\space height\space
                (\the\@tempskipa)\space dangerously\space
                large}\@eha
   \else
     \ifdim \vsize<.5\maxdimen
       \@bsphack
         \insert\@kludgeins{#1\vskip-\@tempskipa}%
       \@esphack
     \else
       \@latexerr{Page\space height\space already\space
                  too\space large}\@eha
     \fi
   \fi
}
\makeatother

\input amssym.def 
\input amssym
\hfuzz=5.0pt
%
%
%
%
\def\vec#1{\mathchoice{\mbox{\boldmath$\displaystyle\bf#1$}}
{\mbox{\boldmath$\textstyle\bf#1$}}
{\mbox{\boldmath$\scriptstyle\bf#1$}}
{\mbox{\boldmath$\scriptscriptstyle\bf#1$}}}
\def\mbf#1{{\mathchoice {\hbox{$\rm\textstyle #1$}}
{\hbox{$\rm\textstyle #1$}} {\hbox{$\rm\scriptstyle #1$}}
{\hbox{$\rm\scriptscriptstyle #1$}}}}
\def\PartialsqPsi#1#2{\frac{\partial^2#1}{\partial{#2^2}}}
\def\operatorname#1{{\mathchoice{\rm #1}{\rm #1}{\rm #1}{\rm #1}}}
\chardef\ii="10
\def\widehat{\mathaccent"0362 }
\def\widetilde{\mathaccent"0365 }
\def\vphi{\varphi}
\def\vrho{\varrho}
\def\vtheta{\vartheta}
\def\ih{{\i\over\hbar}}
\def\hi{\frac{\hbar}{\i}}
\def\CD{{\cal D}}
\def\CE{{\cal E}}
\def\CH{{\cal H}}
\def\CL{{\cal L}}
\def\CP{{\cal P}}
\def\CV{{\cal V}}
\def\half{{1\over2}}
\def\bhalf{\hbox{$\half$}}
\def\viert{{1\over4}}
\def\bviert{\hbox{$\viert$}}
\def\hhbox#1#2{\hbox{$\frac{#1}{#2}$}}
\def\dfrac#1#2{\frac{\displaystyle #1}{\displaystyle #2}}
\def\intT{\ih\int_0^\infty\d\,T\,e^{\i ET/\hbar}}
\def\pathint#1{\int\limits_{#1(t')=#1'}^{#1(t'')=#1''}\CD #1(t)}
\def\hbarm{{\dfrac{\hbar^2}{2m}}}
\def\hbarmq{{\dfrac{\hbar^2}{2mq}}}
\def\mzwei{\dfrac{m}{2}}
\def\overh{\dfrac1\hbar}
\def\ihbar{\dfrac\i\hbar}
\def\intt{\int_{t'}^{t''}}
\def\tn{\tilde n}
\def\pmb#1{\setbox0=\hbox{#1}
    \kern-.025em\copy0\kern-\wd0
    \kern.05em\copy0\kern-\wd0
    \kern-.025em\raise.0433em\box0}
\def\pathintG#1#2{\int\limits_{#1(t')=#1'}^{#1(t'')=#1''}\CD_{#2}#1(t)}
\def\limN{\lim_{N\to\infty}}
\def\Norm{\bigg({m\over2\pi\i\epsilon\hbar}\bigg)}
\def\hbaram{{\hbar^2\over8m}}
\def\bbbr{{\rm I\!R}}                                
\def\bbbn{{\rm I\!N}}                                
\def\bbbz{{\mathchoice {\hbox{$\sf\textstyle Z\kern-0.4em Z$}}
{\hbox{$\sf\textstyle Z\kern-0.4em Z$}}
{\hbox{$\sf\scriptstyle Z\kern-0.3em Z$}}
{\hbox{$\sf\scriptscriptstyle Z\kern-0.2em Z$}}}}    
\def\bbbc{{\mathchoice {\setbox0=\hbox{\rm C}\hbox{\hbox
to0pt{\kern0.4\wd0\vrule height0.9\ht0\hss}\box0}}
{\setbox0=\hbox{$\textstyle\hbox{\rm C}$}\hbox{\hbox
to0pt{\kern0.4\wd0\vrule height0.9\ht0\hss}\box0}}
{\setbox0=\hbox{$\scriptstyle\hbox{\rm C}$}\hbox{\hbox
to0pt{\kern0.4\wd0\vrule height0.9\ht0\hss}\box0}}
{\setbox0=\hbox{$\scriptscriptstyle\hbox{\rm C}$}\hbox{\hbox
to0pt{\kern0.4\wd0\vrule height0.9\ht0\hss}\box0}}}}
\def\CC{{\cal C}}
\def\CP{{\cal P}}
\def\CT{{\cal T}}
\def\CQ{{\cal Q}}
\def\Ai{\operatorname{Ai}} 
\def\Cl{\operatorname{Cl}} 
\def\SU{\operatorname{SU}} 
\def\dt{\d t}
\def\d{\operatorname{d}}
\def\e{\operatorname{e}}
\def\i{\operatorname{i}}
\def\sn{\operatorname{sn}}
\def\cn{\operatorname{cn}}
\def\max{\operatorname{max}}
\def\DI{D_{\,\rm I}}
\def\DII{D_{\,\rm II}}
\def\3dDII{D_{\,3d-\rm II}}
\def\threedDII{D_{\,3d-\rm II}}
\def\DIII{D_{\,\rm III}}
\def\DIV{D_{\,\rm IV}}
\def\vphi{\varphi}
\def\tvphi{{\tilde\varphi}}
\def\tomega{{\tilde\omega}}
\def\ttau{{\tilde\tau}}
\def\hvphi{{\hat\varphi}}
\def\homega{{\hat\omega}}
\def\htau{{\hat\tau}}
\def\ps{\operatorname{ps}}
\def\Ps{\operatorname{Ps}}
\def\Si{\operatorname{Si}}
\def\energyldrei{\e^{-\i\hbar T(p^2+1)/2m}}
\def\ints{\int_0^{s''}}
\def\OO{\operatorname{O}}
\def\SO{\operatorname{SO}}
\def\operatorname#1{{\mathchoice{\rm #1}{\rm #1}{\rm #1}{\rm #1}}}
\def\bbbone{{\mathchoice {\rm 1\mskip-4mu l} {\rm 1\mskip-4mu l}
{\rm 1\mskip-4.5mu l} {\rm 1\mskip-5mu l}}}
\def\pathint#1{\int\limits_{#1(t')=#1'}^{#1(t'')=#1''}\CD #1(t)}
\def\pathints#1{\int\limits_{#1(0)=#1'}^{#1(s'')=#1''}\CD #1(s)}
 
\begin{titlepage}
\centerline{\normalsize DESY 06--139 \hfill ISSN 0418 - 9833}
\centerline{\hfill August 2006}

\vskip.3in
\message{TITLE:}
\begin{center}
{\bf\Large Path Integral Approach for Superintegrable Potentials on Spaces
\\[3mm]
of Non-constant Curvature: II. Darboux Spaces $\DIII$ and $\DIV$.}
\end{center}
\message{Path Integral Approach for Superintegrable Potentials
on Spaces of Non-constant Curvature: II. Darboux Spaces DIII and DIV.}
\vskip.2in
\begin{center}
{\large Christian Grosche}
\vskip.05in
{\normalsize\em II.\,Institut f\"ur Theoretische Physik}
\vskip.025in
{\normalsize\em Universit\"at Hamburg, Luruper Chaussee 149}
\vskip.025in
{\normalsize\em 22761 Hamburg, Germany}
\end{center}
\vskip.1in
\begin{center}
{\large George S.\,Pogosyan}
\vskip.05in
{\normalsize\em Laboratory of Theoretical Physics}
\vskip.025in
{\normalsize\em Joint Institute for Nuclear Research (Dubna)}
\vskip.025in
{\normalsize\em 141980 Dubna, Moscow Region, Russia}
\vskip.025in
{\normalsize\em and}
\vskip.025in
{\normalsize\em Departamento de Matematicas}
\vskip.025in
{\normalsize\em CUCEI, Universidad de Guadalajara}
\vskip.025in
{\normalsize\em Guadalajara, Jalisco, Mexico}
\vskip.1in
{\large Alexei N.\,Sissakian}
\vskip.05in
{\normalsize\em Laboratory of Theoretical Physics}
\vskip.025in
{\normalsize\em Joint Institute for Nuclear Research (Dubna)}
\vskip.025in
{\normalsize\em 141980 Dubna, Moscow Region, Russia}
\end{center}
\normalsize
\vfill
\begin{center}
{\bf Abstract}
\end{center}
This is the second paper on the path integral approach of superintegrable
systems on Darboux spaces, spaces of non-constant curvature.
We analyze in  the spaces $\DIII$ and $\DIV$ five respectively four
superintegrable potentials, which were first given by Kalnins et al.
We are able to evaluate the path integral in most of the separating
coordinate systems, leading to expressions for the Green functions,
the discrete and continuous wave-functions, and the discrete
energy-spectra. In some cases, however, the discrete spectrum cannot be stated
explicitly, because it is determined by a higher order polynomial equation.

We show that also the free motion in Darboux space of type III can
contain bound states, provided the boundary conditions are appropriate. 
We state the energy spectrum and the wave-functions, respectively.
\end{titlepage}
 
 
\thispagestyle{empty}%
\tableofcontents


\newpage\noindent\vspace{-1cm}\noindent%
\setcounter{page}{1}%
\setcounter{equation}{0}%
\thispagestyle{empty}%
\section{Introduction}%
\message{Introduction}%
In a previous publication \cite{GROPOe} we have started to study
superintegrable systems on spaces of non-constant curvature, i.e.
Darboux spaces. These spaces were introduced by Kalnins et al.
\cite{KalninsKMWinter,KalninsKWinter}. In the first paper we have studied the
Darboux spaces $\DI$ and $\DII$, and we continue our study by considering the
two other Darboux spaces $\DIII$ and $\DIV$ with five, respectively four
superintegrable potentials as determined in \cite{KalninsKMWinter}.

We find a rich structure of the spectrum of these potentials
yielding bound and continuous states. As it turns out, already the free motion
on $\DIII$ can give a positive continuous and an infinite 
negative discrete spectrum.
This situation is similar as for the quantum motion on the $SU(1,1)$
manifold \cite{BJb}, respectively on the $\SU(2,2)$ \cite{DORW} and
$SO(2,2)$ manifolds \cite{KAMI}.

The notion of superintegrable systems was introduced by Winternitz and
co-workers in \cite{FMSUW,WSUF}, Wojciechowski \cite{WOJ}, 
and was developed further later on also by Evans \cite{EVA}.
Superintegrable potentials have the 
property that one finds additional constants of motion. In two dimensions one
has in total three functional independent constants of motion and in three
dimensions one has four (minimal superintegrable) and five (maximal
superintegrable) functional independent constants of motion. Well-known
examples are the Coulomb potential with its Lenz--Runge vector 
and the harmonic oscillator with its quadrupole moment.
Another property of superintegrable potentials is that usually the
corresponding equations in classical and quantum mechanics separate in more
than one coordinate system. 

Similar studies of the quantum motion on spaces with and without curvature
have been investigated in \cite{GROPOa} for two- and three-dimensional flat
space, in \cite{GROPOb} for the two- and three-dimensional sphere, and in 
\cite{GROPOc} and \cite{GROPOd} for the two- and three-dimensional
hyperboloid. In all these cases the path integral method
\cite{FH,GRSh,SCHUHd,KLEo} was applied to find
the bound and continuous states, i.e., wave-functions and the explicit form of
the spectrum. We have not considered complexified spaces as in 
\cite{KMP2a} for the two-dimensional complex sphere
or \cite{KMP1}--\cite{KMP7} for the two-dimensional complex Euclidean space.
In particular, in \cite{KMP1} coordinate systems on the two-dimensional complex
sphere and corresponding superintegrable potentials, and
in \cite{KMP7} coordinate systems on the two-dimensional complex
plane and corresponding superintegrable potentials were discussed.
The goal of \cite{KMP1,KMP7} was to extend the notion of superintegrable
potentials of real spaces to the corresponding complexified spaces.
The findings were on the real two-dimensional Euclidean plane that there 
are three more coordinate systems and three more superintegrable potentials.
Similarly, in addition to the two coordinate systems on the real
two-dimensional sphere there are three more coordinate systems on the complex
sphere and four more superintegrable potentials.
This is not surprising because the complex plane contains not only the
Euclidean plane but also the pseudo-Euclidean plane (10 coordinate systems
\cite{GROad,KAL,KALc}) and the complex sphere contains not only 
the real sphere but also the two-dimensional hyperboloid (9 coordinate systems
\cite{GROad,KAL,KAMIb,OLE}). 

However, a complexified space is an abstract object. In order to obtain the
actual spectrum of a given potential formulated in a coordinate system one has
to consider a real version of the complexified space, e.g. the complex sphere:
One has to determine whether one considers the potential on the real sphere or
on the real hyperboloid. The complexification serves only as a tool for a
unified investigation.

Further studies on superintegrability in spaces with constant curvature are due
to \cite{KMP5,KMP4} (hyperboloid with new potentials),  \cite{KMP3}
(sphere and Euclidean space), \cite{KMP2a}, and \cite{KMP6} with a general
theory about the connection of separation in non-subgroup coordinate systems
of superintegrable systems and quasi-exactly-solvable problems \cite{USH}.

An extension of the study of path integration on spaces of constant 
curvature is the investigation of path integral formulations in spaces 
of non-constant curvature. Kalnins et al. 
\cite{KalninsKMWinter,KalninsKWinter} denoted four types of 
two-dimensional spaces of non-constant curvature, labeled by
$\DI$--$\DIV$, which are called Darboux spaces \cite{KOENIGS}. 
In terms of the infinitesimal distance they are described~by
(the coordinates $(u,v)$ will be called the $(u,v)$-system; 
the $(x,y)$-system in turn can be called light-cone coordinates):
\begin{eqnarray}
({\rm I})  \qquad  \d s^2&=& (x+y)\d x\d y
\nonumber\\
&=&2u(\d u^2+ \d v^2)\enspace,\qquad (x=u+\i v,y=u-\i v)\enspace,
\label{DarbouxI}
\\[2mm]
({\rm II}) \qquad \d s^2&=& \bigg(\frac{a}{(x-y)^2}+b\bigg) \d x\d y
\nonumber\\
 &=&\frac{bu^2-a}{u^2}(\d u^2+\d v^2)\enspace,\qquad
\Big(x=\bhalf(v+\i u), y=\bhalf(v-\i u)\Big)\enspace,
\label{DarbouxII}
\\[2mm]
({\rm III})\qquad \d s^2&=& \big(a\,\e^{-(x+y)/2}+b\,\e^{-x-y}\big)
\d x\d y
\nonumber\\
&=&\e^{-2u}(b+a\,\e^u)(\d u^2+\d v^2)\enspace,
\qquad (x=u-\i v,y=u+\i v)\enspace,
\label{DarbouxIII}
\\[2mm]
({\rm IV}) \qquad \d s^2&=& -\frac{a\big(\e^{(x-y)/2}+\e^{(y-x)/2}\big)+b}
{\big(\e^{(x-y)/2}-\e^{(y-x)/2}\big)^2}\d x\d y
\nonumber\\
&=&\left(\frac{a_+}{\sin^2u}+\frac{a_-}{\cos^2u}\right)(\d u^2+\d v^2)\,
\qquad (x=u+\i v,y=u-\i v)\enspace.
\label{DarbouxIV}
\end{eqnarray}
$a$ and $b$ are additional (real) parameters ($a_\pm=(a\pm 2b)/4$).
These surfaces are also called
surfaces of revolution~\cite{DASYPS,KKM,KalninsKMWinter}. Kalnins et
al. \cite{KalninsKMWinter,KalninsKWinter} studied not only the 
solution of the free motion, but also emphasized on the superintegrable
systems in theses spaces. 

The Gaussian curvature in a space with metric
$\d s^2=g(u,v)(\d u^2+\d v^2)$ is given by ($g=\det g(u,v)$)
\begin{equation}
G=-\frac{1}{2g}\bigg(\frac{\partial^2}{\partial u^2}
+\frac{\partial^2}{\partial v^2}\bigg)\ln g\enspace.
 \label{Gaussian-curvature}
\end{equation}
Equation (\ref{Gaussian-curvature}) will be used to discuss shortly the
curvature properties of the Darboux spaces, including their limiting cases of
constant curvature.

In the following sections we discuss superintegrable potentials in 
each of the two Darboux spaces $\DIII$ and $\DIV$, respectively.
We set up the classical Lagrangian and Hamiltonian, the quantum operator, and 
formulate and solve (if this is possible) the corresponding path integral.
We also discuss some of the limiting cases of the Darboux-spaces,
i.e. where we obtain a space of constant (zero or negative) curvature.
For the Darboux-space $\DIII$ the zero-curvature case 
$\bbbr^2$ emerges. In $\DIV$ we find a hyperboloid.

In the last section we summarize our results, where we include the
findings of our previous paper which dealt with superintegrable potentials on 
$\DI$ and $\DII$.

In the first two appendices we add some additional material about the path
integral evaluation of the free motion in $\DIV$ in degenerate elliptic
coordinates. In the third appendix we summarize briefly the path integral
investigation of some remaining superintegrable potentials on the
two-dimensional Euclidean plane. Finally, in the fourth appendix an example of
a potential on the two-dimensional complex sphere will be given.
\newpage\noindent%

\thispagestyle{empty}%
\setcounter{equation}{0}
\section{Superintegrable Potentials on Darboux Space $\DIII$}
\message{Superintegrable Potentials on Darboux Space D_III}
The coordinate systems to be considered in the Darboux space $\DIII$
are as follows:
\begin{eqnarray}
\hbox{($(u,v)$-System)}&& x=v+\i u,\quad y=v-\i u\enspace,
\\
\hbox{(Polar:)}&& \xi=\vrho\cos\vphi,\qquad \eta=\vrho\sin\vphi\enspace,\quad
\qquad\qquad\qquad\,\,\,(\vrho>0,\vphi\in[0,2\pi])\enspace,
\\
\hbox{(Parabolic:)}&& \xi=2\,\e^{-u/2}\cos\frac{v}{2}\qquad
\eta=2\,\e^{-u/2}\sin\frac{v}{2}\enspace,
\nonumber\\
&&u=\ln\frac{4}{\xi^2+\eta^2},\qquad
  v=\arcsin\frac{2\xi\eta}{\xi^2+\eta^2}\enspace,\quad
\quad\,\,\,\,(\xi\in\bbbr,\eta>0)\enspace,
\\
\hbox{(Elliptic:)}&& \xi=d\cosh\omega\cos\vphi,\qquad 
                    \eta=d\sinh\omega\sin\vphi\enspace,\quad
\quad(\omega>0,\vphi\in[-\pi,\pi])\,,\qquad
\\
\hbox{(Hyperbolic:)}&& 
\xi=\frac{\mu-\nu}{2\sqrt{\mu\nu}}+\sqrt{\mu\nu},\quad
\eta=\i\left(\frac{\mu-\nu}{2\sqrt{\mu\nu}}-\sqrt{\mu\nu}\right),
\quad(\mu,\nu>0)\enspace.
\end{eqnarray}
For the line element we get (we also display, where the metric is
rescaled in such a way that we set $a=b=1$ \cite{KalninsKMWinter}):
\begin{eqnarray}
\d s^2&=&\e^{-2u}(b+a\,\e^u)(\d u^2+\d v^2)=(\e^{-u}+\e^{-2u})(\d u^2+\d v^2)
\\
\hbox{(Polar:)}&=&(a+\hbox{$\frac{b}{4}$}\vrho^2)(\d\vrho^2+\vrho^2\d\vphi^2)
=(1+\bviert\vrho^2)(\d\vrho^2+\vrho^2\d\vphi^2)\,,
\\
\hbox{(Parabolic:)}
&=&(a+\hbox{$\frac{b}{4}$}(\xi^2+\eta^2))(\d\xi^2+\d\eta^2)
=(1+\bviert(\xi^2+\eta^2))(\d\xi^2+\d\eta^2)\,,
\\
\hbox{(Elliptic:)}
&=&(a+\hbox{$\frac{b}{4}$}d^2(\sinh^2\omega+\cos^2\vphi))d^2
   (\sinh^2\omega+\sin^2\vphi)(\d\omega^2+\d\vphi^2)\,,\qquad
\\
\hbox{(Hyperbolic:)}
&=&(a+\hbox{$\frac{b}{2}$}(\mu-\nu))(\mu+\nu)
\left(\frac{\d\mu^2}{\mu^2}-\frac{\d\nu^2}{\nu^2}\right)\,.
\end{eqnarray}
For the Gaussian curvature we find
\begin{equation}
G=-\frac{ab\,\e^{-3u}}{(b\,\e^{-2u}+a\,\e^{-u})^4}\enspace.
\end{equation}
For e.g. $a=1, b=0$ we recover two-dimensional flat space with the
corresponding coordinate systems. To assure the positive definiteness
of the metric (1.3), we require $a,b>0$.

\noindent
We introduce the following constants of motion on $\DIII$:
\begin{eqnarray}
X_1&=&\viert\frac{\e^{2u}}{a+b\,\e^u}\cos v\cdot p_u^2
     -\viert\frac{e^u(e^u+2)}{a+b\,\e^u}\cos v\cdot  p_v^2
     +\half\e^u\sin v\cdot p_up_v\enspace,
\\
X_2&=&\viert\frac{\e^{2u}}{a+b\,\e^u}\sin v\cdot  p_u^2
     -\viert\frac{e^u(e^u+2)}{a+b\,\e^u}\sin v\cdot  p_v^2
     +\half\e^u\cos v\cdot p_up_v\enspace,
\\
K&=&p_v\enspace.
\end{eqnarray}
These operators satisfy the Poisson relations
\begin{equation}
\{K,X_1\}=-X_2\enspace,\qquad
\{K,X_2\}=X_1\enspace,\qquad
\{X_1,X_2\}=K \tilde\CH_0\enspace,
\end{equation}
and the functional relation
\begin{equation}
X_1^2+X_2^2-\tilde\CH_0^2-\tilde\CH_0K^2=0\enspace.
\end{equation}
The operators $K,X_1,X_2$ can be used to characterize the separating
coordinate systems on $\DIII$, as indicated in Table \ref{cosytabDIII}.
The corresponding quantum operators are given by
\begin{eqnarray}
X_1&=&\bviert \e^u
     \Bigg[\frac{\e^{u}\cos v}{a+b\,\e^u}\cdot\partial_u^2
     -\frac{e^u+2}{a+b\,\e^u}\cos v \cdot\partial_v^2
     +(2\sin v\cdot\partial_u\partial_v
     +\cos v\cdot\partial_u+\sin v\cdot\partial_v)\Bigg],
\nonumber\\   &\\
X_2&=&\bviert\e^u
     \Bigg[\frac{\e^{u}\sin v}{a+b\,\e^u}\cdot\partial_u^2
     -\frac{e^u+2}{a+b\,\e^u}\sin v\cdot \partial_v^2
     -(2\cos v\cdot\partial_u\partial_v-\sin v\cdot\partial_u
           +\cos v\cdot\partial_v)\Bigg],
\nonumber\\   &\\
K&=&\partial_v\enspace.
\end{eqnarray}
\begin{table}[t!]
\caption{\label{cosytabDIII} 
Constants of Motion and Limiting Cases of Coordinate Systems on $\DIII$}
\begin{eqnarray}\begin{array}{l}\vbox{\small\offinterlineskip
\halign{&\vrule#&$\strut\ \hfil\hbox{#}\hfill\ $\cr
\noalign{\hrule}
height2pt&\omit&&\omit&&\omit&&\omit&\cr
&Metric:&&Constant of Motion &&$\DIII$ && $E_2$ ($a=1,b=0$)&\cr
height2pt&\omit&&\omit&&\omit&&\omit&\cr
\noalign{\hrule}\noalign{\hrule}
height2pt&\omit&&\omit&&\omit&&\omit&\cr
&$\e^{-2u}(b+a\,\e^u)(\d u^2+\d v^2)$  &&$K^2$
        &&$(u,v)$-System     &&Cartesian     &\cr
height2pt&\omit&&\omit&&\omit&&\omit&\cr
\noalign{\hrule}
height2pt&\omit&&\omit&&\omit&&\omit&\cr
&$(a+\hbox{$\frac{b}{4}$}\vrho^2)(\d\vrho^2+\vrho^2\d\vphi^2$
         &&$X_2$      &&Polar       &&Polar         &\cr
height2pt&\omit&&\omit&&\omit&&\omit&\cr
\noalign{\hrule}
height2pt&\omit&&\omit&&\omit&&\omit&\cr
&$(a+\hbox{$\frac{b}{4}$}(\xi^2+\eta^2))(\d\xi^2+\d\eta^2)$
         &&$X_1$      &&Parabolic   &&Parabolic     &\cr
height2pt&\omit&&\omit&&\omit&&\omit&\cr
\noalign{\hrule}
height2pt&\omit&&\omit&&\omit&&\omit&\cr
&$(a+\hbox{$\frac{b}{4}$}d^2(\sinh^2\omega+\cos^2\vphi))d^2$
         && && && &\cr
&$\times(\sinh^2\omega+\sin^2\vphi)(\d\omega^2+\d\vphi^2)$
         &&$d^2 X_1+2K^2$  &&Elliptic   &&Elliptic &\cr
height2pt&\omit&&\omit&&\omit&&\omit&\cr
\noalign{\hrule}}}\end{array}\nonumber\end{eqnarray}
\end{table}
These operators satisfy the commutation relations
\begin{equation}
[\widehat K,\widehat X_1]=-\widehat X_2\enspace,\qquad
[\widehat K,\widehat X_2]=\widehat X_1\enspace,\qquad
[\widehat X_1,\widehat X_2]=\widehat K \widehat H_0\enspace,
\end{equation}
and the relation
\begin{equation}
\widehat X_1^2+\widehat X_2^2
-\widehat H_0^2-\widehat H_0\widehat K^2+\bviert\widehat H_0=0\enspace.
\end{equation}
(Let us note that by $\tilde\CH_0$ the classical Hamiltonian without
the $1/2m$-factor is meant. Keeping this factor is no problem, however, 
in the present form the algebra is simpler).

We now state the superintegrable potentials on $\DIII$:
\begin{eqnarray}
V_1(u,v)&=&
\frac{2k_1\,\e^{-u}\cos\frac{v}{2}+2k_2\,\e^{-u}\sin\frac{v}{2}+k_3}
{a+\frac{b}{4}\,\e^{-u}}\enspace,
\\
V_2(u,v)&=&
\frac{1}{a+b\,\e^{-u}}\left[-\alpha
+\e^u\frac{\hbar^2}{8m}\bigg(\frac{k_1^2-\viert}{\cos^2\frac{v}{2}}
+\frac{k_1^2-\viert}{\cos^2\frac{v}{2}}\bigg)\right]\enspace,
\\
V_3(u,v)&=&
\frac{1}{a+b\,\e^{-u}}
\left[-\alpha+\frac{\hbar^2}{2m}4\e^{u}\Big(
c_1^2\,\e^{-\i v}-2c_2\,\e^{-2\i v}\Big)\right]\enspace,
\\
V_4(\mu,\nu)&=&
\frac{1}{(a+\frac{b}{2}(\mu-\nu))(\mu+\nu)}
\bigg[d_1\mu+d_2\nu+\frac{m}{2}\omega^2(\mu^2-\nu^2)\bigg]\enspace,
\\
V_5(u,v)&=&
\frac{1}{a+b\,\e^{-u}}\frac{\hbar^2v_0^2}{2m}\enspace.
\end{eqnarray}
In Table \ref{PotentialsDIII} we list the properties of these 
potentials on $\DIII$, where the coordinate systems were an explicit path
integral solution is possible are $\underline{\hbox{underlined}}$.
\begin{table}[h]
\caption{Separation of variables for the superintegrable potentials on $\DIII$}
\label{PotentialsDIII}
\begin{eqnarray}\begin{array}{l}\vbox{\small\offinterlineskip
\halign{&\vrule#&$\strut\ \hfil\hbox{#}\hfill\ $\cr
\noalign{\hrule}
height2pt&\omit&&\omit&&\omit&&\omit&&\omit&\cr
&Potential&&Constants of Motion &&Separating &\cr
&         &&                    &&coordinate &\cr
&         &&                    &&system &\cr
height2pt&\omit&&\omit&&\omit&&\omit&&\omit&\cr
\noalign{\hrule}\noalign{\hrule}
height2pt&\omit&&\omit&&\omit&&\omit&&\omit&\cr
&$V_1$    &&$R_1=X_1+
          \dfrac{2k_1\xi(2+\eta^2)-2k_2\eta(2+\xi^2)+k_3(\eta^2-\xi^2)}
               {4a+b(\xi^2+\eta^2)}\vphantom{\Bigg]}$
            &&$\underline{\hbox{Parabolic}}$            &\cr
&         &&$R_2=X_2+
          \dfrac{k_1\eta(\eta^2-\xi^2+4)+k_2\xi(\xi^2-\eta^2+4)-2k_3\xi\eta}
               {4a+b(\xi^2+\eta^2)}$
            &&$\underline{\hbox{Translated}}$ &\cr
&         &&&&$\underline{\hbox{parabolic}}$  &\cr
&         &&&&$\underline{\hbox{($\xi,\eta\to\xi\eta\pm c$)}}$ &\cr
height2pt&\omit&&\omit&&\omit&&\omit&&\omit&\cr
\noalign{\hrule}
height2pt&\omit&&\omit&&\omit&&\omit&&\omit&\cr
&$V_2$    &&$R_1=X_1+
          \dfrac{\hbox{$\frac{\hbar^2}{m}$}\Big((k_1^2-\bviert)\eta^2(\eta^2+2)
                -(k_2^2-\bviert)\xi^2(\xi^2+2)\Big)
                -\alpha(\eta^2-\xi^2)}
               {4a+b(\xi^2+\eta^2)}$
              &&$\underline{\hbox{$(u,v)$-System}}$     &\cr
&         &&$R_2=K^2+\dfrac{\hbar^2}{8m}
                \Big((k_1^2-\viert)\dfrac{\eta^2}{\xi^2}
            +(k_2^2-\viert)\dfrac{\xi^2}{\eta^2}\Big)\vphantom{\Bigg]}$
              &&$\underline{\hbox{Polar}}$              &\cr
&         &&  &&$\underline{\hbox{Parabolic}}$          &\cr
height2pt&\omit&&\omit&&\omit&&\omit&&\omit&\cr
\noalign{\hrule}
height2pt&\omit&&\omit&&\omit&&\omit&&\omit&\cr
&$V_3$    &&$R_1=X_1+\i X_2
  -\dfrac{-\alpha\mu^2\nu^2+c_1^2\mu\nu-2c_2(1+\mu-\nu)}
                  {(a+\hbox{$\frac{b}{2}$}(\mu-\nu))(\mu+\nu)}$
              &&$\underline{\hbox{Polar}}$             &\cr
&         &&$R_2=K^2-c_1^2\dfrac{\mu-\nu}{\mu\nu}
                    +c_2\dfrac{(\mu-\nu)^2}{\mu^2\nu^2}\vphantom{\Bigg]}$
              &&Hyperbolic                             &\cr
height2pt&\omit&&\omit&&\omit&&\omit&&\omit&\cr
\noalign{\hrule}
height2pt&\omit&&\omit&&\omit&&\omit&&\omit&\cr
&$V_4$    &&$R_1=X_1+\i X_2-K^2
               -\dfrac{\mu\nu\Big(d_1(\nu-2)+d_2(\mu+2)
                                 +m\omega^2(\nu-\mu+\mu\nu)\Big)}
                     {(a+\hbox{$\frac{b}{2}$}(\mu-\nu))(\mu+\nu)}$
              &&$\underline{\hbox{Hyperbolic}}$        &\cr
&         &&$R_1=X_1-\i X_2$                      &&   &\cr
&         &&$\qquad-\dfrac{(\mu-\nu)\Big((\mu-\nu)(d_1\mu+d_2\nu)
                     -m\omega^2(\mu^2+\nu^2+\mu\nu(2+\mu-\nu))\Big)}
                     {4(a+\hbox{$\frac{b}{2}$}(\mu-\nu))(\mu+\nu)}$
          &&Elliptic                                   &\cr
height2pt&\omit&&\omit&&\omit&&\omit&&\omit&\cr
\noalign{\hrule}
height2pt&\omit&&\omit&&\omit&&\omit&&\omit&\cr
&$V_5$    &&$R_1=X_1+\dfrac{\hbar^2v_0^2}{8m}
                     \dfrac{\eta^2-\xi^2}{a+\hbox{$\frac{b}{4}$}
                           (\xi^2+\eta^2)}\vphantom{\Bigg]}$
              &&$\underline{\hbox{$(u,v)$-System}}$     &\cr
&         &&$R_2=X_1-\dfrac{\hbar^2v_0^2}{4m}
     \dfrac{\xi\eta}{a+\hbox{$\frac{b}{4}$}(\xi^2+\eta^2)}\vphantom{\Bigg]}$
              &&$\underline{\hbox{Polar}}$              &\cr
&         &&$R_3=K=p_v$
              &&$\underline{\hbox{Parabolic}}$          &\cr
&         &&  &&Elliptic$\vphantom{\Big]}$              &\cr
&         &&  &&$\underline{\hbox{Hyperbolic}}$         &\cr
height2pt&\omit&&\omit&&\omit&&\omit&&\omit&\cr
\noalign{\hrule}}}\end{array}\nonumber\end{eqnarray}
\end{table}
We see that $V_5$ is a special case, and it has three integrals
of motion. We will threat this case in some more detail as in the
other spaces, because on $\DIII$ the free quantum motion can give 
bound state solutions (provided the constant are chosen properly).
This feature has not been discussed in~\cite{GROas}.

\subsection{The Superintegrable Potential $V_1$ on $\DIII$.}
\message{The Superintegrable Potential V_1 on D_III.}
We state the potential $V_1$ in the respective coordinate systems
\begin{eqnarray}
V_1(u,v)&=&
\frac{2k_1\e^{-u}\cos\frac{v}{2}+2k_2\e^{-u}\sin\frac{v}{2}+k_3}
{a+\frac{b}{4}\,\e^{-u}}\enspace,
\\      &=&
\frac{k_1\xi+k_2\eta+k_3}{a+\hbox{$\frac{b}{4}$}(\xi^2+\eta^2)}\enspace,
\\      &=&
\frac{k_1\xi+k_2\eta+(k_1c-k_2c+k_3)}
{a+\frac{b}{4}((\xi+c)^2+(\eta-c)^2)}
\enspace.
\end{eqnarray}
and $V_1$ is also separable in translated parabolic coordinates 
$\xi\to\xi+c,\eta\to\eta-c$. The translated parabolic coordinates 
just modifies the solution of a shifted harmonic
oscillator, and this case we do not discuss separately.

\subsubsection{Separation of $V_1$ in Parabolic Coordinates.}
The classical Lagrangian and Hamiltonian in parabolic coordinates on
$\DIII$ are given by 
\begin{eqnarray}
\CL(\xi,\dot\xi,\eta,\dot\eta)
&=&\frac{m}{2}\Big(a+\hbox{$\frac{b}{4}$}(\xi^2+\eta^2)\Big)
(\dot\xi^2+\dot\eta^2)-V(\xi,\eta),\quad
\\      
\CH(\xi,p_\xi,\eta, p_\xi)&=&
\frac{1}{2m}\frac{1}{a+\hbox{$\frac{b}{4}$}(\xi^2+\eta^2)}
            (p_\xi^2+p_\eta^2)+V(\xi,\eta)\enspace.
\end{eqnarray}
The canonical momenta are given by
\begin{equation}
p_\xi=\hi\bigg(\frac{\partial}{\partial\xi}
+\frac{b\xi}{a+\hbox{$\frac{b}{4}$}(\xi^2+\eta^2)}\bigg),\quad
p_\eta=\hi\bigg(\frac{\partial}{\partial\eta}
+\frac{b\eta}{a+\hbox{$\frac{b}{4}$}(\xi^2+\eta^2)}\bigg)\enspace.
\end{equation}
and for the quantum Hamiltonian (product ordering) we find
\begin{eqnarray}
H&=&-\frac{\hbar^2}{2m}\frac{1}{a+\hbox{$\frac{b}{4}$}(\xi^2+\eta^2)}
\bigg(\frac{\partial^2}{\partial\xi}+\frac{\partial^2}{\partial\eta^2}\bigg)
+V(\xi,\eta)\enspace,
\\
&=&\frac{1}{2m}\sqrt{\frac{1}{a+\hbox{$\frac{b}{4}$}(\xi^2+\eta^2)}}\,
(p_\xi^2+p_\eta^2)\sqrt{\frac{1}{a+\hbox{$\frac{b}{4}$}
(\xi^2+\eta^2)}}+V(\xi,\eta)\enspace.
\end{eqnarray}
Therefore we obtain for the path integral formulation for $V_1$
\begin{eqnarray}
&&\!\!\!\!\!\!\!\!\!\!
K^{(V_1)}(\xi'',\xi',\eta'',\eta';T)
=\pathint{\xi}\pathint{\eta}\Big(a+\hbox{$\frac{b}{4}$}(\xi^2+\eta^2)\Big)
\nonumber\\ &&\!\!\!\!\!\!\!\!\!\!\qquad\qquad\qquad\qquad\times
\exp\left\{\ih\int_0^T\left[
\Big(a+\hbox{$\frac{b}{4}$}(\xi^2+\eta^2)\Big)(\dot\xi^2+\dot\eta^2)
-\frac{k_1\xi+k_2\eta+k_3}{(a+\hbox{$\frac{b}{4}$}(\xi^2+\eta^2)}
 \right]\dt\right\}
\nonumber\\ &&\!\!\!\!\!\!\!\!\!\!
=\int_{-\infty}^\infty\frac{\d E}{2\pi\hbar}\,\e^{-\i ET/\hbar}
\int_0^\infty\d s''\exp\bigg[\ih\bigg(
aE-k_3-\frac{k_1^2+k_2^2}{2m\omega^2}\bigg)s''\bigg]
K^{(V_1)}(\xi'',\xi',\eta'',\eta';s'')\enspace,\qquad
\label{pathintegral-parabolic-1DIIIV1}
\end{eqnarray}
with the time-transformed path integral $K(s'')$ given by 
\begin{eqnarray}
&&K^{(V_1)}(\xi'',\xi',\eta'',\eta';s'')
=\pathints{\xi}\pathints{\eta}
\nonumber\\ &&\qquad\qquad\qquad\qquad\times
\exp\left\{\ih\ints \bigg[\frac{m}{2}\bigg((\dot\xi^2+\dot\eta^2)
-\frac{m}{2}\omega^2(\tilde\xi^2+\tilde\eta^2)\bigg)
\bigg]\d s\right\}\enspace.
\label{pathintegral-parabolic-2DIIIV1}
\end{eqnarray}
The transformed variable $\tilde\xi,\tilde\eta$ are given by
$\tilde\xi=\xi+k_1/m\omega^2,\tilde\eta=\eta+k_2/m\omega^2$, and
$\omega^2=-bE/2m$. Similarly as in \cite{GROas} we can determine the
Green function to have the form
\begin{eqnarray}
&&G^{(V_1)}(\xi'',\xi',\eta'',\eta';E)
=\int\d\CE\frac{m}{\pi\hbar^2b}\sqrt{-\frac{m}{2E}}\,
\Gamma\bigg(\half+\frac{\tilde\CE}{b\hbar}\sqrt{-\frac{m}{2E}}\,\bigg)
\Gamma\bigg(\half+\frac{\CE}{b\hbar}\sqrt{-\frac{m}{2E}}\,\bigg)
\nonumber\\ &&\qquad\qquad\times
D_{-\half+\frac{\tilde\CE}{b\hbar}\sqrt{-\frac{m}{2E}}}
\left(\sqrt[4]{-\frac{8mEb^2}{\hbar^2}}\,\tilde\xi_>\right)
D_{-\half+\frac{\tilde\CE}{b\hbar}\sqrt{-\frac{m}{2E}}}
\left(-\sqrt[4]{-\frac{8mEb^2}{\hbar^2}}\,\tilde\xi_<\right)
\nonumber\\ &&\qquad\qquad\times
D_{-\half+\frac{\CE}{b\hbar}\sqrt{-\frac{m}{2E}}}
\left(\sqrt[4]{-\frac{8mEb^2}{\hbar^2}}\,\tilde\eta_>\right)
D_{-\half+\frac{\CE}{b\hbar}\sqrt{-\frac{m}{2E}}}
\left(-\sqrt[4]{-\frac{8mEb^2}{\hbar^2}}\,\tilde\eta_<\right)\enspace.\qquad
\end{eqnarray}
The $D_\nu(z)$ are parabolic cylinder-functions \cite[p.1064]{GRA},
and  the $\tilde\CE$ is defined by
$\tilde\CE=aE-k_3-(k_1^2+k_2^2)/bE-\CE$.
On the other hand we can insert for the discrete part of the Green
function the harmonic oscillator wave-functions and obtain
\begin{eqnarray}
&&G^{(V_1)}_{\rm discrete}(\xi'',\xi',\eta'',\eta';E)
=\sum_{n_\xi=0}^\infty\sum_{n_\eta=0}^\infty
\frac{N_{n_\xi n_\eta}^2}{E_{n_\xi n_\eta}-E}
\nonumber\\ &&\qquad\qquad\times
\Psi_{n_\xi}^{(HO)}(\tilde\xi'')\Psi_{n_\xi}^{(HO)}(\tilde\xi')
\Psi_{n_\eta}^{(HO)}(\tilde\eta'')\Psi_{n_\eta}^{(HO)}(\tilde\eta')
\enspace.\qquad
\end{eqnarray}
The wave-functions for the harmonic oscillator are given by the
well-known form in terms of Hermite-polynomials \cite{GRA}
\begin{equation}
\Psi_n^{(HO)}(x)=
  \bigg({m\omega\over\pi\hbar}\bigg)^{1/4}\bigg({1\over2^nn!}\bigg)^{1/2}
   H_n\bigg(\sqrt{m\omega\over\hbar}\,x\bigg)
   \exp\bigg(-{m\omega\over2\hbar}x^2\bigg)\enspace.
\end{equation}
$E_{n_\xi n_\eta}$ is determined by the equation
\begin{equation}
aE-k_3-\frac{k_1^2+k_2^2}{2m\omega^2}
-\hbar(n_\xi+n_\eta+1)\sqrt{-\frac{bE}{2m}}=0
\end{equation}
which is actually an equation of fourth order in $E$
\begin{eqnarray}
&&E_{n_\xi n_\eta}^4
+\bigg(\frac{b\hbar^2}{2ma^2}(n_\xi+n_\eta+1)^2-\frac{2k_3}{a}\bigg)
E_{n_\xi n_\eta}^3
\nonumber\\ &&\qquad\qquad
-\bigg(2\frac{k_1^2+k_2^2}{ab}-\frac{k_3^2}{a^2}\bigg)E_{n_\xi n_\eta}^2
+2k_3\frac{k_1^2+k_2^2}{a^2b}E_{n_\xi n_\eta}
-\frac{(k_1^2+k_2^2)^2}{a^2b^2}=0\enspace.
\end{eqnarray}
This equation we dot not solve.
Note that for $k_1=k_2=k_3=0$ a discrete spectrum emerges for the free
motion on $\DIII$, a feature which we will discuss in more detail in the
subsection for $V_5$. For the special case $k_1=k_2=0$ we obtain the solution
($N=n_\xi+n_\eta+1$)
\begin{equation}
E_{n_\xi n_\eta\pm}=
-\frac{b\hbar^2N^2}{4ma^2}+\frac{k_3}{a}\pm
\frac{1}{a}\sqrt{ \bigg(\frac{b\hbar^2N^2}{4am}\bigg)^2
                 -\frac{bk_3\hbar^2N^2}{2am}-k_3^2}\enspace.
\end{equation}
Note that $\omega_{n_\xi n_\eta}$ must be taken on
$\omega_{n_\xi n_\eta}=\sqrt{-bE_{n_\xi n_\eta}/2m}$.
The normalization $N_{n_\xi n_\eta}$ is determined
by the residuum in $G^{(V_1)}(E)$.
If one fixes the parameters $a$ and $b$ and the specific surface of revolution,
a more detailed investigation can be performed (special cases, limiting cases,
which sign of the square-root gives a positive definite Hilbert space, etc.).
Because we do not fix these parameters, we keep both signs of the
square root-expression (recall that the free motion on $\DIII$ allows 
already a discrete spectrum reaching to $-\infty$).

Note that for the translated parabolic coordinates, the variables 
$\tilde\xi,\tilde\eta$ are translated by $\pm c$, respectively, and
the quantity $\CE$ by an additional $Ebc^2/2$.

\subsection{The Superintegrable Potential $V_2$ on $\DIII$.}
\message{The Superintegrable Potential V_2 on D_III.}
We state the potential $V_2$ in the respective coordinate systems
\begin{eqnarray}
V_2(u,v)&=&
\frac{1}{a+b\,\e^{-u}}\left[-\alpha
+\e^u\frac{\hbar^2}{8m}\bigg(\frac{k_1^2-\viert}{\cos^2\frac{v}{2}}
+\frac{k_1^2-\viert}{\cos^2\frac{v}{2}}\bigg)\right]\enspace,
\\      &=&
\frac{1}{a+\frac{b}{4}\vrho^2}
\left[-\alpha+\frac{\hbar^2}{2m\vrho^2}
\bigg(\frac{k_1^2-\viert}{\cos^2\vphi}+\frac{k_2^2-\viert}{\sin^2\vphi}
\bigg)\right]\enspace,
\\      &=&
\frac{1}{a+\frac{b}{4}(\xi^2+\eta^2)}
\left[-\alpha+\frac{\hbar^2}{2m}
\bigg(\frac{k_1^2-\viert}{\xi^2}+\frac{k_2^2-\viert}{\eta^2}
\bigg)\right]\enspace,
\\      &=&
\frac{1}{a+b\,\e^{-u}}
\left[-\alpha+\frac{\hbar^2}{2md^2}
\bigg(\frac{k_1^2-\viert}{\cosh^2\omega\cos^2\vphi}
+\frac{k_2^2-\viert}{\sinh^2\omega\sin^2\vphi}\bigg)\right]\enspace.
\end{eqnarray}
$V_2$ is obviously separable in elliptic coordinates, but the
corresponding path integral is not solvable; this case will be omitted.

\subsubsection{Separation of $V_2$ in the $(u,v)$-System.}
The classical Lagrangian and Hamiltonian are given by:
\begin{eqnarray}
\CL(u,\dot u,v,\dot v)&=&\frac{m}{2}\frac{b+a\,\e^u}{\e^{2u}}
(\dot u^2+\dot v^2)-V(u,v),
\\     
\CH(u,p_u,v,p_v)&=&\frac{1}{2m}\frac{\e^{2u}}{b+a\,\e^u}(p_u^2+p_v^2)
+V(u,v)\enspace.
\end{eqnarray}
The canonical momenta are given by
\begin{equation}
p_u=\hi\bigg(\frac{\partial}{\partial u}
-\half\frac{a\,\e^{-u}+2b\,\e^{-2u}}{a\,\e^{-u}+b\,\e^{-2u}}\bigg),\quad
p_v=\hi\frac{\partial}{\partial v}\enspace,
\end{equation}
and for the quantum Hamiltonian we find
\begin{eqnarray}
H&=&-\frac{\hbar^2}{2m}\frac{1}{a\,\e^{-u}+b\,\e^{-2u}}
\bigg(\frac{\partial^2}{\partial u^2}+\frac{\partial^2}{\partial v^2}\bigg)
+V(u,v)\enspace,
\\
&=&\frac{1}{2m}\sqrt{ \frac{1}{a\,\e^{-u}+b\,\e^{-2u}} }\,
\Big(p_u^2+p_v^2\Big)
\sqrt{ \frac{1}{a\,\e^{-u}+b\,\e^{-2u}} }+V(u,v)\enspace.
\end{eqnarray}
Therefore we obtain for the path integral
($f(u)=a\,\e^{-u}+b\,\e^{-2u}$)
\begin{eqnarray}
&&K^{(V_2)}(u'',u',v'',v';T)=\pathint{u}\pathint{v}(a\,\e^{-u}+b\,\e^{-2u})
\nonumber\\ &&\qquad\qquad\times
\exp\Bigg(\ih\int_0^T\Bigg\{(a\,\e^{-u}+b\,\e^{-2u})(\dot u^2+\dot v^2) 
\nonumber\\ &&\qquad\qquad\qquad\qquad\qquad\qquad
-\frac{1}{a+b\,\e^{-u}}\Bigg[-\alpha
+\e^u\frac{\hbar^2}{8m}\Bigg(\frac{k_1^2-\viert}{\cos^2\frac{v}{2}}
+\frac{k_1^2-\viert}{\cos^2\frac{v}{2}}\Bigg)\Bigg]\Bigg\}\dt\Bigg)
\nonumber\\ &&
=\frac{1}{[f(u')f(u'')]^{1/4}}
\sum_{l=0}^\infty\Phi_l^{(k_2,k_1)}(\hbox{$\frac{v''}{2}$})
\Phi_l^{(k_2,k_1)}(\hbox{$\frac{v'}{2}$})
\nonumber\\ &&\qquad\times
\pathint{u}(a\,\e^{-u}+b\,\e^{-2u})^{1/2}
\exp\Bigg(\ih\int_0^T\Bigg\{(a\,\e^{-u}+b\,\e^{-2u})\dot u^2 
\nonumber\\ &&\qquad\qquad\qquad\qquad\qquad\qquad
-\frac{1}{a+b\,\e^{-u}}\Bigg[-\alpha
+\e^u\frac{\hbar^2}{8m}\big(2l+1+|k_1|+|k_2|\big)\Bigg]
\Bigg\}\dt\Bigg)\qquad\qquad
\nonumber\\ &&
=\int_{-\infty}^\infty\frac{\d E}{2\pi\hbar}\,\e^{-\i ET/\hbar}
\int_0^\infty\d s''
\exp\bigg[-\ih\frac{\hbar^2}{8m}\big(2l+1+|k_1|+|k_2|\big)^2s''\bigg]
K_l^{(V_2)}(u'',u';s'')\enspace,
\end{eqnarray}
with the time-transformed path integral $K_l(s'')$ given by 
\begin{eqnarray}
K_l^{(V_2)}(u'',u';s'')
=\pathints{u}\exp\left[\ih\ints \bigg(\frac{m}{2}\dot u^2
+Eb\,\e^{-2u}+(aE-\alpha)\,\e^{-u}\bigg)\d s\right]\,.\qquad\quad
\label{pathintegral-DIIIV2u-v}
\end{eqnarray}
The $\Phi_n^{(k_1,k_2)}(\beta)$ are the wave-functions of the 
P\"oschl--Teller potential, which are given by 
\begin{eqnarray}
  V(x)&=&\hbarm\bigg(
  {\alpha^2-{1\over4}\over\sin^2x}+{\beta^2-{1\over4}\over\cos^2x}\bigg)
           \\  
  \Phi_n^{(\alpha,\beta)}(x)
  &=&\bigg[2(\alpha+\beta+2l+1)
  {l!\Gamma(\alpha+\beta+l+1)\over\Gamma(\alpha+l+1)\Gamma(\beta+l+1)}
  \bigg]^{1/2}
  \nonumber\\   &&\qquad\qquad\times
  (\sin x)^{\alpha+1/2}(\cos x)^{\beta+1/2}
  P_n^{(\alpha,\beta)}(\cos2x)\enspace.
\end{eqnarray}
Equation (\ref{pathintegral-DIIIV2u-v}) is a path integral for the Morse
potential. Inserting the corresponding solution \cite{GRSh} we obtain
\begin{eqnarray}
&&G^{(V_2)}(u'',u',v'',v';E)
=\sum_{l=0}^\infty\Phi_l^{(k_2,k_1)}(\hbox{$\frac{v''}{2}$})
\Phi_l^{(k_2,k_1)}(\hbox{$\frac{v'}{2}$})
\nonumber\\ &&\qquad\times
\sqrt{-\frac{m}{2bE}}\,
\frac{m\Gamma\Big(\half+\lambda+\frac{aE-\alpha}{\hbar}\sqrt{-m/2bE}\,\Big)}
{\hbar\Gamma(1+2\lambda)\,\e^{(u'+u'')/2}}
\nonumber\\ &&\qquad\times
W_{\frac{aE-\alpha}{\hbar}\sqrt{-m/2bE},\lambda}
\bigg(\frac{\sqrt{-8mbE}}{\hbar}\,\e^{-u_<}\bigg)
M_{\frac{aE-\alpha}{\hbar}\sqrt{-m/2bE},\lambda}
\bigg(\frac{\sqrt{-8mbE}}{\hbar}\,\e^{-u_>}\bigg)\enspace.\qquad\qquad
\label{Green-DIIIV2u-v}
\end{eqnarray}
Inserting the bound state wave-functions for the Morse-potential gives
the bound state contribution of $G^{(V_2)}(E)$
\begin{eqnarray}
&&G_{\rm discrete}^{(V_2)}(u'',u',v'',v';E)
=\sum_{l=0}^\infty\Phi_l^{(k_2,k_1)}(\hbox{$\frac{v''}{2}$})
\Phi_l^{(k_2,k_1)}(\hbox{$\frac{v'}{2}$})
\sum_{l=0}^\infty\frac{N_{nl}^2}{E_{nl}-E}
\Psi^{(MP)}_n(u'')\Psi^{(MP)}_n(u')\enspace,
\\    &&
\Psi^{(MP)}_n(u)=N_{nl}\left[\Bigg(-\frac{2mbE_{nl}}{\hbar^2}\Bigg)
^{\frac{aE_{nl}-\alpha}{\hbar}\sqrt{-{m}/{2bE_{nl}}}-n-1/2}\cdot
\frac{\Big(\frac{aE_{nl}-\alpha}{\hbar}\sqrt{-\frac{m}{2bE_{nl}}}-2n-1\Big)}
     {\Gamma\Big(\frac{aE_{nl}-\alpha}{\hbar}\sqrt{-\frac{2m}{bE_{nl}}}-n\Big)}
\right]^{1/2}
\nonumber\\ &&\qquad\times
\exp\left[(u'+u'')\left(
\frac{aE_{nl}-\alpha}{\hbar}\sqrt{-\frac{m}{2bE_{nl}}}-n-\half\right)
-\frac{\sqrt{-2mbE_{nl}}}{\hbar}\,\e^u\right]
\nonumber\\ &&\qquad\times
L_n^{(\frac{aE_{nl}-\alpha}{\hbar}\sqrt{-{2m}/{bE_{nl}}}-2n-1)}
\left(\frac{-8mbE_{nl}}{\hbar}\,\e^u\right)\enspace.
\end{eqnarray}
The $L_n^{(\alpha)}(z)$ are Laguerre polynomials \cite{GRA}.
Here, the spectrum $E_{nl}$ is determined by
\begin{equation}
aE_{nl}-\alpha-\hbar\sqrt{-\frac{bE_{nl}}{2m}}\,
(2n+2l+|k_1|+|k_2|+2)\enspace,
\end{equation}
which is a quadratic equation in $E_{nl}$ with solution
$(N=2n+2l+|k_1|+|k_2|+2$)
\begin{equation}
E_{nl\pm}=\frac{1}{2a^2}\left[
-\left(\frac{b\hbar^2}{2m}N^2-2a\alpha\right)
\pm\frac{b\hbar^2}{2m}N^2
\sqrt{1-\frac{8a\alpha m}{b\hbar^2N^2}}\,\right]\enspace,
\label{DIIIV2Enl}
\end{equation}
and the  the normalization constants $N_{nl}$ are determined by the residuum of
(\ref{Green-DIIIV2u-v}).
For large $n,l$ we have
\begin{eqnarray}
E_{nl-}&\simeq&-\frac{b\hbar^2}{m}(2n+2l+|k_1|+|k_2|+2)^2\enspace,
\\
E_{nl+}&\simeq&-\frac{m\alpha^2}{2b\hbar^2(2n+2l+|k_1|+|k_2|+2)^2}\enspace,
\end{eqnarray}
with $E_{nl+}$ showing a Coulomb-like behavior.

\subsubsection{Separation of $V_2$ in Polar Coordinates}
In the coordinates $(\vrho,\vphi)$ the classical Lagrangian and Hamiltonian
take on the form
\begin{eqnarray}
\CL(\vrho,\dot\vrho,\vphi,\dot\vphi)
&=&\frac{m}{2}(a+\hbox{$\frac{b}{4}$}\vrho^2)
(\dot \vrho^2+\vrho^2\dot\vphi^2)-V(\vrho,\vphi),
\\
\CH(\vrho,p_\vrho,\vphi,p_\vphi)&=&\frac{1}{2m}
\frac{1}{a+\hbox{$\frac{b}{4}$}\vrho^2}
\bigg(p_\vrho^2+\frac{1}{\vrho^2}p_\vphi^2\bigg)+V(\vrho,\vphi)\enspace.
\end{eqnarray}
The canonical momenta are given by
\begin{equation}
p_\vrho=\hi\bigg(\frac{\partial}{\partial\vrho}
+\frac{b\vrho}{4a+b\vrho^2}+\frac{1}{2\vrho}\bigg),\quad
p_\vphi=\hi\frac{\partial}{\partial\vphi}\enspace.
\end{equation}
Therefore the quantum Hamiltonian is given by:
\begin{eqnarray}
H&=&-\frac{\hbar^2}{2m}\frac{1}{a+\hbox{$\frac{b}{4}$}\vrho^2}
\bigg(\frac{\partial^2}{\partial \vrho^2}
+\frac{1}{\vrho}\frac{\partial}{\partial\vrho}
+\frac{1}{\vrho^2}\frac{\partial^2}{\partial\vphi^2}\bigg)+V(\vrho,\vphi)
\\  &=&
\frac{1}{2m}\sqrt{\frac{1}{a+\hbox{$\frac{b}{4}$}\vrho^2}}\bigg(p_\vrho^2
+\frac{1}{\vrho^2}p_\vphi^2\bigg)  
\sqrt{\frac{1}{a+\hbox{$\frac{b}{4}$}\vrho^2}}+V(\vrho,\vphi)
      -(a+\hbox{$\frac{b}{4}$}\vrho^2)^{-1}\frac{\hbar^2}{8m\vrho^2}\enspace,
\end{eqnarray}
and in this case we have an additional quantum potential $\propto\hbar^2$.
This gives for the path integral 
($f(\vrho)=a+\hbox{$\frac{b}{4}$}\vrho^2$)
\begin{eqnarray}
&&K^{(V_2)}(\vrho'',\vrho',\vphi'',\vphi';T)
=\pathint{\vrho}\pathint{\vphi}f(\vrho)\vrho
\nonumber\\ &&\qquad\times
\exp\Bigg(\ih\int_0^T\Bigg\{\frac{m}{2}
f(\vrho)(\dot \vrho^2+\vrho^2\dot\vphi^2)
-\frac{1}{f(\vrho)}\Bigg[-\alpha+\frac{\hbar^2}{2m\vrho^2}
\bigg(\frac{k_1^2-\viert}{\cos^2\vphi}+\frac{k_2^2-\viert}{\sin^2\vphi}
-\viert\bigg)\Bigg]\Bigg\}\dt\Bigg)
\nonumber\\ &&
=\sum_{l=0}^\infty\Phi_l^{(k_2,k_1)}(\vphi'')\Phi_l^{(k_2,k_1)}(\vphi')
\frac{1}{[(\vrho'\vrho'')^2 f(\vrho')f(\vrho'')]^{1/4}}
\nonumber\\ &&\qquad\times
\pathint{\vrho}f^{1/2}(\vrho)
\exp\Bigg\{\ih\int_0^T\Bigg[\frac{m}{2}f(\vrho)\dot \vrho^2
-\frac{1}{f(\vrho)}
\Bigg(-\alpha+\frac{\hbar^2}{2m}\frac{\lambda^2-\viert}{\vrho^2}
\Bigg)\Bigg]\dt\Bigg\}
\nonumber\\ &&
=\frac{1}{\sqrt{\vrho'\vrho''}}
\sum_{l=0}^\infty\Phi_l^{(k_2,k_1)}(\vphi'')\Phi_l^{(k_2,k_1)}(\vphi')
\nonumber\\ &&\qquad\times
\int_{-\infty}^\infty\frac{\d E}{2\pi\hbar}\,\e^{-\i ET/\hbar}
\int_0^\infty\d s''\exp\bigg[\ih(aE-\alpha)s''\bigg]
K_l^{(V_2)}(\vrho'',\vrho';s'')\enspace,
\end{eqnarray}
with the time-transformed path integral $K_l(s'')$ given by 
($\lambda=2l+|k_1|+|k_2|+1)$
\begin{eqnarray}
&&K_l^{(V_2)}(\vrho'',\vrho';s'')
=\pathints{\vrho}\exp\left[\ih\ints \bigg(\frac{m}{2}\dot\vrho^2
+\frac{Eb}{4}\vrho^2-\frac{\hbar^2}{2m}\frac{\lambda^2-\viert}{\vrho^2}
\bigg)\d s\right]
\nonumber\\ 
&&=\frac{m\omega\sqrt{\vrho'\vrho''}}{\i\hbar\sin\omega s''}
\exp\bigg[-\frac{m\omega}{2\i\hbar}({\vrho'}^2+{\vrho''}^2)\cot\omega s''\bigg]
I_\lambda\bigg(\frac{m\omega \vrho'\vrho''}{\i\hbar\sin\omega s''}\bigg)
\enspace.
\label{pathintegral-DIIIV2rho-phi}
\end{eqnarray}
Performing the $s''$-integration yields the Green function
\begin{eqnarray}
&&G^{(V_2)}(\vrho'',\vrho',\vphi'',\vphi';E)
=\sum_{l=0}^\infty\Phi_l^{(k_2,k_1)}(\vphi'')\Phi_l^{(k_2,k_1)}(\vphi')
\nonumber\\ &&\qquad\times
\sqrt{-\frac{2m}{Eb}}\,
\frac{\Gamma\Big[\half\Big(1+\lambda-\frac{1}{\hbar}(aE-\alpha)\sqrt{-2m/bE}\,
\Big)\Big]}{\Gamma(1+\lambda)\,\sqrt{\vrho'\vrho''}}
\nonumber\\ &&\qquad\times
M_{\frac{aE-\alpha}{2\hbar}\sqrt{-\frac{2m}{bE}},\frac{\lambda}{2}}
\left(\frac{m}{\hbar}\sqrt{-\frac{bE}{2m}}\,\vrho_<^2\right)
M_{\frac{aE-\alpha}{2\hbar}\sqrt{-\frac{2m}{bE}},\frac{\lambda}{2}}
\left(\frac{m}{\hbar}\sqrt{-\frac{bE}{2m}}\,\vrho_>^2\right)
\enspace.\qquad\qquad
\end{eqnarray}
Inserting the expansion into Laguerre polynomial 
yields the discrete contribution of the Green-function
\begin{eqnarray}
&&G_{disc.}^{(V_2)}(\vrho'',\vrho',\vphi'',\vphi';E)
\nonumber\\ &&
=\frac{1}{\sqrt{\vrho'\vrho''}}
\sum_{l=0}^\infty\Phi_l^{(k_2,k_1)}(\vphi'')\Phi_l^{(k_2,k_1)}(\vphi')
\sum_{n=0}^\infty\frac{N_{nl}^2}{E_{nl}-E}
\Psi_n^{(RHO,\lambda)}(\vrho'')\Psi_n^{(RHO,\lambda)}(\vrho')\enspace.\qquad
\end{eqnarray}
The wave-functions for the radial harmonic oscillator 
$V(r)=\frac{m}{2}\omega^2-\frac{\hbar^2}{2m}\frac{\lambda^2-1/4}{r^2}$
have the form \cite{GRSh,PI}
\begin{equation}
\Psi_n^{(RHO,\lambda)}(r)
=\sqrt{\frac{2m}{\hbar}\frac{n!}{\Gamma(n+\lambda+1)}\,r}
  \bigg({m\omega\over\hbar}r\bigg)^{\lambda/2}
  \exp\bigg(-{m\omega\over2\hbar}r^2\bigg)
  L_n^{(\lambda)}\bigg({m\omega\over\hbar}r^2\bigg)
\end{equation}
The spectrum $E_{nl}$ is determined by
\begin{equation}
aE_{nl}-\alpha-\hbar\sqrt{-\frac{bE_{nl}}{2m}}\,
(2n+2l+|k_1|+|k_2|+2)\enspace,
\end{equation}
which is the same as in (\ref{DIIIV2Enl}).
In the wave-functions $\Psi_n^{(RHO,\lambda)}(\vrho)$ the quantity 
$\omega$ has to be taken on $\omega=\sqrt{-bE_{nl}/2m}$, and the 
the normalization constants $N_{nl}$ are determined by the residuum of
(\ref{pathintegral-DIIIV2rho-phi}).

\subsubsection{Separation of $V_2$ in Parabolic Coordinates}
We insert the potential $V_2$ into the path integral and obtain
($f=a+\hbox{$\frac{b}{4}$}(\xi^2+\eta^2)$)
\begin{eqnarray}
&&\!\!\!\!\!\!\!\!\!\!
K^{(V_2)}(\xi'',\xi',\eta'',\eta';T)
=\pathint{\xi}\pathint{\eta}f(\xi,\eta)
\nonumber\\ &&\!\!\!\!\!\!\!\!\!\!\qquad\qquad\times
\exp\left(\ih\int_0^T\left\{f(\xi,\eta)(\dot\xi^2+\dot\eta^2)
-\frac{1}{f(\xi,\eta)} \left[-\alpha+\frac{\hbar^2}{2m}
\bigg(\frac{k_1^2-\viert}{\xi^2}+\frac{k_2^2-\viert}{\eta^2}
\bigg)\right]\right\}\dt\right)
\nonumber\\ &&\!\!\!\!\!\!\!\!\!\!
=\int_{-\infty}^\infty\frac{\d E}{2\pi\hbar}\,\e^{-\i ET/\hbar}
\int_0^\infty\d s''\exp\bigg[\ih\big(aE-\alpha\big)s''\bigg]
K^{(V_2)}(\xi'',\xi',\eta'',\eta';s'')\enspace,\qquad
\label{pathintegral-parabolic-1DIIIV2}
\end{eqnarray}
with the time-transformed path integral $K^{(V_2)}(s'')$ given by 
($\omega^2=-bE/2m$)
\begin{eqnarray}
&&K^{(V_2)}(\xi'',\xi',\eta'',\eta';s'')
=\pathints{\xi}\pathints{\eta}
\nonumber\\ &&\qquad\qquad\times
\exp\left\{\ih\ints \bigg[\frac{m}{2}\bigg((\dot\xi^2+\dot\eta^2)
-\frac{m}{2}\omega^2(\xi^2+\eta^2)
-\frac{\hbar^2}{2m}
\bigg(\frac{k_1^2-\viert}{\xi^2}+\frac{k_2^2-\viert}{\eta^2}\bigg)
\bigg]\d s\right\}
\nonumber\\ &&
=\frac{m\omega\sqrt{\xi'\xi''}}{\i\hbar\sin\omega s''}
\exp\bigg[-\frac{m\omega}{\i\hbar\sin\omega s''}({\xi'}^2+{\xi''}^2
\cot\omega s''\bigg]
I_{k_2}\bigg(\frac{m\omega\xi'\xi''}{\i\hbar\sin\omega s''}\bigg)
\nonumber\\ &&\qquad\qquad\times
\frac{m\omega\sqrt{\eta'\eta''}}{\i\hbar\sin\omega s''}
\exp\bigg[-\frac{m\omega}{\i\hbar\sin\omega s''}({\eta'}^2+{\eta''}^2
\cot\omega s''\bigg]
I_{k_1}\bigg(\frac{m\omega\eta'\eta''}{\i\hbar\sin\omega s''}\bigg)
\enspace.
\label{pathintegral-parabolic-2DIIIV2}
\end{eqnarray}
Performing the $s''$-integration yields the Green function
($\tilde\CE=aE-\alpha-\CE$)
\begin{eqnarray}
&&G^{(V_2)}(\xi'',\xi',\eta'',\eta';E)
=\int\d\CE\sqrt{-\frac{2m}{bE}}\,
\frac{\Gamma[\half(1+|k_1|-\CE\sqrt{-2m/bE}/\hbar)]}
     {\hbar\Gamma(1+|k_1|)\sqrt{\xi'\xi''}}
\nonumber\\ &&\qquad\qquad\times
W_{\CE\sqrt{-2m/bE}/2\hbar,|k_1|/2}
\left(\frac{m}{\hbar}\sqrt{-\frac{bE}{2m}}\,\xi_>^2\right)
M_{\CE\sqrt{-2m/bE}/2\hbar,|k_1|/2}
\left(\frac{m}{\hbar}\sqrt{-\frac{bE}{2m}}\,\xi_<^2\right)
\nonumber\\ &&\qquad\times
\sqrt{-\frac{2m}{bE}}
\frac{\Gamma[\half(1+|k_2|-\tilde\CE\sqrt{-2m/bE}/\hbar)]}
     {\hbar\Gamma(1+|k_2|)\sqrt{\eta'\eta''}}
\nonumber\\ &&\qquad\qquad\times
W_{\tilde\CE\sqrt{-2m/bE}/2\hbar,|k_2|/2}
\left(\frac{m}{\hbar}\sqrt{-\frac{bE}{2m}}\,\eta_>^2\right)
M_{\tilde\CE\sqrt{-2m/bE}/2\hbar,|k_2|/2}
\left(\frac{m}{\hbar}\sqrt{-\frac{bE}{2m}}\,\eta_<^2\right)
\enspace.
\label{Green-DIIIV2xieta}\nonumber\\ &&
\end{eqnarray}
On the other we insert the expansion of the bound states of the
radial harmonic oscillator and obtain for the discrete spectrum
contribution of the Green function:
\begin{eqnarray}
&&G^{(V_2)}(\xi'',\xi',\eta'',\eta';E)
=\sum_{n_\xi=0}^\infty\sum_{n_\eta=0}^\infty
\frac{N_{n_\xi,n_\eta}^2}{E_{n_\xi,n_\eta}-E}
\nonumber\\ &&\qquad\qquad\times
\Psi_{n_\xi}^{(RHO,|k_1|)}(\xi'')\Psi_{n_\xi}^{(RHO,|k_2|)}(\xi')
\Psi_{n_\eta}^{(RHO,|k_2|)}(\eta'')\Psi_{n_\eta}^{(RHO,|k_1|)}(\eta')\,,
\end{eqnarray}
where the energy $E_{n_\xi,n_\eta}$ is determined by the equation
\begin{equation}
2n_\xi+2n_\eta+|k_1|+|k_2|+2=
\frac{aE_{n_\xi,n_\eta}-\alpha}{\hbar}
\sqrt{-\frac{2m}{bE_{n_\xi,n_\eta}}}\enspace,
\end{equation}
which is equivalent with (\ref{DIIIV2Enl}).
The  normalization constants $N_{n_\xi n_\eta}$ 
are determined by the residuum of (\ref{Green-DIIIV2u-v}), 
and $\omega$ in the $\Psi_{n_\xi}^{(RHO,|k_2|)}\Psi_{n_\eta}^{(RHO,|k_1|)}$
has to be taken on $\omega_{n_\xi,n_\eta}=
\sqrt{-bE_{n_\xi,n_\eta}/2m}$.

\subsection{The Superintegrable Potential $V_3$ on $\DIII$.}
\message{The Superintegrable Potential V_3 on D_III.}
First we state the potential $V_3$ in the respective coordinate systems
\begin{eqnarray}
V_3(u,v)&=&
\frac{1}{a+b\,\e^{-u}}
\left[-\alpha+\frac{\hbar^2}{2m}4\e^{u}\Big(
c_1^2\,\e^{-\i v}-2c_2\,\e^{-2\i v}\Big)\right]\enspace,
           \\   &=&
\frac{1}{a+\frac{b}{4}\vrho^2}
\left[-\alpha+\frac{\hbar^2}{2m\vrho^2}4\Big(
c_1^2\,\e^{-2\i\vphi}-2c_2\,\e^{-4\i\vphi}\Big)\right]
           \\   &=&
\frac{-\alpha(\mu+\nu)+c_1^2\dfrac{\mu+\nu}{\mu\nu}
-c_2\dfrac{\mu^2-\nu^2}{\mu^2\nu^2}}{(a+\frac{b}{2}(\mu-\nu))(\mu+\nu)}
\enspace.
\end{eqnarray}
In hyperbolic coordinates no closed solution can be obtained due to
the involved mixture of linear, quadratic, inverse-linear and 
inverse-quadratic terms. In polar coordinates the path integral in
$\vrho$ turns out to be a path integral for the radial harmonic oscillator.
Note that the $(u,v)$-system is equivalent to polar coordinates.

\subsubsection{Separation of $V_3$ in Polar Coordinates}
We insert the potential $V_3$ into the path integral 
and get ($f(\vrho)=a+\hbox{$\frac{b}{4}$}\vrho^2=\sqrt{g}$)
\begin{eqnarray}
&&\!\!\!\!\!\!\!\!\!\!\!
K^{(V_3)}(\vrho'',\vrho',\vphi'',\vphi';T)
=\pathint{\vrho}\pathint{\vphi}f(\vrho)\vrho
\nonumber\\ &&\!\!\!\!\!\!\!\!\!\!\!\qquad\times
\exp\Bigg(\ih\int_0^T\Bigg\{\frac{m}{2}
f(\vrho)(\dot \vrho^2+\vrho^2\dot\vphi^2)
-\frac{1}{f(\vrho)}\Bigg[
-\alpha+\frac{\hbar^2}{2m\vrho^2}4c_1^2\Big(
\e^{-4\i\vphi}-2\frac{c_2}{c_1^2}\,\e^{-2\i\vphi}-\bviert\Big)
\Bigg]\Bigg\}\dt\Bigg)
\nonumber\\ &&\!\!\!\!\!\!\!\!\!\!\!
=\sum_{l=0}^\infty\Phi_{[cMP],l}^{(c_1,c_2)}(\vphi'')
\Phi_{[cMP],l}^{(c_1,c_2)}(\vphi')
\frac{1}{[(\vrho'\vrho'')^2 f(\vrho')f(\vrho'')]^{1/4}}
\nonumber\\ &&\!\!\!\!\!\!\!\!\!\!\!\qquad\times
\pathint{\vrho}f^{1/2}(\vrho)
\exp\Bigg\{\ih\int_0^T\Bigg[\frac{m}{2}f(\vrho)\dot \vrho^2
-\frac{1}{f(\vrho)}
\Bigg(-\alpha
+\frac{\hbar^2}{2m}\frac{(l+\frac{2c_2}{c_1}+\half)^2-\viert}{\vrho^2}
\Bigg)\Bigg]\dt\Bigg\}
\nonumber\\ &&\!\!\!\!\!\!\!\!\!\!\!
=\frac{1}{\sqrt{\vrho'\vrho''}}
\sum_{l=0}^\infty
\Phi_{[cMP],l}^{(c_1,c_2)}(\vphi'')\Phi_{[cMP],l}^{(c_1,c_2)}(\vphi')
\nonumber\\ &&\!\!\!\!\!\!\!\!\!\!\!\qquad\times
\int_{-\infty}^\infty\frac{\d E}{2\pi\hbar}\,\e^{-\i ET/\hbar}
\int_0^\infty\d s''\exp\bigg[\ih(aE-\alpha)s''\bigg]
K_l^{(V_3)}(\vrho'',\vrho';s'')\enspace,
\end{eqnarray}
with the time-transformed path integral $K_l(s'')$ given by 
\begin{eqnarray}
&&K_l^{(V_3)}(\vrho'',\vrho';s'')
=\pathints{\vrho}\exp\Bigg[\ih\ints \Bigg(\frac{m}{2}\dot\vrho^2
+\frac{Eb}{4}\vrho^2-\frac{\hbar^2}{2m}
\frac{(l+\frac{2c_2}{c_1}+\half)^2-\viert}{\vrho^2}
\Bigg)\d s\Bigg]
\nonumber\\ 
&&=\frac{m\omega\sqrt{\vrho'\vrho''}}{\i\hbar\sin\omega s''}
\exp\bigg[-\frac{m\omega}{2\i\hbar}({\vrho'}^2+{\vrho''}^2)\cot\omega s''\bigg]
I_{l+\frac{2c_2}{c_1}+\half}
\bigg(\frac{m\omega \vrho'\vrho''}{\i\hbar\sin\omega s''}\bigg) 
\end{eqnarray}
By $\Phi_{[cMP],l}^{(c_1,c_2)}(\vphi)$ we denote the wave-functions of the
complex periodic Morse potential in the variable $\vphi$ with 
spectrum $E_l=\hbar^2(l+2\frac{c_2}{c_1}+\half)^2/2m$ 
\cite{BBRR,BQZ,MOAH,KMP7,Znojila,Znojil}, c.f. Appendix C:
\begin{eqnarray}
\Phi_{[cMP],l}^{(c_1,c_2)}(\vphi)
&=&\frac{(4\frac{c_2}{c_1}-2n-1)n!}{\Gamma(4\frac{c_2}{c_1}-2n)}
 \bigg(4\frac{c_2}{c_1}\bigg)^{4\frac{c_2}{c_1}-2n-1}
\nonumber\\ &&\qquad\times
 \exp\bigg[-2\i\bigg(2\frac{c_2}{c_1}-n-\bhalf\bigg)\vphi
  -2c_1\,\e^{-2\i\vphi}\bigg]
  L_n^{(4\frac{c_2}{c_1}-2n-1)}\big(4c_1\e^{-2\i\vphi}\big)\enspace.\qquad
\end{eqnarray}
Performing the $s''$-integration gives the Green function
\begin{eqnarray}
&&\!\!\!\!\!\!\!\!\!\!\!
G^{(V_3)}(\vrho'',\vrho',\vphi'',\vphi';E)
=\sum_{l=0}^\infty
\Phi_{[cMP],l}^{(c_1,c_2)}(\vphi'')\Phi_{[cMP],l}^{(c_1,c_2)}(\vphi')
\nonumber\\ &&\!\!\!\!\!\!\!\!\!\!\!\quad\times
\sqrt{-\frac{2m}{Eb}}\,
\frac{\Gamma\Big[\half\Big(l+2\frac{c_2}{c_1}+\hbox{$\frac{3}{2}$}
                 -\frac{1}{\hbar}(aE-\alpha)\sqrt{-2m/bE}
\Big)\Big]}{\Gamma(l+2\frac{c_2}{c_1}+\hbox{$\frac{3}{2}$})
\,\sqrt{\vrho'\vrho''}}
\nonumber\\ &&\!\!\!\!\!\!\!\!\!\!\!\quad\times
M_{\frac{aE-\alpha}{2\hbar}\sqrt{-\frac{2m}{bE}},
                  \half(l+2\frac{c_2}{c_1}+\half)}
\left(\frac{m}{\hbar}\sqrt{-\frac{bE}{2m}}\,\vrho_<^2\right)
M_{\frac{aE-\alpha}{2\hbar}\sqrt{-\frac{2m}{bE}},
                 \half(l+2\frac{c_2}{c_1}+\half)}
\left(\frac{m}{\hbar}\sqrt{-\frac{bE}{2m}}\,\vrho_>^2\right)
\,.\qquad\quad
\label{Green-DIIIV3rho-phi}
\end{eqnarray}
Inserting the expansion into Laguerre polynomials 
yields the discrete contribution of the Green-function
($\lambda=l+\frac{2c_2}{c_1}+\half$)
\begin{eqnarray}
&&\!\!\!\!\!\!\!\!\!\!\!
G_{disc.}^{(V_3)}(\vrho'',\vrho',\vphi'',\vphi';E)
\nonumber\\ &&\!\!\!\!\!\!\!\!\!\!\! 
=\frac{1}{\sqrt{\vrho'\vrho''}}
\sum_{l=0}^\infty
\Phi_{[cMP],l}^{(c_1,c_2)}(\vphi'')\Phi_{[cMP],l}^{(c_1,c_2)}(\vphi')
\sum_{n=0}^\infty\frac{N_{nl}^2}{E_{nl}-E}
\Psi_n^{(RHO,\lambda)}(\vrho'')\Psi_n^{(RHO,\lambda)}(\vrho')\,.\qquad
\end{eqnarray}
and the  the normalization constants $N_{nl}$ are determined by the residuum of
(\ref{Green-DIIIV3rho-phi}).
Here, the spectrum $E_{nl}$ is determined by
\begin{equation}
aE_{nl}-\alpha-\hbar\sqrt{-\frac{bE_{nl}}{2m}}\,
                \bigg(2n+2l+\frac{c_2}{c_1}+1\bigg)\enspace.
\end{equation}
which is quadratic equation in $E_{nl}$ with solution 
$(N=2n+2l+\frac{c_2}{c_1}+1$)
\begin{equation}
E_{nl\pm}=\frac{1}{2a^2}\left[
-\left(\frac{b\hbar^2}{2m}N^2-2a\alpha\right)
\pm\frac{b\hbar^2}{2m}N^2
\sqrt{1-\frac{8a\alpha m}{b\hbar^2N^2}}\,\right]\enspace,
\label{DIIIV3Enl}
\end{equation}
In the wave-functions $\Psi_n^{(RHO,\lambda)}(\vrho)$ the quantity 
$\omega$ has to be taken on $\omega=\sqrt{-bE_{nl}/2m}$.
For large $n,l$ we have
\begin{eqnarray}
E_{nl-}&\simeq&-\frac{b\hbar^2}{m}(2n+2l+1)^2\enspace,
\\
E_{nl+}&\simeq&-\frac{m\alpha^2}{2b\hbar^2(2n+2l+1)^2}\enspace,
\end{eqnarray}
with $E_{nl+}$ showing a Coulomb-like behavior.

\subsection{The Superintegrable Potential $V_4$ on $\DIII$.}
\message{The Superintegrable Potential V_4 on D_III.}
\begin{eqnarray}
V_4(\mu,\nu)&=&
\frac{1}{(a+\frac{b}{2}(\mu-\nu))(\mu+\nu)}
\bigg[d_1\mu+d_2\nu+\frac{m}{2}\omega^2(\mu^2-\nu^2)\bigg]
         \\   &=&
\frac{1}{a+b\,\e^{-u}}\bigg[2(d_1+d_2)(\cos2\vphi-\cosh2\omega)+
2(d_1-d_2)(2\i\sin2\vphi+\sinh2\omega)
\nonumber\\   &&\qquad\qquad\qquad\qquad\qquad\qquad
+2d_3(2\i\sin2\vphi+\sinh4\omega)\bigg]\enspace.
\end{eqnarray}
We can evaluate the path integral in hyperbolic coordinates (application of
the Morse potential); in elliptic coordinates no closed solution  can be found.

\subsubsection{Separation of $V_4$ in Hyperbolic Coordinates}
The classical Lagrangian and Hamiltonian have the form
\begin{eqnarray}
\CL(\mu,\dot\mu,\nu,\dot\nu)&=&
\frac{m}{2}\Big(a+\hbox{$\frac{b}{2}$}(\mu-\nu)\Big)(\mu+\nu)
\bigg(\frac{\dot\mu^2}{\mu^2}-\frac{\dot\nu^2}{\nu^2}\bigg)-V(\mu,\nu),
\\
\CH(\mu,p_\mu,\nu,p_\nu&=&\frac{1}{2m}
\frac{\mu^2p_\mu^2-\nu^2p_\nu^2}
     {\Big(a+\hbox{$\frac{b}{2}$}(\mu-\nu)\Big)(\mu+\nu)}
+V(\mu,\nu)\enspace.
\end{eqnarray}
The canonical momentum operators are given by
\begin{eqnarray}
p_\mu&=&\hi\bigg[\frac{\partial}{\partial\mu}+\half\bigg(
+\frac{1}{\mu+\nu}+\frac{b}{a+\hbox{$\frac{b}{2}$}
              (\mu-\nu)}-\frac{1}{\mu}\bigg)\bigg],\\
p_\nu&=&\hi\bigg[\frac{\partial}{\partial\mu}+\half\bigg(
+\frac{1}{\mu+\nu}-\frac{b}{a+\hbox{$\frac{b}{2}$}(\mu-\nu)}
-\frac{1}{\nu}\bigg)\bigg]\enspace,
\end{eqnarray}
and the quantum Hamiltonian has the form
\begin{eqnarray}
H&=&-\frac{\hbar^2}{2m}\frac{1}{\Big(a+\hbox{$\frac{b}{2}$}(\mu-\nu)\Big)(\mu+\nu)}
\left[\mu^2\bigg(\frac{\partial^2}{\partial\mu^2}-\frac{1}{\mu}
\frac{\partial}{\partial\mu}\bigg)
-\nu^2\bigg(\frac{\partial^2}{\partial\nu^2}-\frac{1}{\nu}
\frac{\partial}{\partial\nu}\bigg)\right]+V(\mu,\nu)\qquad
\\
&=&\frac{1}{2m}\left[
\frac{\mu}{\sqrt{\Big(a+\hbox{$\frac{b}{2}$}(\mu-\nu)\Big)(\mu+\nu)}}p_\mu^2
\frac{\mu}{\sqrt{\Big(a+\hbox{$\frac{b}{2}$}(\mu-\nu)\Big)(\mu+\nu)}}\right.
\nonumber\\  &&\qquad\qquad\left.
-\frac{\nu}{\sqrt{\Big(a+\hbox{$\frac{b}{2}$}(\mu-\nu)\Big)(\mu+\nu)}}p_\nu^2
\frac{\nu}{\sqrt{\Big(a+\hbox{$\frac{b}{2}$}(\mu-\nu)\Big)(\mu+\nu)}}\right]
+V(\mu,\nu)\enspace.
\end{eqnarray}
Note that from each coordinate there comes a quantum potential
$\Delta V=\hbar^2/8m$, however they are canceling each other due 
to the minus-sign in the metric in $\nu$. 

We insert the potential $V_4$ into the path integral which has the
form ($f(\mu,\nu)=\Big(a+\hbox{$\frac{b}{2}$}(\mu-\nu)\Big)(\mu+\nu)$)
\begin{eqnarray}
&&\!\!\!\!\!\!\!\!\!\!
K^{(V_4)}(\mu'',\mu',\nu'',\nu';T)
=\pathint{\mu}\pathint{\nu}\frac{f(\mu,\nu)}{\mu\nu}
\nonumber\\  &&\!\!\!\!\!\!\!\!\!\!\qquad\times
\exp\left\{\ih\int_0^T\left[\frac{m}{2}f(\mu,\nu)
\bigg(\frac{\dot\mu^2}{\mu^2}-\frac{\dot\nu^2}{\nu^2}\bigg)
-\frac{1}{f(\mu,\nu)}
\bigg(d_1\mu+d_2\nu+\frac{m}{2}\omega^2(\mu^2-\nu^2)\bigg)\right]
\dt\right\}
\nonumber\\  &&\!\!\!\!\!\!\!\!\!\!
=\int_{-\infty}^\infty\frac{\d E}{2\pi\hbar}\,\e^{-\i ET/\hbar}
\int_0^\infty\d s''K^{(V_4)}(\mu'',\mu',\nu'',\nu';s'')\enspace,
\end{eqnarray}
and the path integral $K^{(V_4)}(s'')$ is given by
\begin{eqnarray}
&&\!\!\!\!\!\!\!\!\!\!
K^{(V_4)}(\mu'',\mu',\nu'',\nu';s'')
=\pathints{\mu}\pathints{\nu}\frac{1}{\mu\nu}
\nonumber\\  &&\!\!\!\!\!\!\!\!\!\!\qquad\times
\exp\Bigg\{\ih\ints \Bigg[\frac{m}{2}
\bigg(\frac{\dot\mu^2}{\mu^2}-\frac{\dot\nu^2}{\nu^2}\bigg)
\nonumber\\  &&\!\!\!\!\!\!\!\!\!\!
\qquad\qquad\qquad\qquad
+aE(\mu+\nu)+\half bE(\mu^2-\nu^2)
-\bigg(d_1\mu+d_2\nu+\frac{m}{2}\omega^2(\mu^2-\nu^2)\bigg)
\Bigg]\d s\Bigg\}\,.\qquad
\end{eqnarray}
Each of the last path integrals has a similar form as the one discussed in
\cite{GROas}. One can perform the transformation $\mu=\e^x$,
$\nu=\e^y$. Then the path-integration in ($\mu,\nu$) gives a
path-integration in ($x,y$) of the following form

\newpage\noindent%
\begin{eqnarray}
&&\!\!\!\!\!\!\!\!\!\!
K^{(V_4)}(x'',x',y'',y';s'')
\nonumber\\  &&\!\!\!\!\!\!\!\!\!\!\qquad
=\pathints{x}\exp\left\{\ih\ints \bigg[\frac{m}{2}\dot x^2-\half
(m\omega^2-bE)\,\e^{2x}-(d_1-aE)\,\e^x)\bigg]\d s\right\}
\nonumber\\  &&\!\!\!\!\!\!\!\!\!\!\qquad\times
\pathints{y}
\exp\left\{-\ih\ints \bigg[\frac{m}{2}\dot y^2-\half
(m\omega^2-bE)\,\e^{2y}-(d_2+aE)\,\e^y\bigg]\d s\right\}\,,\qquad
\end{eqnarray}
and we find the product of two path integrals for the Morse potential.
This can be evaluated as follows. We introduce the abbreviations
\begin{equation}
V_0^2=\frac{m}{\hbar^2}(m\omega^2-bE),\quad
\alpha_{x,y}=-\frac{d_{1,2}\mp aE}{m\omega^2-bE}\enspace.
\end{equation}
We expand each path integral into the discrete spectrum contribution by
means of the known solution of the Morse potential in terms of Laguerre
polynomials with the quantum numbers $n$ and $l$, respectively, and the
corresponding energy-spectra. The $s''$-integration gives 
the energy-spectrum
\begin{equation}
E_{n,l}=\frac{m\omega^2}{b}-\frac{m}{4b\hbar^2}
        \frac{(d_1+d_2)^2}{(n+l+1)^2}\enspace,
\end{equation}
together with the wave-functions ($N_{n,l}$ is determined by the
corresponding residuum)
\begin{eqnarray}
\Psi_{n,l}(x,y)&=&N_{n,l}\Psi_n^{(MP)}(x)\cdot\Psi_n^{(MP)}(y)
\\    
\Psi_k^{(MP)}(z)&=&
\bigg(\frac{2\alpha_z V_0-2k-1}{k!\Gamma(2\alpha_z V_0-k)}\bigg)^{1/2}
(2V_0)^{\alpha_z V_0-k-1/2}
\e^{(\alpha_z V_0-k-1/2)z-V_0\e^z}
L_k^{(2\alpha_z V_0-2k-1)}(2V_0\,\e^z)\enspace,
\nonumber\\ &&
\end{eqnarray}
for $z=x,y$ with $k=n,l$.
The continuous spectrum is examined in an analogous way yielding
\begin{equation}
E=\frac{\hbar^2p^2}{2m}\enspace,
\end{equation}
with the wave-functions
\begin{eqnarray}
\Psi_{p,\lambda}(x,y)&=&\Psi_{p,\lambda}^{(MP)}(x)\cdot
                        \Psi_{p,\lambda}^{(MP)}(y)
\\    
\Psi_{p,\lambda}^{(MP)}(z)&=&
\bigg(\frac{p_\pm\sinh2\pi p_\pm}{2\pi^2V_0}\bigg)^{1/2}
\Big|\Gamma(\i p_\pm-\alpha_z+\bhalf)\Big|\e^{-z}
W_{\alpha_zV_0,\i p_\pm}(2V_0\,\e^{x})\enspace.
\end{eqnarray}
with $p_\pm=p\pm\lambda$ for $z=x,y$. The entire Green function has the form
\begin{equation}
G(\mu'',\mu',\nu'',\nu';E)
=\sum_{n,l}\frac{\Psi_{n,l}(\mu'',\nu'')\Psi_{n,l}(\mu',\nu')}{E_{n,l}-E}
+\int\d p\int\d\lambda
\frac{\Psi_{p,\lambda}(\mu'',\nu'')\Psi_{p,\lambda}^*(\mu',\nu')}
{\frac{\hbar^2p^2}{2m}-E}\enspace,
\end{equation}
together with the replacement $\mu=\e^x$, $\nu=\e^y$. This concludes the
discussion. 

\subsection{The Superintegrable Potential $V_5$ on $\DIII$.}
\message{The Superintegrable Potential V_5 on D_III.}
We display the potential $V_5$ in the respective coordinate systems
\begin{eqnarray}
V_5(u,v)&=&\frac{1}{a+b\,\e^{-u}}\frac{\hbar^2v_0^2}{2m}
         \\ &=&
\frac{1}{a+\frac{b}{4}\vrho^2}\frac{\hbar^2v_0^2}{2m}
         \\ &=&
\frac{1}{a+\frac{b}{4}(\xi^2+\eta^2)}\frac{\hbar^2v_0^2}{2m}
         \\ &=&
\frac{1}{a+\frac{b}{4}d^2(\sinh^2\omega+\cos^2\vphi)}\frac{\hbar^2v_0^2}{2m}
         \\ &=&
\frac{1}{(a+\frac{b}{2}(\mu-\nu))(\mu+\nu)}\frac{\hbar^2v_0^2}{2m}\enspace.
\end{eqnarray}
We discuss the path integral solution of $V_5$ in some extend, where
the case of elliptic coordinates is omitted due to intractability of
this system in the path integral. Provided that $b>0$, there
is in the case of the free motion a discrete spectrum 
\begin{equation}
E_N=-\frac{\hbar^2}{2m}\frac{b}{a^2}(2N+1)^2\enspace,
\end{equation}
with the principal quantum number $N\in\bbbn$.

\subsubsection{Separation of $V_5$ in the $(u,v)$-System.}
We insert the potential $V_5$ into the path integral for the
$(u,v)$-system and obtain 
\begin{eqnarray}
&&K^{(V_5)}(u'',u',v'',v';T)
=\pathint{u}\pathint{v}(a\,\e^{-u}+b\,\e^{-2u})
\nonumber\\   &&\qquad\times
\exp\left\{\ih\int_0^T\bigg[\frac{m}{2}
(a\,\e^{-u}+b\,\e^{-2u})(\dot u^2+\dot v^2)
-\frac{1}{a+b\,\e^{-u}}\frac{\hbar^2v_0^2}{2m}\bigg]\dt\right\}
\nonumber\\
&&=\int_{-\infty}^\infty\frac{\d E}{2\pi\hbar}\,\e^{-\i ET/\hbar}
\int_0^\infty\d s''\e^{-\i\hbar v_0^2s''/2m}
K^{(V_5)}(u'',u',v'',v';s'')\enspace,
\end{eqnarray}
with the time-transformed path integral $K^{(V_5)}(s'')$ given by 
\begin{eqnarray}
&&\!\!\!\!\!\!\!\!\!\!\!
K^{(V_5)}(u'',u',v'',v';s'')
=\pathints{u}\pathints{v}
\nonumber\\   &&\!\!\!\!\!\!\!\!\!\!\!\qquad\times
\exp\left(\ih\ints \left\{\frac{m}{2}(\dot u^2+\dot v^2)
+Eb\left[\e^{-2u}+\bigg(\frac{aE-\hbar^2v_0^2/2m}{Eb}
\bigg)\e^{-u}\right]\right\}\d s\right)
\nonumber\\   &&\!\!\!\!\!\!\!\!\!\!\!
=\sum_{l=0}^\infty\frac{\e^{\i l(v''-v')}}{2\pi}\e^{-\i\hbar l^2s''/2m}
\nonumber\\   &&\!\!\!\!\!\!\!\!\!\!\!\qquad\times
\pathints{u}\exp\left(\ih\ints \left\{\frac{m}{2}\dot u^2
+Eb\left[\e^{-2u}+\bigg(\frac{aE-\hbar^2v_0^2/2m}{Eb}
\bigg)\e^{-u}\right]\right\}\d s\right)\,.\qquad
\end{eqnarray}
The path integral in $u$ is a path integral for the Morse potential.
Performing the $s''$-integration gives,  c.f.\cite{GROas}, the Green
function as follows ($\CE=[Ea-(\hbar^2v_0^2/2m)]\sqrt{-2m/bE}/2\hbar$)
\begin{eqnarray}
&&G^{(V_5)}(u'',u',v'',v';E)=\sum_{l=-\infty}^\infty
\frac{\e^{\i l(v''-v')}}{2\pi}
\frac{m\Gamma(\half+l-\CE)}
{\hbar\sqrt{-2mbE}\,\Gamma(1+2l)}\,\e^{(u'+u'')/2}
\nonumber\\ &&\qquad\qquad\qquad\qquad\qquad\times
W_{\CE,l}\bigg(\frac{\sqrt{-8mbE}}{\hbar}\,\e^{-u_<}\bigg)
M_{\CE,l}\bigg(\frac{\sqrt{-8mbE}}{\hbar}\,\e^{-u_>}\bigg)
\enspace.\qquad\qquad
\label{GDIIIV2uv}
\end{eqnarray}
The corresponding continuous part of the Green function is evaluated as
\cite{GROas}
\begin{eqnarray}
&&G_{cont.}^{(V_5)}(u'',u',v'',v';E)
=\sum_{l=-\infty}^\infty\frac{\e^{\i l(v''-v')}}{2\pi}\e^{(u'+u'')/2}
\nonumber\\ &&\qquad\qquad\times
\int_0^\infty\frac{\e^{\pi p/2}\d p}{\frac{\hbar^2p^2}{2m}-E}
\frac{|\Gamma(\half+l+\i p)|^2}{2\pi\Gamma^2(1+2l)}
M_{ \i p/2,l}\Big(-2\i p\,\e^{-u'}\Big)
M_{-\i p/2,l}\Big( 2\i p\,\e^{-u''}\Big)\enspace.\qquad
\end{eqnarray}
In addition, we have a discrete spectrum. This is found by analyzing the
poles of the Green function (\ref{GDIIIV2uv}):
\begin{equation}
\half+l-\frac{aE_{nl}-\frac{\hbar^2v_0^2}{2m}}{2\hbar}
\sqrt{-\frac{2m}{bE_{nl}}}=-n\enspace,
\end{equation}
In the case of $v_0=0$ this simplifies to
\begin{equation}
n+l+\bhalf-\frac{a}{2\hbar}\sqrt{-\frac{2m}{bE_{nl}}}=0\enspace,
\end{equation}
with the solution 
\begin{equation}
E_{nl}=-\frac{\hbar^2}{2m}\frac{b}{a^2}(2n+2l+1)^2\enspace.
\end{equation}
yielding for $b>0$ an infinite number of bound states.
For $v_0\not=0$ the equation for $E_{nl}$ is a quadratic equation in
$E$ with solution
\begin{eqnarray}
E_{nl\pm}&=&-\frac{\hbar^2}{2m}\frac{1}{2a^2}
\left[b(2n+2l+1)^2-2av_0^2\pm b(2n+2l+1)^2
\sqrt{1-\frac{4av_0^2}{b(2n+2l+1)^2}}\,\right]\enspace,
\nonumber\\ &&
         \\ 
E_{nl+}& \stackrel{(n,l)\to\infty}{=}&-\frac{\hbar^2}{2m}\frac{b}{a^2}
          \Bigg[(2n+2l+1)^2-2\frac{a}{b}v_0^2\Bigg]\enspace,
         \\  
E_{nl-}& \stackrel{(n,l)\to\infty}{=}&
        -\frac{\hbar^2}{2bm}\frac{v_0^4}{(2n+2l+1)^2}\enspace.
\end{eqnarray}
For $v_0=0$, there is only $E_{nl+}$. For $(2n+2l+1)^2<4av_0^2/b$ there 
are semi-bound states located approximately around $E_0=-\hbar^2v_0^2/2ma$.

\noindent
Therefore we have for the discrete spectrum contribution
\begin{eqnarray}
&&G_{\rm discrete}^{(V_5)}(u'',u',v'',v';E)
=\sum_{l=-\infty}^\infty\frac{\e^{\i l(v''-v')}}{2\pi} 
\sum_{n=0}^\infty\frac{1}{E_{nl}-E}
\Psi_{nl}^{(V_5)}(u'')\Psi_{nl}^{(V_5)}(u')\enspace,
\end{eqnarray}
with the functions $\Psi_{nl}^{(V_5)}(u)$ given by
($\CE$ as in (\ref{GDIIIV2uv}))
\begin{eqnarray}
&&\Psi_{nl}^{(V_5)}(u)
=N_{nl}\frac{(2\CE-2n-1)n!}{\Gamma(2\CE-n)}
\bigg(\frac{\sqrt{-8mbE_{nl}}}{\hbar}\bigg)^{\CE-n-1/2}
\nonumber\\ &&\qquad\qquad\qquad\qquad\qquad\times
\e^{(\CE-n-1/2)u-\sqrt{-8mbE_{nl}}\,\e^u/\hbar}
L_n^{(2\CE-2n-1)}\bigg(\frac{\sqrt{-8mbE_{nl}}}{\hbar}\,\e^u\bigg)
\enspace.\qquad
\end{eqnarray}
The constant $N_{nl}$ is determined by taking the Green function at
the residuum $E_{nl}$.
The wave-functions vanish for $u\to\infty$ due to 
$\e^{-\sqrt{-8mbE_{nl}}\,\e^u/\hbar}=\e^{-2b\hbar(2n+2l+1)\e^u/a}\to0$ 
for $u\to\infty$, provided $b/a>0$ for all $n\in\bbbn$, which shows that the
discrete spectrum is indeed infinite.\footnote{%
The feature that an homogeneous space with curvature has at the same
time a discrete and a continuous spectrum is already know from the
path integration on the $\SU(1,1)$ group manifold
\cite{GRSh}. Actually, this property allows the analysis of
the modified P\"oschl--Teller potential with its continuous and
(finite) discrete spectrum.}

\subsubsection{Separation of $V_5$ in Polar Coordinates.}
We insert the potential $V_5$ into the path integral in polar
coordinates and obtain:
\begin{eqnarray}
&&K^{(V_5)}(\vrho'',\vrho',\vphi'',\vphi';T)
=\pathint{\vrho}\pathint{\vphi}(a+\hbox{$\frac{b}{4}$}\vrho^2)\vrho
\nonumber\\ &&\qquad\qquad\times
\exp\left\{\ih\int_0^T\left[\frac{m}{2}
(a+\hbox{$\frac{b}{4}$}\vrho^2)(\dot \vrho^2+\vrho^2\dot\vphi^2)
+(a+\hbox{$\frac{b}{4}$}\vrho^2)^{-1}
\frac{\hbar^2}{2m}\bigg(v_0^2+\frac{1}{4\vrho^2}\bigg)\right]\dt\right\}
\qquad
\nonumber\\ &&
=\int_{-\infty}^\infty\frac{\d E}{2\pi\hbar}\,\e^{-\i ET/\hbar}
G^{(V_5)}(\vrho'',\vrho',\vphi'',\vphi';E)\enspace,
\end{eqnarray}
and the Green function is evaluated to have the form \cite{GROas}
($\CE=(aE-\frac{\hbar^2v_0^2}{2m})/\hbar\omega$, $\omega^2=-bE/2m$)
\begin{eqnarray}
&&G^{(V_5)}(\vrho'',\vrho',\vphi'',\vphi';E)
\sum_{l=-\infty}^\infty\frac{\e^{\i l(\vphi''-\vphi')}}{2\pi}
\frac{1}{\vrho'\vrho''}\cdot\sqrt{-\frac{2m}{2E}}\,
\frac{\Gamma[\half(1+l-\CE)]}{\Gamma(1+l)}
\nonumber\\  &&\qquad\qquad\qquad\times
W_{\CE/2,\frac{l}{2}}\left(\sqrt{-\frac{2mbE}{\hbar^2}}
\,\vrho_>\right)
M_{\CE/2,\frac{l}{2}}\left(\sqrt{-\frac{2mbE}{\hbar^2}}
\,\vrho_<\right)\enspace. 
\end{eqnarray}
The Green function has poles which are determined by
\begin{equation}
2n+l+1-\frac{1}{\hbar}\bigg(aE_{nl}-\frac{v_0^2\hbar^2}{2m}\bigg)
\sqrt{-\frac{2m}{Eb_{nl}}}=0\enspace.
\end{equation}
In the case of $v_0=0$ this simplifies to
\begin{equation}
(2n+l+1)-\frac{a}{\hbar}\sqrt{-\frac{2m}{E_{nl}b}}=0\enspace,
\label{DIIIV5Enl}
\end{equation}
with the solution 
\begin{equation}
E_{nl}=-\frac{\hbar^2}{2m}\frac{b}{a^2}(2n+l+1)^2\enspace.
\end{equation}
yielding for $b>0$ an infinite number of bound states.
For $v_0\not=0$ the equation for $E_{nl}$ is a quadratic equation in
$E$ with solution
\begin{equation}
E_{nl\pm}=-\frac{\hbar^2}{2m}\frac{1}{2a^2}
\left[b(2n+l+1)^2-2av_0^2\pm b(2n+l+1)^2
\sqrt{1-\frac{4av_0^2}{b(2n+l+1)^2}}\,\right]\enspace.
\end{equation}
The limit of $N,l\to\infty$ yields
\begin{eqnarray}
E_{nl+}&\simeq&-\frac{\hbar^2}{2m}\left[
\frac{b}{a^2}(2n+l+1)^2+\frac{v_0^2}{a}\right]\enspace,
\\
E_{nl-}&\simeq&-\frac{\hbar^2}{2m}\frac{v_0^2}{4b(2n+l+1)^2}\enspace,
\end{eqnarray}
and $E_{nl+}$ corresponds in this limit to the spectrum of the free motion.

\subsubsection{Separation of $V_5$ in Parabolic Coordinates.}
We insert the potential $V_5$ into the path integral in parabolic
coordinates and obtain:
\begin{eqnarray}
&&K^{(V_5)}(\xi'',\xi',\eta'',\eta';T)
=\pathint{\xi}\pathint{\eta}\Big(a+\hbox{$\frac{b}{4}$}(\xi^2+\eta^2)\Big)
\nonumber\\   &&\qquad\times
\exp\left\{\ih
\int_0^T\left[\frac{m}{2}(a+\hbox{$\frac{b}{4}$}
(\xi^2+\eta^2))(\dot\xi^2+\dot\eta^2)-
\frac{1}{a+\frac{b}{4}(\xi^2+\eta^2)}\frac{\hbar^2v_0^2}{2m}\right]\dt\right\}
\nonumber\\   &&
=\int_{-\infty}^\infty\frac{\d E}{2\pi\hbar}\,\e^{-\i ET/\hbar}
G^{(V_5)}(\xi'',\xi',\eta'',\eta';E)\enspace,
\end{eqnarray}
with the time-transformed path integral $K(s'')$ given by 
\begin{eqnarray}
&&K^{(V_5)}(\xi'',\xi',\eta'',\eta';s'')
=\pathints{\xi}\pathints{\eta}
\nonumber\\ &&\qquad\qquad\times
\exp\left\{\ih\ints \bigg[\frac{m}{2}(\dot\xi^2+\dot\eta^2)
+E\frac{b}{4}(\xi^2+\eta^2)\bigg]\d s+
\ih\bigg(aE-\frac{\hbar^2v_0^2}{2m}\bigg)\d s\right\}\enspace.\qquad
\end{eqnarray}
The only difference in comparison the the result in \cite{GROas} is
the the additional $\frac{\hbar^2v_0^2}{2m}$ term in the $s''$-integration.
In order to find the discrete spectrum we insert the solution for the
harmonic oscillator, and get
\begin{eqnarray}
&&G^{(V_5)}_{disc.}(\xi'',\xi',\eta'',\eta';E)
=\sum_{n_\xi=0}^\infty\sum_{n_\eta=0}^\infty
\frac{N_{n_\xi n_\eta}^2}{E_{n_\xi n_\eta}-E}
\Psi_{n_\xi}^{(HO)}(\xi'')\Psi_{n_\xi}^{(HO)}(\xi'')
\Psi_{n_\eta}^{(HO)}(\eta'')\Psi_{n_\eta}^{(HO)}(\eta'')\enspace,\qquad
\nonumber\\ &&
\end{eqnarray}
where $E_{n_\xi n_\eta}$ is determined by the equation
\begin{equation}
(n_\xi+n_\eta+1)-\frac{1}{\hbar}\bigg(aE-\frac{\hbar^2v_0^2}{2m}\bigg)
\sqrt{-\frac{bE}{2m}}=0\enspace.
\end{equation}
which is (up to a different counting in the quantum numbers) identical with
(\ref{DIIIV5Enl}). The normalization $N_{n_\xi n_\eta}$ is determined
by the residuum in $G^{(V_5)}(E)$.
The continuous spectrum part we do not state, it can be derived from
\cite{GROas} by the replacement $aE\to aE-{\hbar^2v_0^2}/{2m}$

\subsubsection{Separation of $V_5$ in Hyperbolic Coordinates.}
We insert the potential $V_5$ into the path integral in hyperbolic
coordinates and obtain:
The path integral has the form
\begin{eqnarray}
&&K^{(V_5)}(\mu'',\mu',\nu'',\nu';T)
=\pathint{\mu}\pathint{\nu}
\frac{\Big(a+\hbox{$\frac{b}{2}$}(\mu-\nu)\Big)(\mu+\nu)}{\mu\nu}
\nonumber\\  &&\qquad\times
\exp\left\{\ih\int_0^T\left[\Big(a+\hbox{$\frac{b}{2}$}(\mu-\nu)\Big)(\mu+\nu)
\bigg(\frac{\dot\mu^2}{\mu^2}-\frac{\dot\nu^2}{\nu^2}\bigg)
-\frac{1}{(a+\frac{b}{2}(\mu-\nu))(\mu+\nu)}\frac{\hbar^2v_0^2}{2m}\right]
\dt\right\}
\nonumber\\  
&=&\int_{-\infty}^\infty\frac{\d E}{2\pi\hbar}\,\e^{-\i ET/\hbar}
\int_0^\infty\d s''K^{(V_5)}(\mu'',\mu',\nu'',\nu';s'')\enspace,
\end{eqnarray}
and the path integral $K^{(V_5)}(s'')$ is given by
\begin{eqnarray}
&&\!\!\!\!\!\!\!\!\!\!
K^{(V_5)}(\mu'',\mu',\nu'',\nu';s'')
=\pathints{\mu}\pathints{\nu}\frac{1}{\mu\nu}
\nonumber\\  &&\!\!\!\!\!\!\!\!\!\!\qquad\times
\exp\left\{\ih\ints \left[\frac{m}{2}
\bigg(\frac{\dot\mu^2}{\mu^2}-\frac{\dot\nu^2}{\nu^2}\bigg)
+(\mu+\nu)\bigg(aE-\frac{\hbar^2v_0^2}{2m}\bigg)
+\half bE(\mu^2-\nu^2)\right]\d s\right\}\,.\qquad
\end{eqnarray}
Each of the last path integrals has a similar form as the one discussed in
\cite{GROb}. One can perform the transformation $\mu=\e^x$,
$\nu=\e^y$ yielding
\begin{eqnarray}
&&K^{(V_5)}(x'',x',y'',y';s'')=\pathints{x}
\exp\left\{\ih\ints \bigg[\frac{m}{2}\dot x^2
+\Big(E\hbox{$\frac{b}{2}$}\,\e^{2x}
+(aE-\hbox{$\frac{\hbar^2v_0^2}{2m}$})\,\e^x\Big)\bigg]\d s\right\}
\nonumber\\  &&\qquad\qquad\times
\pathints{y}\exp\left\{-\ih\ints \bigg[\frac{m}{2}\dot y^2
+\Big(E\hbox{$\frac{b}{2}$}\,\e^{2y}
-(aE-\hbox{$\frac{\hbar^2v_0^2}{2m}$})\,\e^y\Big)\bigg]\d s\right\}\qquad
\end{eqnarray}
and we find the product of two path integrals for the Morse potential,
however more complicated as in \cite{GROas}.
The continuous part of the spectrum can be analyzed similarly as in 
\cite{GROas} yielding products of $M$-Whittaker functions. 
Analyzing the discrete spectrum contribution from the Morse potential
we find the quantization condition
\begin{equation}
(n_\xi+n_\eta+1)-\frac{1}{\hbar}\bigg(aE-\frac{\hbar^2v_0^2}{2m}\bigg)
\sqrt{-\frac{4m}{E_{nl}b}}=0\enspace,
\end{equation}
which is up to a different counting in the quantum numbers equivalent
with (\ref{DIIIV5Enl}). This concludes the discussion.


\newpage\noindent%

\thispagestyle{empty}%
\setcounter{equation}{0}
\setcounter{equation}{0}
\section{Superintegrable Potentials on Darboux Space $\DIV$}
\message{Superintegrable Potentials on Darboux Space D_IV}
Finally, we consider the Darboux space $\DIV$.  We have the coordinate systems:
\begin{eqnarray}
\hbox{($(u,v)$-System:)}&& x=v+\i u,\quad y=v-\i u\enspace,
\qquad\qquad\qquad\quad\,\,
(u\in(0,\hbox{$\frac{\pi}{2}$}),v\in\bbbr)\enspace,
\\
\hbox{(Equidistant:)}&& 
u=\arctan(\e^\alpha),\quad v=\frac{\beta}{2}\enspace,
\qquad\qquad\qquad\quad\,\,
(\alpha\in\bbbr,\beta\in\bbbr)\enspace,
\\
\hbox{(Horospherical:)}&& 
x=\log\frac{\mu-\i\nu}{2},\quad 
y=\log\frac{\mu+\i\nu}{2}\enspace,\qquad\qquad
(\mu,\nu>0)\enspace,
\\   &&
\mu=2\e^v\cos u,\quad \nu=-2\e^v\sin u\enspace,
\\
\hbox{(Elliptic:)}&& 
\mu=d\cosh\omega\cos\vphi,\quad \nu=d\sinh\omega\sin\vphi\enspace,\quad\,\,\,
(\omega>0,\vphi\in(0,\hbox{$\frac{\pi}{2}$}))\enspace.\qquad
\end{eqnarray}
We obtain the following forms of the line-element ($a>2b$,
$a_\pm=(a\pm 2b)/4$):
\begin{eqnarray}
\d s^2&=&\frac{2b\cos u +a}{4\sin^2u}(\d u^2+\d v^2)
\nonumber\\
&=&
\left(\frac{a_+}{\sin^2u}+\frac{a_-}{\cos^2u}\right)(\d u^2+\d v^2)
\quad\hbox{(rescaling ${u\over2}\to u$:)}\enspace,\quad
\\
\hbox{(Equidistant:)}
&=&\frac{a-2b\tanh\alpha}{4}(\d\alpha^2+\cosh^2\alpha\d\beta^2)\enspace,
\\
\hbox{(Horospherical:)}
&=&\left(\frac{a_+}{\nu^2}+\frac{a_-}{\mu^2}\right)
(\d\mu^2+\d\nu^2)\enspace,
\\
\hbox{(Elliptic:)}
&=&\left(\frac{a_-}{\cosh^2\omega\cos^2\vphi}
         +\frac{a_+}{\sinh^2\omega\sin^2\vphi}\right)
(\cosh^2\omega-\cos^2\vphi)(\d\omega^2+\d\vphi^2)\enspace,
\nonumber\\
&=&\left(\frac{a_+}{\sin^2\vphi}+\frac{a_-}{\cos^2\vphi}
+\frac{a_+}{\sinh^2\omega}-\frac{a_-}{\cosh^2\omega}
\right)(\d\omega^2+\d\vphi^2)\,,
\\
\hbox{(Degenerate Elliptic I:)}
&=&\Bigg[a_-\bigg(\frac{1}{\sinh^2\homega}+\frac{1}{\sin^2\hvphi}\bigg)
-a_+\bigg(\frac{1}{\cosh^2\homega}-\frac{1}{\cos^2\hvphi}\bigg)\Bigg]
   (\d\homega^2+\d\hvphi^2),
\nonumber\\
&&(\gamma=1)\enspace.
\\
\hbox{(Degenerate Elliptic II:)}
&=&\viert\left(\frac{a_-}{\sinh^2\tomega}+\frac{a_+}{\sin^2\tvphi}\right)
   (\d\tomega^2+\d\tvphi^2)\enspace,\qquad(\gamma=2)\,. 
\end{eqnarray}
We observe that the diagonal term in the metric corresponds in most cases
to a combination of a P\"oschl--Teller potential and a modified
P\"oschl--Teller, respectively.  In particular, the $(u,v)$ and the
equidistant systems are the same, they just differ in the parameterization.
The limiting cases $a=2b$ and $b=0$ give particular cases for the metric on 
the two-dimensional hyperboloid. We have also displayed two versions of
degenerate elliptic coordinates. They come from the observation
that for the representatives
\begin{equation}
K^2,\qquad X_2,\qquad \gamma X_2+K2,\qquad X_1+X_2+\gamma K^2\enspace,
\end{equation}
one can distinguish the cases $\gamma=0$, $\gamma=2$, and $\gamma\not=0,2$.
For $\gamma\not=0,2$, one has coordinate systems which can be
explicitly formulated in terms of the elliptic functions
$\sn(\alpha,k),\cn(\beta,k)$, and only for a special choice of the 
parameter $k$ they can be simplified in trigonometric and hyperbolic functions.
Then the line element has the form
\begin{equation}
\d s^2=\bviert[a_+k^4\sn^2(\alpha,k)-\sn^2(\beta,k)+k^2a_-]
(\d\alpha^2+\d\beta^2)\enspace,
\end{equation}
and separated equations are versions of Lame's equation, if we assume an
Ansatz of the form $\Psi=A(\alpha)B(\beta)$ \cite{KalninsKWinter}:
\begin{eqnarray}
\PartialsqPsi{A(\alpha)}{\alpha}
+\left(-\bviert k^4Ea_+\sn^2(\alpha,k)-\lambda_1\right)A(\alpha) = 0\,, \\
\PartialsqPsi{B(\beta)}{\beta}
+\left(-\bviert k^4Ea_+\sn^2(\beta,k)-\lambda_2\right)B(\beta) = 0\,,
\end{eqnarray}
where $\lambda_1-\lambda_2=-Ea_-k^2/4$.  
$k$ denotes the modulus of the elliptic functions.

In particular for the potential $V_2$ one has the possibilities taking
$\gamma=0$, and $\gamma=2$. For $\gamma=0$, the modulus $k$ of the
elliptic functions equals $k=-\i$. We do not treat $V_2$ in these elliptic
coordinates, but only the degenerate case of $\gamma=2$.

For the potential $V_3$, however, the elliptic systems with $\gamma=1$ can be 
explicitly worked out. We have stated the respective line elements for these
two cases. Note that for $\gamma=2$ the coordinate transformation can be
put into
\begin{equation}
x=\ln\Big[\tan(\tvphi-\i\tomega)\Big],\quad
y=\ln\Big[\tan(\tvphi+\i\tomega)\Big],\quad
(\tomega>0,\tvphi\in(0,\hbox{$\frac{\pi}{4}$}))\enspace.
\end{equation}
We do not dwell into a discussion of elliptic systems any further, for details
we refer to \cite{KalninsKMWinter}.
Let us finally note that the notion $elliptic$ is also used for the
$(\omega,\vphi)$-system, and they must not be confused with the general
elliptic coordinates just discussed.

Because we have not worked out the path integral for the free motion 
in these two further coordinates systems, this will be done in the appendix, 

\noindent
For the Gaussian curvature we obtain e.g. in the $(u,v)$-system
\begin{equation}
G=-\dfrac{\frac{a_+^2}{\sin^6u}+\frac{a_-^2}{\cos^6u}
     +\frac{a_-a_+}{\sin^4u\cos^4u}}
    {\bigg(\frac{a_+}{\sin^2u}+\frac{a_-}{\cos^2u}\bigg)^3}
\enspace.
\end{equation}
The case $a=2b$ yields $a_-=0$, and
\begin{equation}
G=-\frac{1}{b}\enspace,
\end{equation}
and therefore again a space of constant curvature, the hyperboloid
$\Lambda^{(2)}$ is given for $b>0$. We have set the sign in the 
metric (1.4) in such a way that from $a=2b>0$ the hyperboloid
$\Lambda^{(2)}$ emerges. We could also choose the  metric (1.4) with
the opposite sign, then $a=2b<0$ would 
give the same result. In the following it is understood that we make
this restriction of positive definiteness of the metric and we do not
dwell into the problem of continuation into non-positive definiteness.
Because the $(u,v)$-coordinates and the equidistant system are the
same, we do not  evaluate the path integral in the equidistant system.
In the following we assume $a_+>0$ and $a_+>a_-$.

\begin{table}[t!]
\caption{\label{cosytab2} 
Constants of Motion and Limiting Cases of Coordinate Systems on $\DIV$}
\begin{eqnarray}\begin{array}{l}\vbox{\small\offinterlineskip
\halign{&\vrule#&$\strut\ \hfil\hbox{#}\hfill\ $\cr
\noalign{\hrule}
height2pt&\omit&&\omit&&\omit&&\omit&&\omit&\cr
&Metric:   &&Constant  &&$\DIV$   &&$\Lambda^{(2)},\ (a=2b$)          
                       &&$\Lambda^{(2)},\ (b=0$)             &\cr
&          &&of Motion &&         &&  &&                    &\cr
height2pt&\omit&&\omit&&\omit&&\omit&&\omit&\cr
\noalign{\hrule}\noalign{\hrule}
height2pt&\omit&&\omit&&\omit&&\omit&&\omit&\cr
&$\dfrac{2b\cos u +a}{4\sin^2u}(\d u^2+\d v^2)$
 &&$K^2$   &&$(u,v)$-System    &&Equidistant  &&Equidistant &\cr
height2pt&\omit&&\omit&&\omit&&\omit&&\omit&\cr
\noalign{\hrule}
height2pt&\omit&&\omit&&\omit&&\omit&&\omit&\cr
&$\bigg(\dfrac{a_+}{\nu^2}+\dfrac{a_-}{\mu^2}\bigg)
(\d\mu^2+\d\nu^2)$
  &&$X_2$  &&Horospherical   &&Horicyclic  &&Semi-circular  &\cr
& && &&                      &&            &&parabolic      &\cr
height2pt&\omit&&\omit&&\omit&&\omit&&\omit&\cr
\noalign{\hrule}
height2pt&\omit&&\omit&&\omit&&\omit&&\omit&\cr
&$\left(\dfrac{a_-}{\cosh^2\omega\cos^2\vphi}
         +\dfrac{a_+}{\sinh^2\omega\sin^2\vphi}\right)$
 &&$K^2+d^2 X_2$  
           &&Elliptic&&Elliptic- &&Hyperbolic-              &\cr
&$\quad\times(\cosh^2\omega-\cos^2\vphi)(\d\omega^2+\d\vphi^2)
 \vphantom{\Big]}$
 &&        &&          &&Parabolic  &&parabolic             &\cr
height2pt&\omit&&\omit&&\omit&&\omit&&\omit&\cr
\noalign{\hrule}\noalign{\hrule}
height2pt&\omit&&\omit&&\omit&&\omit&&\omit&\cr
&$\Big[a_+k^2\Big(\sn^2(\alpha,k)-\sn^2(\beta,k)\Big)+a_-\Big]$
 &&$X_1+X_2+\gamma K^2$
 &&Elliptic    &&Elliptic           &&Elliptic              &\cr
&$\quad\times\dfrac{k^2}{4}(\d^2\alpha+\d^2\beta)\vphantom{\Big]}$ 
 &&        &&          &&           &&                      &\cr
height2pt&\omit&&\omit&&\omit&&\omit&&\omit&\cr
\noalign{\hrule}}}\end{array}\nonumber\end{eqnarray}
\end{table}

We introduce the following three constants of motion on $\DIV$:
\begin{eqnarray}
X_1&=&\e^{2v}(-\tilde\CH_0+\cos2u\cdot p_u^2+\sin2u\cdot p_up_v)\enspace,
\\
X_2&=&\e^{2v}(-\tilde\CH_0+\cos2u\cdot p_u^2-\sin2u\cdot p_up_v)\enspace.
\\
K&=&p_v\enspace.
\end{eqnarray}
These integrals of motion satisfy the Poisson relations
\begin{equation}
\{K,X_1\}=2X_1\enspace,\qquad
\{K,X_2\}=-2X_2\enspace,\qquad
\{X_1,X_2\}=-K^3-4aK H_0\enspace,
\end{equation}
and satisfy the relation
\begin{equation}
X_1X_2-K^4-aK^2H_0-H_0^2=0\enspace.
\end{equation}
The corresponding quantum operators have the form
\begin{eqnarray}
\widehat H_0&=&\frac{\sin^22u}{2\cos 2u+a}(\partial_u^2+\partial_v^2)\enspace,
\\
\widehat X_1&=&\e^{2v}(-\widehat H_0+\cos2u\cdot (\partial_u^2+\partial_v)
+\sin2u\cdot (\partial_u\partial_v+\partial_u)\enspace,
\\
\widehat X_2&=&\e^{2v}(-\widehat H_0+\cos2u\cdot (\partial_u^2-\partial_v)
-\sin2u\cdot (\partial_u\partial_v-\partial_u)\enspace,
\end{eqnarray}
and the commutation relations read
\begin{equation}
[\widehat K,\widehat X_1]= 2\widehat X_1\enspace,\quad
[\widehat K,\widehat X_2]=-2\widehat X_2\enspace,\quad
[\widehat X_1,\widehat X_2]=-8\widehat K^3
-4a\widehat K \widehat H_0-4\widehat K\,,
\end{equation}
and satisfy the operator relation
\begin{equation}
\bhalf\{\widehat X_1,\widehat X_2\}-\widehat K^4
-a\widehat H_0\widehat K^2-5\widehat K^2-\widehat H_0^2
-a\widehat  H_0=0\enspace.
\end{equation}
In Table \ref{cosytab2} we list the connection with these operators and the
corresponding coordinate systems on $\DIV$.

\begin{table}[t]
\caption{Separation of variables for the superintegrable potentials on $\DIV$}
\label{PotentialsDIV}
\begin{eqnarray}\!\!\!\!\!\!\!\!
\begin{array}{l}\vbox{\small\offinterlineskip
\halign{&\vrule#&$\strut\ \hfil\hbox{#}\hfill\ $\cr
\noalign{\hrule}
height2pt&\omit&&\omit&&\omit&&\omit&&\omit&\cr
&Potential&&Constants of Motion &&Separating &\cr
&         &&                    &&coordinate &\cr
&         &&                    &&system &\cr
height2pt&\omit&&\omit&&\omit&&\omit&&\omit&\cr
\noalign{\hrule}\noalign{\hrule}
height2pt&\omit&&\omit&&\omit&&\omit&&\omit&\cr
&$V_1$    &&$R_1=K^2-\alpha(\mu^2+\nu^2)+
          \frac{m}{2}\omega^2(\mu^2+\nu^2)$
            &&$\underline{\hbox{$(u,v)$-System}}$            &\cr
&         &&$R_2=X_2+\dfrac{-2\alpha(a_+\mu^2-a_-\nu^2)
                    +8(k^2-\bviert)\hbox{$\frac{\hbar^2}{m}$}
                    +2m\omega^2(a_+\mu^4-a_-\nu^4)}
                  {a_+\mu^2+a_-\nu^2}$
            &&$\underline{\hbox{Horospherical}}$            &\cr
&         &&&&Elliptic                                      &\cr
height2pt&\omit&&\omit&&\omit&&\omit&&\omit&\cr
\noalign{\hrule}
height2pt&\omit&&\omit&&\omit&&\omit&&\omit&\cr
&$V_2$    &&$R_1=X_1+X_2+(2\cos u+a)^{-1}\dfrac{\hbar^2}{2m}
             \bigg[(k_1^2+k_2^2-\half)-2(k_3^2-\half)\cosh2v$
              &&$\underline{\hbox{$(u,v)$-System}}$         &\cr  
&         &&$\qquad\qquad+(\cos4u+2a\cos2u+3)
               \bigg(\dfrac{k_1^2-\bviert}{\sinh^2v}
                    -\dfrac{k_2^2-\bviert}{\cosh^2v}\bigg)\bigg]$
              &&$\underline{\hbox{Degenerate}}$                &\cr
&         &&$R_2=K^2+\dfrac{\hbar^2}{2m}
  \bigg(\dfrac{k_1^2-\bviert}{\sinh^2v}+\dfrac{k_2^2-\bviert}{\cosh^2v}\bigg)$
              &&$\underline{\hbox{elliptic I}}$    &\cr
height2pt&\omit&&\omit&&\omit&&\omit&&\omit&\cr
\noalign{\hrule}
height2pt&\omit&&\omit&&\omit&&\omit&&\omit&\cr
&$V_3$    &&$R_1=X_1+X_2+2K^2+a H+\dfrac{\hbar^2}{2m}            
  \bigg(\dfrac{a_+}{\sinh^22\tomega}+\dfrac{a_-}{\sinh^2\tomega}\bigg)^{-1}$  
           &&$\underline{\hbox{Degenerate}}$    &\cr
&          &&$\qquad\times\bigg[\dfrac{a_+}{\sinh^22\tomega}
     \Big(\dfrac{c_3}{\sin^2\tvphi}+\dfrac{c_1}{\sin^2\tvphi}\Big)
     +\dfrac{a_-}{\sinh^22\tvphi}
     \Big(\dfrac{c_3}{\sinh^2\tomega}-\dfrac{c_2}{\cos^2\tomega}\Big)\bigg]$  
              &&$\underline{\hbox{elliptic I \& II}}$    &\cr
&         &&$R_2=X_1-X_2
   +\dfrac{\hbar^2}{2m}
     \bigg(\dfrac{a_+}{\sinh^22\tomega}+\dfrac{a_-}{\sinh^2\tomega}\bigg)^{-1}$
                                                        &&  &\cr 
&         &&$\times\bigg[\dfrac{a_+}{\sinh^22\tomega}
            \bigg(c_1\cosh2\tomega\tan^2\tvphi-c_2\cos2\tvphi
            -\dfrac{c_3(2\cos^2\tvphi(\sinh^2\tomega-\sin^2\tvphi)+1}
                  {\sin^2\tvphi}\bigg)$                 &&  &\cr
&         &&$+\dfrac{a_-}{\sin^22\tvphi}
            \bigg(c_2\cos2\tvphi\tanh^2\tomega+c_1\cosh2\tomega
            -\dfrac{c_3(2\cosh^2\tomega(\sinh^2\tomega-\sin^2\tvphi)+1}
             {\sinh^2\tomega}\bigg)\bigg]\vphantom{\Bigg]}$ && &\cr
height2pt&\omit&&\omit&&\omit&&\omit&&\omit&\cr
\noalign{\hrule}
height2pt&\omit&&\omit&&\omit&&\omit&&\omit&\cr
&$V_4$    &&$R_1=X_1
      +\dfrac{2\hbox{$\frac{\hbar^2}{m}$}(k_0^2-\bviert)(\mu^2+\nu^2)}
                          {a_+\mu^2+a_-\nu^2}$
              &&$\underline{\hbox{$(u,v)$-System}}$         &\cr
&         &&$R_2=X_2+\dfrac{32\hbox{$\frac{\hbar^2}{m}$}(k_0^2-\bviert)}
                          {a_+\mu^2+a_-\nu^2}\vphantom{\Bigg]}$
              &&$\underline{\hbox{Horospherical}}$          &\cr
&         &&$R_3=\mu p_\mu+\nu p_\nu$
              &&$\underline{\hbox{Elliptic}}$               &\cr
height2pt&\omit&&\omit&&\omit&&\omit&&\omit&\cr
\noalign{\hrule}}}\end{array}\nonumber\end{eqnarray}
\end{table}

We state the superintegrable potentials on $\DIV$:
\begin{eqnarray}
V_1(u,v)&=&
\bigg(\frac{a_+}{\sin^2u}+\frac{a_-}{\cos^2u}\bigg)^{-1}
\left[\frac{\hbar^2}{2m}
\bigg(\frac{k^2-\viert}{\cos^2u}+\frac{k^2-\viert}{\sin^2u}\bigg)
-4\alpha\e^{2v}+8m\omega^2\e^{4v}\right]
\\
V_2(u,v)&=&
\bigg(\frac{a_+}{\sin^2u}+\frac{a_-}{\cos^2u}\bigg)^{-1}
\left[\frac{\hbar^2}{2m}\bigg(
\frac{k_1^2-\viert}{\sinh^2v}-\frac{k_2^2-\viert}{\cosh^2v}\bigg)
-\frac{\alpha}{4}\bigg(\frac{1}{\sin^2u}+\frac{1}{\cos^2u}\bigg)\right]\qquad
\\
V_3(\tomega,\tvphi)&=&
\frac{\hbar^2}{2m}\left( 
\frac{a_+}{\sinh^2\tomega}-\frac{a_+}{\cosh^2\tomega}
   +\frac{a_-}{\sin^2\tvphi}+\frac{a_-}{\cos^2\tvphi}\right)^{-1}
\nonumber\\   &&\qquad\qquad\times
\Bigg[\frac{c_3}{\sin^2\tvphi}+\frac{c_2}{\cos^2\tvphi}
-\frac{c_3}{\sinh^2\tomega}+\frac{c_2}{\cosh^2\tomega}\Bigg]\enspace,
\\
V_4(\mu,\nu)&=&
\bigg(\frac{a_+}{\nu^2}+\frac{a_-}{\mu^2}\bigg)^{-1}
\frac{\hbar^2}{2m}(k_0^2-\bviert)
\bigg(\frac{1}{\mu^2}+\frac{1}{\nu^2}\bigg)\enspace.
\end{eqnarray}

In Table \ref{PotentialsDIV} we list the properties of these 
potentials on $\DIV$. We see that $V_4$ is a special case, and it has
three integrals of motion.  The variables $\tomega,\tvphi$ are defined by 
\begin{equation}
x=\log[\tan(\tvphi-\i\tomega)]\enspace,\qquad
y=\log[\tan(\tvphi+\i\tomega)]\enspace.
\label{degenerate}
\end{equation}
In terms of these coordinates the line element is given by
\begin{eqnarray}
\d s^2&=&\frac{a+2b}{\sinh^22\tomega}+\frac{a+2b}{\sin^22\tvphi}
\nonumber\\   &=&
 \frac{a_+}{\sinh^2\tomega}-\frac{a_+}{\cosh^2\tomega}
-\frac{a_-}{\sin^2\tvphi}+\frac{a_-}{\cos^2\tvphi}\enspace.
\end{eqnarray}

\subsection{The Superintegrable Potential $V_1$ on $\DIV$.}
\message{The Superintegrable Potential V_1 on D_IV.}
We start by stating the potential $V_1$ in the respective 
coordinate systems
\begin{eqnarray}
V_1(u,v)&=&
\bigg(\frac{a_+}{\sin^2u}+\frac{a_-}{\cos^2u}\bigg)^{-1}
\left[\frac{\hbar^2}{2m}
\bigg(\frac{k^2-\viert}{\cos^2u}+\frac{k^2-\viert}{\sin^2u}\bigg)
-4\alpha\e^{2v}+8m\omega^2\e^{4v}\right]
\\      &=&
\bigg(\frac{a_+}{\nu^2}+\frac{a_-}{\mu^2}\bigg)^{-1}
\left[-\alpha+\frac{\hbar^2}{2m}\bigg(
\frac{k^2-\viert}{\mu^2}+\frac{k^2-\viert}{\nu^2}\bigg)
+\frac{m}{2}\omega^2(\mu^2+\nu^2)\right]
\\      &=&
d^2\bigg(\frac{a_+}{\sinh^2\omega\sin^2\vphi}+
\frac{a_-}{\cosh^2\omega\cos^2\vphi}\bigg)^{-1}
\nonumber\\      &&\qquad\times
\Bigg[-\alpha+\frac{\hbar^2}{2m}
\bigg(\frac{k^2-\bviert}{\sinh^2\omega\sin^2\vphi}
+\frac{k^2-\bviert}{\cosh^2\omega\cos^2\vphi}
\bigg)
+\frac{m}{2}\omega^2d^2(\cosh^2\omega-\sin^2\vphi)\Bigg].
\nonumber\\      &&
\end{eqnarray}
The path integral for the potential $V_1$ can be solved in the
$(u,v)$-system and in horospherical coordinates. We also keep the 
parameters $k_1$ and $k_2$ different in comparison to Kalnins et al.

\subsubsection{Separation of $V_1$ in the $(u,v)$-System.}
The classical Lagrangian and Hamiltonian are given by
\begin{eqnarray}
\CL(u,\dot u,v,\dot v)&=&\frac{m}{2}\frac{2b\cos2u+a}{\sin^22u}
(\dot u^2+\dot v^2)+V(u,v)\enspace,
\\
\CH(u,p_u,v,p_v)&=&\frac{1}{2m}\frac{\sin^22u}{2b\cos2u+a}(p_u^2+p_v^2)
+V(u,v)\enspace.
\end{eqnarray}
The canonical momentum operators are given by
\begin{equation}
p_u=\hi\bigg(\frac{\partial}{\partial u}+2\cot2u
-\frac{2b\sin2u}{2b\cos2u+a}\bigg),\quad 
p_v=\hi\frac{\partial}{\partial v}\enspace,
\end{equation}
and the Hamiltonian operator has the form
\begin{eqnarray}
H&=&-\frac{\hbar^2}{2m}\frac{\sin^22u}{2b\cos2u+a}
\bigg(\frac{\partial^2}{\partial u^2}
+\frac{\partial^2}{\partial v^2}\bigg)+V(u,v)
\\
&=&\frac{1}{2m}\frac{\sin2u}{\sqrt{2b\cos2u+a}}
(p_u^2+p_v^2)\frac{\sin2u}{\sqrt{2b\cos2u+a}}+V(u,v)\enspace.
\end{eqnarray}
We insert $V_1$ into the path integral and obtain 
($f={a_+}/{\sin^2u}+{a_-}/{\cos^2u}$)
\begin{eqnarray}
&&\!\!\!\!\!\!\!\!\!\!\!\!
K^{(V_1)}(u'',u',v'',v';T)
=\pathint{u}\pathint{v}f(u)
\nonumber\\  &&\!\!\!\!\!\!\times
\exp\Bigg(\ih\int_0^T\Bigg\{\frac{m}{2}f(\dot u^2+\dot v^2)
-\frac{1}{f}\Bigg[\frac{\hbar^2}{2m}
\bigg(\frac{k_1^2-\viert}{\cos^2u}-\frac{k_2^2-\viert}{\sin^2u}\bigg)
+8m\omega^2\bigg(\e^{4v}-\frac{\alpha}{2m\omega^2}\e^{2v}\bigg)
\Bigg]\Bigg\}\dt\Bigg).
\nonumber\\ 
\label{DIV-V1-uv}
\end{eqnarray}
We see that the $v$-dependence has the form of a Morse-potential
($\tilde\alpha=\alpha/4m\omega^2$):
\begin{equation}
V^{(MP)}(x)
=\frac{\hbar^2V_0^2}{2M}\Big(\e^{2x}-2\tilde\alpha\e^{x}\Big)\enspace,
\end{equation}
where the (finite) discrete energy spectrum is given by
\begin{equation}
E_l=-\frac{\hbar^2}{2M}(\tilde\alpha-l-\bhalf)^2\enspace.
\end{equation}
Proceeding in the usual way we obtain for the time-transformed path integral
\begin{eqnarray}
&&\!\!\!\!\!\!\!\!\!\!\!\!
K^{(V_1)}(u'',u',v'',v';s'')
=\pathints{u}\pathints{v}
\nonumber\\  &&\!\!\!\!\!\!\times
\exp\Bigg\{\ih\int_0^T\Bigg[\frac{m}{2}(\dot u^2+\dot v^2)
-\frac{\hbar^2}{2m}\bigg(
\frac{\lambda_1^2-\viert}{\cos^2u}-\frac{\lambda_2^2-\viert}{\sin^2u}\bigg)
-8m\omega^2\bigg(\e^{4v}-\frac{\alpha}{2m\omega^2}\e^{2v}\bigg)
\Bigg]\d s\Bigg\}.
\nonumber\\  &&\!\!\!\!\!\!\!\!\!\!\!\!
=\sum_{n}\Phi_n^{(\lambda_2,\lambda_1)}(u'')\Phi_n^{(\lambda_2,\lambda_1)}(u')
\exp\bigg[-\ih\frac{\hbar^2}{2m}(\lambda_1+\lambda_2+2n+1)2s''\bigg]
\nonumber\\  &&\!\!\!\!\!\!\qquad\times
\Bigg\{\int\d\kappa 
\Psi_\kappa^{(MP)}(v'')\Phi_\kappa^{(MP)\,*}(u')\e^{-\i\hbar\kappa^2s''/2m}
\nonumber\\  &&\!\!\!\!\!\!\qquad\qquad\qquad\qquad
+\sum_{l}\Psi_l^{(MP)}(v'')\Phi_l^{(MP\,*)}(u')
\exp\bigg[\ih\frac{\hbar^2}{2m}(\tilde\alpha-l-\bhalf)^2\bigg]\Bigg\}.
\end{eqnarray}
Here, $\lambda_{1,2}^2=k_{1,2}^2-2ma_{-,+}E/\hbar^2$, and in the variable 
$v$ we have used the solution of the Morse potential
and in the variable $u$ the solution of the P\"oschl--Teller potential,
respectively. This form of the solution is convenient to obtain the bound
state solutions. The bound state energy-levels are determined by:
\begin{equation}
2(n+l+1)+\lambda_1+\lambda_2-\frac{\alpha}{\hbar\omega}=0\enspace.
\label{energy-V1-DIV}
\end{equation}
By denoting
\begin{equation}
N_{n,l}=\bigg(2(n+l+1)-\frac{\alpha}{\hbar\omega}\bigg)^2-(k_1^2+k_2^2)
\end{equation}
the quadratic equation in $E$ can be solved to give
(with the further abbreviation $K_a=4(a_+k_1^2+a_-k_2^2)$)
\begin{equation}
E_{n,l}=\frac{\hbar^2}{4mb^2}\left\{
\pm\sqrt{(aN_{n,l}+K_a)^2-4b^2(N_{n,l}^2-4k_1^2k_2^2)}-(aN_{n,l}+K_a)\right\}.
\end{equation}
We keep the $\pm$-sign to allow for different boundary conditions which may 
depend on the parameters $a$ and $b$. For instance, for $a=2b$ we get the
the limiting case:
\begin{equation}
E_{n,l}=-\frac{\hbar^2}{2ma}\bigg[
\bigg(2(n+l+1)+k_1^2-\frac{\alpha}{\hbar\omega}\bigg)^2-k_2^2\bigg].
\end{equation}
For $k_2=\pm\half$ it has the form of
the usual zero-energy on the two-dimensional hyperboloid.

In order to obtain the continuous spectrum, the formulation in 
$(u,v)$-coordinates is inconvenient. Following \cite{GROf} we 
perform the coordinate transformation $\cos u=\tanh\tau$,
and additionally we make a time-transformation with the 
time-transformation function $f={a_+}/{\sin^2u}+{a_-}/{\cos^2u}$. 
Due to the coordinate transformation 
$\cos u=\tanh\tau$ additional quantum terms appear according to 
\begin{equation}
\exp\left({\i m\over2\epsilon\hbar}
{\big(\Delta u^{(j)}\big)^2\over\cos u^{(j-1)}\cos u^{(j)}}\right)
\dot=
\exp\left[{\i m\over2\epsilon\hbar}\big(\Delta\tau^{(j)}\big)^2
-\i\frac{\hbar}{8m}\left(1+{1\over\cosh^2\tau^{(j)}}\right)\right]\enspace.
\end{equation}
We get for the path integral (\ref{DIV-V1-uv})
\begin{equation}
K^{(V_1)}(u'',u',v'',v';T)
=\int_{-\infty}^\infty\frac{\d E}{2\pi\hbar}\,\e^{-\i ET/\hbar}
\int_0^\infty\d s''\exp\bigg[\ih\bigg(a_+E-\frac{\hbar^2k_2^2}{2m}\bigg)\bigg]
K^{(V_1)}(\tau'',\tau',v'',v';s'')\,,
\end{equation}
and the time-transformed path integral $K^{(V_1)}(s'')$ is given by
\begin{eqnarray}
&&K^{(V_1)}(\tau'',\tau',v'',v';s'')
=(\cosh\tau'\cosh\tau'')^{-1/2}
\nonumber\\   &&\qquad\times
\Bigg[
\sum_{l}\Psi_{l}^{(MP)}(v')\Psi_{l}^{(MP)}(v'')K_l(\tau'',\tau';s'')
+\int\d\kappa\Psi_{\kappa}^{(MP)\,*}(v')\Psi_{\kappa}^{(MP)}(v'')
K_\kappa(\tau'',\tau';s'')\Bigg]
\nonumber\\   &&
         \\   &&
K^{(V_1)}_{l,\kappa}(\tau'',\tau';s'')
\nonumber\\   &&
=\pathints{\tau}\exp\left\{\ih\ints \left[\frac{m}{2}\dot\tau^2
-\frac{\hbar^2}{2m}\left(\frac{\lambda_1^2-\viert}{\sinh^2\tau}
-\frac{\nu_{l,\kappa}^2-\viert}{\cosh^2\tau}\right)\right]\d s\right\}.
\label{DIV-V1-uv-tau}
\end{eqnarray}
The parameters $\lambda_{1,2}$ are the same as in the previous paragraph
and $\nu$ is given be
\begin{equation}
\nu_l=\bigg\vert 2l+1-\frac{\alpha}{\hbar\omega}\bigg\vert\qquad
\hbox{(discrete)},\qquad \nu_{\kappa}=\i\kappa\qquad\hbox{(continuous)},
\end{equation}
where discrete and continuous means the discrete and continuous contribution
of the Morse potential. Of course, the analysis of the discrete spectrum 
gives the same result as before. The kernel
$K^{(V_1)}_{l,\kappa}(s'')$ now allow us to write down the
entire kernel $K^{(V_1)}(T)$ in terms of Morse wave-functions and modified
P\"oschl--Teller wave-functions in the following form:
\begin{eqnarray}
&&K^{(V_1)}(u'',u',v'',v';T)=(\cosh\tau'\cosh\tau'')^{-1/2}
\nonumber\\   &&\qquad\times
\Bigg\{\sum_{ln}N_{ln}^2
\Psi_{l}^{(MP)\,*}(v')\Psi_{l}^{(MP)}(v'')
\Psi_{n}^{(\lambda_1,\nu_l)\,*}(\tau')\Psi_{n}^{(\lambda_1,\nu_l)}(\tau'')
\e^{-\i E_{ln}T/\hbar}
\nonumber\\   &&\qquad\qquad
+\int\d p\sum_{l}N_{lp}^2
\Psi_{l}^{(MP)\,*}(v')\Psi_{l}^{(MP)}(v'')
\Psi_{p}^{(\lambda_1,\nu_l)\,*}(\tau')\Psi_{p}^{(\lambda_1,\nu_l)}(\tau'')
\e^{-\i E_{p}T/\hbar}
\nonumber\\   &&\qquad\qquad
+\int\d p\int\d\kappa N_{\kappa p}^2
\Psi_{\kappa}^{(MP)\,*}(v')\Psi_{\kappa}^{(MP)}(v'')
\Psi_{p}^{(\lambda_1,\i\kappa)\,*}(\tau')
\Psi_{p}^{(\lambda_1,\i\kappa)}(\tau'')
\e^{-\i E_{p}T/\hbar}\Bigg\},\qquad
\end{eqnarray}
with the proper normalization constants
 $N_{ln},N_{lp}, N_{\kappa p}$, where e.g. $N_{ln}$ is determined by the
residuum corresponding to $E_{ln}$ in the Green function, and with the 
continuous spectrum
\begin{equation}
E_p=\frac{\hbar^2}{2ma_+}(p^2+k_2^2)\enspace.
\end{equation}
Note that for $k_2=1/2$ we obtain the well-known zero-energy on the
two-dimensional hyperboloid, which appears here in a natural way after
performing the coordinate transformation $\cos u=\tanh\tau$.

The $\Psi_p^{(\mu,\nu)}(\omega)$ are the modified  P\"oschl--Teller
functions, which are given by
\begin{eqnarray}
  \Psi_n^{(\eta,\nu)}(r)
  &=&N_n^{(\eta,\nu)}(\sinh r)^{2k_2-\half}
                    (\cosh r)^{-2k_1+{3\over2 }}
  \nonumber\\   &&\qquad\times
  {_2}F_1(-k_1+k_2+\kappa,-k_1+k_2-\kappa+1;2k_2;-\sinh^2r)
  \\
  N_n^{(\eta,\nu)}
  &=&{1\over\Gamma(2k_2)}
  \bigg[{2(2\kappa-1)\Gamma(k_1+k_2-\kappa)
                     \Gamma(k_1+k_2+\kappa-1)\over
    \Gamma(k_1-k_2+\kappa)\Gamma(k_1-k_2-\kappa+1)}\bigg]^{1/2}\enspace.
\end{eqnarray}
The scattering states are given by:
 \begin{eqnarray}
  V(r)&=&\hbarm \bigg({\eta^2-{1\over4}\over\sinh^2r}
   -{\nu^2-{1\over4}\over\cosh^2r}\bigg)
  \nonumber\\   
  \Psi_p^{(\eta,\nu)}(r)
  &=&N_p^{(\eta,\nu)}(\cosh r)^{2k_1-\half}(\sinh r)^{2k_2-\half}
  \nonumber\\   &&\qquad\qquad\times
  {_2}F_1(k_1+k_2-\kappa,k_1+k_2+\kappa-1;2k_2;-\sinh^2r)
  \\
  N_p^{(\eta,\nu)}
  &=&{1\over\Gamma(2k_2)}\sqrt{p\sinh\pi p\over2\pi^2}
  \Big[\Gamma(k_1+k_2-\kappa)\Gamma(-k_1+k_2+\kappa)
  \nonumber\\   &&\qquad\qquad\times
  \Gamma(k_1+k_2+\kappa-1)\Gamma(-k_1+k_2-\kappa+1)\Big]^{1/2}\enspace,
\end{eqnarray}
$k_1,k_2$ defined by:
$k_1=\half(1\pm\nu)$, $k_2=\half(1\pm\eta)$, where the correct sign
depends on the boundary-conditions for $r\to0$ and $r\to\infty$,
respectively. The number $N_M$ denotes the maximal number of
states with $0,1,\dots,N_M<k_1-k_2-\half$. $\kappa=k_1-k_2-n$ for the
bound states and $\kappa=\half(1+\i p)$ for the scattering states.
${_2}F_1(a,b;c;z)$ is the hypergeometric function \cite[p.1057]{GRA}.

\subsubsection{Separation of $V_1$ in Horospherical Coordinates.}
We evaluate the path integral for $V_1$ in horospherical coordinates.
The classical Lagrangian and Hamiltonian are given by
\begin{eqnarray}
\CL(\mu,\dot\mu,\nu,\dot\nu)&=&
\frac{m}{2}\bigg(\frac{a_+}{\nu^2}+\frac{a_-}{\mu^2}\bigg)
(\dot\mu^2+\dot\nu^2)-V(\mu,\nu)\enspace,
\\
\CH(\mu,p_\mu,\nu,p_\nu)&=&
\frac{1}{2m}\frac{\mu^2\nu^2(p_\mu^2+p_\nu^2)}{a_+\mu^2+a_-\nu^2}
+V(\mu,\nu)\enspace.
\end{eqnarray}
For the canonical momentum operators we have
\begin{eqnarray}
p_\mu&=&\hi\bigg(\frac{\partial}{\partial\mu}
-\frac{\nu^2a_-/\mu}{a_+\mu^2+a_-\nu^2}\bigg)\enspace,
\\
p_\nu&=&\hi\bigg(\frac{\partial}{\partial\nu}
-\frac{\mu^2a_+/\nu}{a_+\mu^2+a_-\nu^2}\bigg)\enspace,
\end{eqnarray}
and for the quantum Hamiltonian we get
\begin{eqnarray}
H&=&-\frac{\hbar^2}{2m}\frac{\mu^2\nu^2}{a_+\mu^2+a_-\nu^2}
\bigg(\frac{\partial^2}{\partial\mu^2}+
\frac{\partial^2}{\partial\nu^2}\bigg)+V(\mu,\nu)
\\
&=&\frac{1}{2m}\sqrt{\frac{\mu^2\nu^2}{a_+\mu^2+a_-\nu^2}}\,
(p_\mu^2+p_\nu^2)\,\sqrt{\frac{\mu^2\nu^2}{a_+\mu^2+a_-\nu^2}}
+V(\mu,\nu)\enspace.
\end{eqnarray}
We insert $V_1$ into the path integral and obtain
($f={a_+}/{\nu^2}+{a_-}/{\mu^2}$ and keeping to constants $k_{1,2}$)
\begin{eqnarray}
&&K^{(V_1)}(\mu'',\mu',\nu'',\nu';T)
=\pathint{\mu}\pathint{\nu}f(\mu,\nu)
\nonumber\\  &&\qquad\times
\exp\Bigg\{\ih\int_0^T\Bigg[\frac{m}{2}f(\mu,\nu)(\dot\mu^2+\dot\nu^2)
\nonumber\\  &&\qquad\qquad\qquad\qquad
-\frac{1}{f(\mu,\nu)}
\Bigg(\frac{m}{2}\omega^2(\mu^2+\nu^2)-\alpha+\frac{\hbar^2}{2m}\bigg(
\frac{k_1^2-\bviert}{\mu^2}+\frac{k_2^2-\bviert}{\nu^2}\bigg)
\Bigg)\Bigg]\dt\Bigg\}
\nonumber\\  &=&
\int_{-\infty}^\infty\frac{\d E}{2\pi\hbar}\,\e^{-\i ET/\hbar}
\int_0^\infty\d s''\e^{\i\alpha s''/\hbar}
K^{(V_1)}(\mu'',\mu',\nu'',\nu';s'')\enspace,
\end{eqnarray}
and the time-transformed path integral $K^{(V_1)}(s'')$ is given by
\begin{eqnarray}
&&\!\!\!\!\!\!\!\!\!\!
K^{(V_1)}(\mu'',\mu',\nu'',\nu';s'')
\nonumber\\   &&\!\!\!\!\!\!\!\!\!\!
=\pathints{\mu}
\exp\left\{\ih\ints \left[\frac{m}{2}(\dot\mu^2-\omega^2\mu^2)
-\frac{\hbar^2}{2m}\frac{k_1^2-2ma_-E/\hbar^2-\viert}{\mu^2}
\right]\d s\right\}
\nonumber\\   &&\!\!\!\!\!\!\!\!\!\!\qquad\times
\pathints{\nu}
\exp\left\{\ih\ints \left[\frac{m}{2}(\dot\nu^2-\omega^2\nu^2)
-\frac{\hbar^2}{2m}\frac{k_2^2-2ma_+E/\hbar^2-\viert}{\nu^2}
\right]\d s\right\}
\nonumber\\   &&\!\!\!\!\!\!\!\!\!\!
=\frac{m^2\omega^2\sqrt{\mu'\mu''\nu'\nu''}}{\i^2\hbar^2\sin^2\omega s''}
\exp\bigg[-\frac{m\omega}{2\i\hbar}({\mu'}^2+{\mu''}^2+{\nu'}^2+{\nu''}^2)
\cot\omega s''\bigg]
I_{\lambda_1}\bigg(\frac{m\omega\mu'\mu''}{\i\hbar\sin\omega s''}\bigg)
I_{\lambda_2}\bigg(\frac{m\omega\nu'\nu''}{\i\hbar\sin\omega s''}\bigg)
\enspace,
\nonumber\\   &&\!\!\!\!\!\!\!\!\!\!
\label{DIV-V1-munu}
\end{eqnarray}
where $\lambda_{1,2}=k_{1,2}^2-2ma_{\mp}E/\hbar^2$.
We can extract the bound state wave-functions for the bound state
contribution of the Green function according to:
\begin{eqnarray}
&&\!\!\!\!\!\!\!\!\!\!
G^{(V_1)}(\mu'',\mu',\nu'',\nu';E)
\nonumber\\   &&\!\!\!\!\!\!\!\!\!\!
=\sum_{n_\mu=0}^\infty\sum_{n_\nu=0}^\infty
\frac{N_{n_\mu n_\nu}^2}{E_{n_\mu n_\nu}-E}
\Psi_{n_\mu}^{(RHO,\lambda_1)}(\mu')\Psi_{n_\mu}^{(RHO,\lambda_1)}(\mu'')
\Psi_{n_\nu}^{(RHO,\lambda_2)}(\nu')\Psi_{n_\nu}^{(RHO,\lambda_2)}(\nu'')
\enspace.
\end{eqnarray}
The bound states are determined by the equation
\begin{equation}
\frac{\alpha}{\hbar\omega}-2(n_\mu+n_\nu+1)
=\sqrt{k_1^2-\frac{2ma_-E}{\hbar^2}}+\sqrt{k_2^2-\frac{2ma_+E}{\hbar^2}}
\enspace.
\end{equation}
This quadratic equation in $E$ is identical with (\ref{energy-V1-DIV}).

\subsection{The Superintegrable Potential $V_2$ on $\DIV$.}
\message{The Superintegrable Potential V_2 on D_IV.}
We state the potential in the respective coordinate systems
\begin{eqnarray}
V_2(u,v)&=&
\bigg(\frac{a_+}{\sin^2u}+\frac{a_-}{\cos^2u}\bigg)^{-1}
\frac{\hbar^2}{2m}\left[
\frac{k_1^2-\viert}{\sinh^2v}-\frac{k_2^2-\viert}{\cosh^2v}
+(k_3^2-\bviert)\bigg(\frac{1}{\sin^2u}+\frac{1}{\cos^2u}\bigg)\right]\qquad
\\      &=&
4\bigg(\frac{a_+}{\sinh^22\tomega}+\frac{a_-}{\sin^22\tvphi}\bigg)^{-1}
\frac{\hbar^2}{2m}\Bigg[(k_3^2-\bviert)
\bigg(\frac{1}{\sinh^22\tomega}+\frac{1}{\sin^22\tvphi}\bigg)
\nonumber \\   &&\qquad\qquad\qquad\qquad\qquad\qquad\qquad\qquad
+\bigg(\frac{k_2^2-\viert}{\cos^22\tvphi}-\frac{k_1^2-\viert}{\cosh^22\tomega}
\bigg)\Bigg]\enspace.\qquad
\end{eqnarray}
It is possible to evaluate the path integral for $V_2$ in the $(u,v)$ and the
degenerate elliptic system with $\gamma=2$. The elliptic
system with $\gamma=0$ is not treated.

\subsubsection{Separation of $V_2$ in the ($u,v$)-System.}
We insert $V_2$ into the path integral and obtain 
($f={a_+}/{\sin^2u}+{a_-}/{\cos^2u}$)
\begin{eqnarray}
&&\!\!\!\!\!\!
K^{(V_2)}(u'',u',v'',v';T)
=\pathint{u}\pathint{v}f(u)
\nonumber\\  &&\!\!\!\!\!\!\times
\exp\Bigg\{\ih\int_0^T\Bigg[\frac{m}{2}f
(\dot u^2+\dot v^2)
\!-\!\frac{\hbar^2}{2mf}
\Bigg(\frac{k_1^2-\viert}{\sinh^2v}-\frac{k_2^2-\viert}{\cosh^2v}
+(k_3^2-\bviert)\bigg(\frac{1}{\sin^2u}+\frac{1}{\cos^2u}\bigg)\Bigg)
\Bigg]\dt\Bigg\}.
\nonumber\\ 
\label{DIV-V2-uv}
\end{eqnarray}
This formulation in $(u,v)$-coordinates is inconvenient. Following 
the procedure as for $V_1$ in the $(u,v)$-system
we perform the coordinate transformation $\cos u=\tanh\tau$, 
and get for the path integral (\ref{DIV-V2-uv})
\begin{equation}
K^{(V_2)}(u'',u',v'',v';T)
=\int_{-\infty}^\infty\frac{\d E}{2\pi\hbar}\,\e^{-\i ET/\hbar}
\int_0^\infty\d s''\exp\bigg[\ih\bigg(a_+E-\frac{\hbar^2k_3^2}{2m}\bigg)\bigg]
K(\tau'',\tau',v'',v';s'')\enspace,
\end{equation}
and the time-transformed path integral $K^{(V_2)}(s'')$ is given by
\begin{eqnarray}
&&K^{(V_2)}(\tau'',\tau',v'',v';s'')
=(\cosh\tau'\cosh\tau'')^{-1/2}\sum_{n_v=0}^{N_{\rm max}}
\Psi_{n_v}^{(k_1,k_2)}(v')\Psi_{n_v}^{(k_1,k_2)}(v'')
 \nonumber\\   &&\qquad\qquad\times
\pathints{\tau}\exp\left\{\ih\ints \left[\frac{m}{2}\dot\tau^2
-\frac{\hbar^2}{2m}\left(\frac{\lambda_2^2-\viert}{\sinh^2\tau}
-\frac{\lambda_1^2-\viert}{\cosh^2\tau}\right)
\right]\d s\right\}
\nonumber\\   &&\qquad
+(\cosh\tau'\cosh\tau'')^{-1/2}
\int \d k_v\Psi_{k_v}^{(k_1,k_2)}(v')\Psi_{k_v}^{(k_1,k_2)}(v'')
 \nonumber\\   &&\qquad\qquad\times
\pathints{\tau}\exp\left\{\ih\ints \left[\frac{m}{2}\dot\tau^2
-\frac{\hbar^2}{2m}\left(\frac{\lambda_2^2-\viert}{\sinh^2\tau}
-\frac{-k_v^2-\viert}{\cosh^2\tau}\right)
\right]\d s\right\}\enspace.\qquad\qquad
\label{DIV-V2-uv-tau}
\end{eqnarray}
($\lambda_1^2=(2n_v+|k_1|-|k_2|+1)^2$, $\lambda_2^2=k_3^2-2ma_-E/\hbar^2$).

The $v$-path integration gives a discrete and a continuous spectrum,
thus two different parts for the $\tau$-path integration.
We therfore find for the Green function
\begin{eqnarray}
&&\!\!\!\!\!\!\!\!\!\!\!\!
G^{(V_2)}(\tau'',\tau',v'',v';E)
=(\cosh\tau'\cosh\tau'')^{-1/2}\sum_{n_v=0}^{N_{\rm max}}
\Psi_{n_v}^{(k_1,k_2)}(v')\Psi_{n_v}^{(k_1,k_2)}(v'')
\nonumber\\   &&\!\!\!\!\!\!\!\!\!\!\!\!\quad\times
  {m\over\hbar^2}{\Gamma(m_1-L_{\lambda_1})\Gamma(L_{\lambda_1}+m_1+1)\over
   \Gamma(m_1+m_2+1)\Gamma(m_1-m_2+1)}
  (\cosh\tau'\cosh\tau'')^{-(k_1-k_2)}(\tanh\tau'\tanh\tau'')^{m_1+m_2+1/2}
\nonumber\\   &&\!\!\!\!\!\!\!\!\!\!\!\!\quad\times
  {_2}F_1\bigg(-L_{\lambda_1}+m_1,L_{\lambda_1}+m_1+1;m_1-m_2+1;
                          {1\over\cosh^2\tau_<}\bigg)
\nonumber\\   &&\!\!\!\!\!\!\!\!\!\!\!\!\quad\times
  {_2}F_1\bigg(-L_{\lambda_1}+m_1,L_{\lambda_1}+m_1+1;m_1+m_2+1;
                          \tanh^2\tau_>\bigg)
\nonumber\\   &&\!\!\!\!\!\!\!\!\!\!\!\!\qquad
+(\cosh\tau'\cosh\tau'')^{-1/2}
\int \d k_v\Psi_{k_v}^{(k_1,k_2)}(v')\Psi_{k_v}^{(k_1,k_2)}(v'')
\nonumber\\   &&\!\!\!\!\!\!\!\!\!\!\!\!\qquad\quad\times
{m\over\hbar^2}{\Gamma(m_1-L_{k_v})\Gamma(L_{k_v}+m_1+1)\over
   \Gamma(m_1+m_2+1)\Gamma(m_1-m_2+1)}
  (\cosh\tau'\cosh\tau'')^{-(k_1-k_2)}(\tanh\tau'\tanh\tau'')^{m_1+m_2+1/2}
\nonumber\\   &&\!\!\!\!\!\!\!\!\!\!\!\!\qquad\quad\times
  {_2}F_1\bigg(-L_{k_v}+m_1,L_{k_v}+m_1+1;m_1-m_2+1;
                          {1\over\cosh^2\tau_<}\bigg)
\nonumber\\   &&\!\!\!\!\!\!\!\!\!\!\!\!\qquad\quad\times
  {_2}F_1\bigg(-L_{k_v}+m_1,L_{k_v}+m_1+1;m_1+m_2+1;
                          \tanh^2\tau_>\bigg)\enspace.
\label{Green-DIV-V2-uv-tau}
\end{eqnarray}
($m_{1,2}=\half(\lambda_2\pm\sqrt{2m\CE}/\hbar)$,
$L_{\lambda_1}=\half(\lambda_1-1)$,
$L_{k_v}=\half(\i k_v-1)$, $\CE=a_+E-\hbar^2k_3^2/2m$.

A discrete spectrum is only possible for the first summand in 
(\ref{DIV-V2-uv-tau}).
First, we can analyze the discrete spectrum by looking at the poles in
(\ref{Green-DIV-V2-uv-tau}) which gives the equation
\begin{equation}
2(n_\tau+n_v)+\lambda_++\lambda_-+|k_2|-|k_1|=0
\label{E-DIV-V_2}
\end{equation}
($\lambda_{\pm}^2=k_3^2-2ma_{\pm}E/\hbar^2$).
This gives a quadratic equation in $E$ with solution
($N_k=2n_\tau-2n_v-|k_1|+|k_2|$)
\begin{equation}
E_{n_\tau n_v}=-\frac{a\hbar^2N_k^2}{4b^2}
\left(1\mp\sqrt{1+
    \frac{4b^2}{a^2}\bigg(\frac{k_3^2}{N_k^2}-1\bigg)}\,\right)\enspace.
\end{equation}
The entire Green function in terms of the wave-functions is given by 
\begin{eqnarray}
&&G^{(V_2)}(\tau'',\tau',v'',v';E)
=(\cosh\tau'\cosh\tau'')^{-1/2}
\int\d p\frac{N_{p k_v}^2}{E_p-E}
\int \d k_v
\nonumber\\   &&\qquad\qquad\times
\Psi_{k_v}^{(k_1,k_2)}(v')\Psi_{k_v}^{(k_1,k_2)}(v'')
\Psi_{p}^{(\lambda_2,\i k_v)}(\tau')
\Psi_{p}^{(\lambda_2,\i k_v)\,*}(\tau'')
\nonumber\\   &&
+(\cosh\tau'\cosh\tau'')^{-1/2}\sum_{n_\tau=0}^\infty\sum_{n_v=0}^\infty
\Psi_{n_v}^{(k_1,k_2)}(v')\Psi_{n_v}^{(k_1,k_2)}(v'')
\nonumber\\   &&\qquad\times
\Bigg\{\sum_{n_\tau=0}^{N_{\rm max}}
\frac{N_{n_\tau n_v}^2}{E_{n_\tau n_v}-E}
\Psi_{n_\tau}^{(\lambda_2,\lambda_1)}(\tau')
\Psi_{n_\tau}^{(\lambda_2,\lambda_1)}(\tau'')
+\int\d p\frac{N_{p n_v}^2}{E_p-E}
\Psi_{p}^{(\lambda_2,\lambda_1)}(\tau')
\Psi_{p}^{(\lambda_2,\lambda_1)\,*}(\tau'')\Bigg\}
\enspace,
\nonumber\\   &&
\end{eqnarray}
where $N_{n_\tau n_v},N_{k_\tau n_v}$ is determined by the residuum in
(\ref{Green-DIV-V2-uv-tau}). The continuous spectrum has the form
\begin{equation}
E_p=\frac{\hbar^2}{2ma_+}(p^2+k_3^2)\enspace.
\end{equation}
For $k_3=\pm\half$ we obtain the usual zero-point energy on the 
two-dimensional hyperboloid.
Re-inserting $\cos u=\tanh v$ gives the Green function in the $(u,v)$-system.

\subsubsection{Separation of $V_2$ in Degenerate Elliptic Coordinates.}
We insert the potential $V_2$ in degenerate elliptic coordinates into
the path integral and obtain
$(f(\tomega,\tvphi)=4(a_+/\sinh^22\tomega+a_-/\sin^22\tvphi))$
\begin{eqnarray}
&&K^{(V_2)}(\tomega'',\tomega',\tvphi'',\tvphi';T)
\nonumber\\   &&\qquad
=\pathint{\tomega}\pathint{\tvphi}f(\tomega,\tvphi)
\exp\Bigg\{\ih\int_0^T\Bigg[
\frac{m}{2}f(\tomega,\tvphi)(\dot\tomega^2+\dot\tvphi^2)
\nonumber\\   &&\qquad\qquad\qquad\qquad
-4\frac{\hbar^2}{2mf(\tomega,\tvphi)}
\Bigg(\frac{k_1^2-\viert}{\sinh^22\tomega}
-\frac{k_2^2-\viert}{\cosh^22\tomega}
+\frac{k_3^2-\viert}{\sin^22\tvphi}
+\frac{k_2^2-\viert}{\cos^22\tvphi}\Bigg)\Bigg]\dt\Bigg\}.\qquad
\end{eqnarray}
The calculation is similar as in the case of the $(u,v)$-system:
First, we re-scale $2\tomega\to\tomega,2\tvphi\to\tvphi$, then we
perform the transformation $\cos\tvphi=\tanh\ttau$. Finally, we perform
a time-transformation in the path integral with the
time-transformation $f(\tomega,\tvphi)\to f(\tomega,\ttau)$ yielding
\begin{equation}
G^{(V_2)}(\ttau'',\ttau',\tomega'',\tomega';E)
=\int_0^\infty\d s''
\exp\bigg[\ih s''\bigg(Ea_--\frac{\hbar^2k_3^2}{2m}\bigg)\bigg] 
K^{(V_2)}(\ttau'',\ttau',\tomega'',\tomega';s'')
\end{equation}
with the transformed path integral $K^{(V_2)}(s'')$ given by 
\begin{eqnarray}
&&K^{(V_2)}(\ttau'',\ttau',\tomega'',\tomega';s'')
=\pathints{\ttau}\pathints{\tomega}
\exp\Bigg\{\ih\int_0^T\Bigg[
\frac{m}{2}(\dot\ttau^2+\cosh^2\ttau\dot\tomega^2)
\nonumber\\   &&\qquad\qquad
-\frac{\hbar^2}{2m}\Bigg(\frac{k_1^2-\viert}{\sinh^2\ttau}
+\frac{1}{\cosh^2\ttau}\bigg(\frac{\lambda_+^2-\viert}{\sinh^2\tomega}
-\frac{k_2^2-\viert}{\cosh^2\tomega}+\viert\bigg)\Bigg)\Bigg]\d s\Bigg\}
\enspace.\qquad
\end{eqnarray}
Again we evaluate this path integral by a successive $\tomega$- and 
$\ttau$-path integration. Performing finally the $s''$-integration we obtain 
\begin{eqnarray}
&&\!\!\!\!\!\!\!\!\!\!
G^{(V_2)}(\ttau'',\ttau',\tomega'',\tomega';E)
=(\cosh\ttau'\cosh\ttau'')^{-1/2}
\nonumber\\   &&\!\!\!\!\!\!\!\!\!\!\qquad\times
\Bigg\{
\int\d p\frac{N^2_{k_\tomega p}}{E_p-E}\int\d k_\tomega
\Psi_{p}^{(k_1,\i k_\tomega)}(\ttau')
\Psi_{p}^{(k_1,\i k_\tomega)\,*}(\ttau'')
\Psi_{k_\tomega}^{(\lambda_1,k_2)}(\tomega')
\Psi_{k_\tomega}^{(\lambda_1,k_2)\,*}(\tomega'')
\nonumber\\   &&\!\!\!\!\!\!\!\!\!\!\qquad\qquad
+\int\d p\sum_{n_\tomega=0}^{N_{\rm max}}
\frac{N^2_{n_\tomega p}}{E_p-E}
\Psi_{p}^{(k_1,\epsilon_{n_\tomega})}(\ttau')
\Psi_{p}^{(k_1,\epsilon_{n_\tomega})\,*}(\ttau'')
\Psi_{n_\tomega}^{(\lambda_1,k_2)}(\tomega')
\Psi_{n_\tomega}^{(\lambda_1,k_2)}(\tomega'')
\nonumber\\   &&\!\!\!\!\!\!\!\!\!\!\qquad\qquad
+\sum_{n_\ttau=0}^{N_{\rm max}}
\sum_{n_\tomega=0}^{N_{\rm max}}
\frac{N^2_{n_\ttau n_\tomega}}{E_{n_\ttau n_\tomega}-E}
\Psi_{n_\ttau}^{(k_1,\epsilon_{n_\tomega})}(\ttau')
\Psi_{n_\ttau}^{(k_1,\epsilon_{n_\tomega})}(\ttau'')
\Psi_{n_\tomega}^{(\lambda_1,k_2)}(\tomega')
\Psi_{n_\tomega}^{(\lambda_1,k_2)}(\tomega'')\Bigg\}.\qquad
\end{eqnarray}
The normalization constants $N_{k_\tomega p},N_{k_\tomega p},
N_{n_\ttau n_\tomega}$ are determined by the respective residuum in 
$G^{(V_2)}(E)$ and the discrete spectrum is determined by the quadratic
equation (\ref{E-DIV-V_2}). The continuous spectrum has
the form
\begin{equation}
E_p=\frac{\hbar^2}{2ma_-}(p^2+k_3^2)\enspace.
\end{equation}
The difference of $E_p$ in comparison to the $(u,v)$-system can be
resolved by making in the $(u,v)$-system the transformation
$\sin u=\tanh\tau$ which changes the sign in the energy term.
This concludes the discussion of $V_2$ on $\DIV$.

\subsection{The Superintegrable Potential $V_3$ on $\DIV$.}
\message{The Superintegrable Potential V_3 on D_IV.}
We state the potential in the respective coordinate systems
\begin{eqnarray}
V_3(\tomega,\tvphi)&=&\frac{\hbar^2}{2m}
\bigg(\frac{4a_+}{\sinh^22\tomega}+\frac{4a_-}{\sinh^2\tvphi}\bigg)^{-1}
\Bigg[\frac{c_1}{\cos^2\tvphi}+\frac{c_2}{\cosh^2\tomega}
+c_3\bigg(\frac{1}{\sin^2\tvphi}-\frac{1}{\sinh^2\tomega}\bigg)\Bigg].\qquad
\\      &=&\frac{\hbar^2}{2m}
   \left[a_+\bigg(\frac{1}{\cosh^2\homega}-\frac{1}{\cos^2\hvphi}\bigg)
   -a_-\bigg(\frac{1}{\sinh^2\homega}+\frac{1}{\sin^2\hvphi}\bigg)
   \right]^{-1}
\nonumber\\   &&\qquad\qquad\times
\Bigg[\frac{c_3}{\sinh^2\homega}+\frac{c_2}{\cosh^2\homega}
+c_3\bigg(\frac{1}{\sin^2\hvphi}-\frac{1}{\cos^2\hvphi}\bigg)\Bigg]\enspace,
\end{eqnarray}
It is possible to evaluate the path integral for $V_3$ in both
separating coordinate systems. However, due to the similarity in the
evaluations, only the degenerate elliptic II case will be presented.

\subsubsection{Separation of $V_3$ in Degenerate Elliptic Coordinates II.}
We insert the potential $V_3$ in the path integral formulation for
degenerate elliptic coordinates on $\DIV$ and obtain
$f(\tomega,\tvphi)=4(a_+/\sinh^22\tomega+a_-/\sin^22\tvphi)$)
\begin{eqnarray} 
&&K^{(V_3)}(\tomega'',\tomega',\tvphi'',\tvphi';T)
=\pathint{\tomega}\pathint{\tvphi}f(\tomega,\tvphi)
\nonumber\\   &&\qquad\times
\exp\Bigg\{\ih\int_0^T\Bigg[\frac{m}{2}
f(\tomega,\tvphi)(\dot\tomega^2+\dot\tvphi^2)
\nonumber\\   &&\qquad\qquad\qquad
-\frac{\hbar^2}{2mf(\tomega,\tvphi)}
\Bigg(\frac{c_1}{\cos^2\tvphi}+\frac{c_2}{\cosh^2\tomega}
+c_3\bigg(\frac{1}{\sin^2\tvphi}-\frac{1}{\sinh^2\tomega}\bigg)\Bigg)\Bigg]
\dt\Bigg\}\enspace.\qquad\quad
\label{pathintegralDIV-V3-degelliptic}
\end{eqnarray}
In order to obtain a convenient form to evaluate 
(\ref{pathintegralDIV-V3-degelliptic}) we perform the coordinate transformation
$\cos\tvphi=\tanh\ttau$ in the same way as for $V_2$.
Performing also the corresponding time-transformation gives
\begin{eqnarray} 
&&K^{(V_3)}(\tomega'',\tomega',\ttau'',\ttau';T)
=\int_{-\infty}^\infty\frac{\d E}{2\pi\hbar}\,\e^{-\i ET/\hbar}
\nonumber\\   &&\qquad\times
\int_0^\infty\d s''
\exp\bigg[\ih\bigg(\frac{\hbar^2}{2m}\lambda^2_{3^+_{a_-}}\bigg)\bigg]
K^{(V_3)}(\tomega'',\tomega',\ttau'',\ttau';s'')\enspace,
\end{eqnarray}
and the time-transformed path integral $K^{(V_3)}(s'')$ is given by
\begin{eqnarray}
&&\!\!\!\!\!\!\!\!\!\!\!\!
K^{(V_3)}(\tomega'',\tomega',\ttau'',\ttau';s'')
=\pathints{\tomega}\pathints{\ttau}\cosh\ttau
\nonumber\\   &&\!\!\!\!\!\!\!\!\!\!\!\!\qquad\times
\exp\left\{\ih\ints \left[
\frac{m}{2}(\dot\ttau^2+\cosh^2\ttau\dot\tomega^2)
-\frac{\hbar^2}{2m}\frac{\lambda_{1^+_{a_-}}^2-\viert}{\sinh^2\ttau}
\right.\right.\nonumber\\   &&\qquad\qquad\qquad\qquad\left.\left.
-\frac{\hbar^2}{2m\cosh^2\ttau}
\left(\frac{\lambda_{3^-_{a^+}}^2-\viert}{\sinh^2\tomega}
     -\frac{\lambda_{2^-_{a^+}}^2-\viert}{\cosh^2\tomega}+\viert\right)
\right]\d s\right\}\qquad
\end{eqnarray} 
($\lambda^2_{i^{\pm}_{a_{\pm}}}=\viert\mp c_i-2ma_{\pm}E/\hbar^2$, $i=1,2,3$).
The latter path integral has the form of two successive modified
P\"oschl--Teller path integrations in $\tomega$ and $\ttau$. In the
$\omega$-path integration we get a contribution form the continuous and
from the discrete spectrum. The continuous contribution gives in the
$\ttau$-path integration only a continuous part, whereas the other
gives a discrete and continuous contribution in $\ttau$.
We denote the continuous parameter in $\tomega$ by $p_\tomega$, the
discrete  parameter in $\tomega$ by
$\epsilon_{n_\tomega}=2n_\tomega+\lambda_{3^-_{a^+}}-\lambda_{2^-_{a^+}}-1$,
the continuous parameter in $\ttau$ by $p$, the
discrete  parameter in $\ttau$ by
$\epsilon_{n_\ttau}=2n_\ttau+\lambda_{1^+_{a^-}}-\epsilon_{n_\tomega}-1$, 
therefore:
\begin{eqnarray}
&&\!\!\!\!\!\!\!\!
K^{(V_3)}(\tomega'',\tomega',\ttau'',\ttau';s'')
=(\cosh\ttau'\cosh\ttau'')^{-1/2}\int_0^\infty\d p_\omega
\Psi_{p_\tomega}^{(\lambda_{3^-_{a^+}},\lambda_{2^-_{a^+}})}(\tomega')
\Psi_{p_\tomega}^{(\lambda_{3^-_{a^+}},\lambda_{2^-_{a^+}})\,*}(\tomega'')
\nonumber\\[-3mm]   &&\qquad\qquad\times
\pathints{\ttau}\exp\left\{\ih\ints \left[
\frac{m}{2}\dot\ttau^2-\frac{\hbar^2}{2m}\left(
\frac{\lambda_{1^+_{a_-}}^2-\viert}{\sinh^2\ttau}+
\frac{p_\tomega^2+\viert}{\cosh^2\ttau}\right)\right]\d s\right\}
\nonumber\\   &&\quad
+(\cosh\ttau'\cosh\ttau'')^{-1/2}\sum_{n_\tomega=0}^{N_{\rm max}}
\Psi_{n_\tomega}^{(\lambda_{3^-_{a^+}},\lambda_{2^-_{a^+}})}(\tomega')
\Psi_{n_\tomega}^{(\lambda_{3^-_{a^+}},\lambda_{2^-_{a^+}})}(\tomega'')
\nonumber\\   &&\qquad\qquad\times
\pathints{\ttau}\exp\left\{\ih\ints \left[
\frac{m}{2}\dot\ttau^2-\frac{\hbar^2}{2m}\left(
\frac{\lambda_{1^+_{a_-}}^2-\viert}{\sinh^2\ttau}-
\frac{\epsilon_{n_\tomega}^2-\viert}{\cosh^2\ttau}\right)\right]\d s\right\}
\nonumber\\   &&
=(\cosh\ttau'\cosh\ttau'')^{-1/2}\int_0^\infty\d p_t\omega
\Psi_{p_\tomega}^{(\lambda_{3^-_{a^+}},\lambda_{2^-_{a^+}})}(\tomega')
\Psi_{p_\tomega}^{(\lambda_{3^-_{a^+}},\lambda_{2^-_{a^+}})\,*}(\tomega'')
\nonumber\\   &&\qquad\qquad\times
\int_0^\infty\d p
\Psi_{p}^{(\lambda_{1^+_{a_-}},\i p_\tomega)}(\tomega')
\Psi_{p}^{(\lambda_{1^+_{a_-}},\i p_\tomega)\,*}(\tomega'')
\,\e^{-\i s''\hbar p^2/2m}
\nonumber\\   &&\quad
+(\cosh\ttau'\cosh\ttau'')^{-1/2}\sum_{n_\tomega=0}^{N_{\rm max}}
\Psi_{n_\tomega}^{(\lambda_{3^-_{a^+}},\lambda_{2^-_{a^+}})}(\tomega')
\Psi_{n_\tomega}^{(\lambda_{3^-_{a^+}},\lambda_{2^-_{a^+}})}(\tomega'')
\nonumber\\   &&\qquad\qquad\times
\left\{\int_0^\infty\d p
\Psi_{p}^{(\lambda_{1^+_{a_-}},\epsilon_{n_\tomega})}(\tomega')
\Psi_{p}^{(\lambda_{1^+_{a_-}},\epsilon_{n_\tomega})\,*}(\tomega'')
\,\e^{-\i s''\hbar p^2/2m}\right.
\nonumber\\[-3mm]   &&\qquad\qquad\qquad\qquad\qquad\qquad\left.
+\sum_{n_\omega=0}^{N_{\rm max}}
\Psi_{n_\tau}^{(\lambda_{1^+_{a_-}},\epsilon_{n_\tomega})}(\ttau')
\Psi_{n_\tau}^{(\lambda_{1^+_{a_-}},\epsilon_{n_\tomega})}(\ttau'')
\,\e^{-\i\hbar s''\epsilon_{n_\ttau}^2/2m}\right\}\enspace.
\end{eqnarray}
Performing the $s''$-integration gives the spectrum. For the
continuous spectrum we obtain
\begin{equation}
E_p=\frac{\hbar^2}{2ma_-}\big(p^2+\bviert-c_3\big)\enspace.
\end{equation}
The discrete spectrum is determined by
\begin{equation}
2(n_\tomega+n_\ttau)+\lambda_{1^+_{a_-}}
+\lambda_{3^-_{a_+}}-\lambda_{2^-_{a_-}}-2=\lambda_{3^+_{a_-}}\enspace.
\end{equation}
\nobreak
This is an equation in $E$ in eighth order which we will not solve.

\goodbreak\noindent%
\subsection{The Superintegrable Potential $V_4$ on $\DIV$.}
\message{The Superintegrable Potential V_4 on D_IV.}
We state the potential in the respective coordinate systems
\begin{eqnarray}\!\!\!\!
V_4(\mu,\nu)&=&
\bigg(\frac{a_+}{\sin^2u}+\frac{a_-}{\cos^2u}\bigg)^{-1}
\frac{\hbar^2}{2m}(k_0^2-\bviert)
\bigg(\frac{1}{\sin^2u}+\frac{1}{\cos^2u}\bigg)\qquad
\\      \!\!\!\!&=&
\bigg(\frac{a_+}{\nu^2}+\frac{a_-}{\mu^2}\bigg)^{-1}
\frac{\hbar^2}{2m}(k_0^2-\bviert)
\bigg(\frac{1}{\nu^2}+\frac{1}{\mu^2}\bigg)
\\      \!\!\!\!&=&
\frac{\hbar^2}{2md^2}\bigg(\frac{a+2b}{\sinh^22\omega'}
+\frac{a-2b}{\sin^22\vphi'}\bigg)^{-1}(k_0^2-\bviert)
\bigg(\frac{1}{\cosh^2\omega\cos^2\vphi}
     +\frac{1}{\sinh^2\omega\sin^2\vphi}\bigg).\qquad
\end{eqnarray}
It is possible to evaluate the path integral for $V_4$ is all the
separating coordinate systems. However, we evaluate the path integral
for $V_4$ only in the $(u,v)$-system because $V_4$ is trivial.

\subsubsection{Separation of $V_4$ in the ($u,v$)-System.}
We insert $V_4$ into the path integral and obtain 
($f={a_+}/{\sin^2u}+{a_-}/{\cos^2u}$)
\begin{eqnarray}
&&\!\!\!\!\!\!
K(u'',u',v'',v';T)
=\pathint{u}\pathint{v}f(u)
\nonumber\\  &&\!\!\!\!\!\!\qquad\times
\exp\left\{\ih\int_0^T\left[\frac{m}{2}f(u)
(\dot u^2+\dot v^2)
-\frac{\hbar^2}{2m}
\frac{k_0^2-\viert}{f(u)}\bigg(\frac{1}{\sin^2u}+\frac{1}{\cos^2u}\bigg)
\right]\dt\right\}\enspace.\qquad
\label{DIV-V4-uv}
\end{eqnarray}
We proceed similarly as in \cite{GROas}. Because the formulation in
$(u,v)$-coordinates is inconvenient, we perform  following 
\cite{GROf} the coordinate transformation $\cos u=\tanh\tau$. 
Further, we separate off the $v$-path integration, and additionally we 
make a time-transformation with the time-transformation function 
$f={a_+}/{\sin^2u}+{a_-}/{\cos^2u}$. Due to the coordinate transformation 
$\cos u=\tanh\tau$ additional quantum terms appear according to 
\begin{equation}
\exp\left({\i m\over2\epsilon\hbar}
{\big(\Delta u^{(j)}\big)^2\over\cos u^{(j-1)}\cos u^{(j)}}\right)
\dot=
\exp\left[{\i m\over2\epsilon\hbar}\big(\Delta\tau^{(j)}\big)^2
-\i\frac{\hbar}{8m}\left(1+{1\over\cosh^2\tau^{(j)}}\right)\right]\enspace.
\end{equation}
We get for the path integral (\ref{DIV-V4-uv})
\begin{equation}
K(u'',u',v'',v';T)
=\int_{-\infty}^\infty\frac{\d E}{2\pi\hbar}\,\e^{-\i ET/\hbar}
\int_0^\infty\d s''\exp\bigg[\ih\bigg(a_+E-\frac{\hbar^2k_0^2}{2m}\bigg)\bigg]
K(\tau'',\tau',v'',v';s'')\enspace,
\end{equation}
and the time-transformed path integral $K(s'')$ is given by
\begin{eqnarray}
&&K(\tau'',\tau',v'',v';s'')
=\int_{-\infty}^\infty\d k_v\frac{\e^{\i k_v(v''-v')}}{2\pi}
 (\cosh\tau'\cosh\tau'')^{-1/2}
\nonumber\\   &&\qquad\times
\pathints{\tau}\exp\left\{\ih\ints \left[\frac{m}{2}\dot\tau^2
-\frac{\hbar^2}{2m}\left(\frac{\lambda_0^2-\viert}{\sinh^2\tau}
-\frac{-k_v^2-\viert}{\cosh^2\tau}\right)\right]\d s\right\}
\enspace.\qquad\qquad\qquad
\label{DIV-V4-uv-tau}
\end{eqnarray}
Inserting the solution for the modified
P\"oschl--Teller potential and evaluating the Green function on the cut
yields for the path integral solution on $\DIV$ as follows
($K(u'',u',v'',v';T)=K(\tau'',\tau',v'',v';T)$):
\begin{eqnarray}
K(u'',u',v'',v';T)&=&
\int_{-\infty}^\infty\d k_v\int_0^\infty \d p\,
\e^{-\i TE_p/\hbar} \Psi_{p,k_v}(\tau'',v'')\Psi_{p,k_v}^*(\tau',v')\enspace,
\qquad\qquad\\   
\Psi_{p,k_v}(\tau,v)&=&
\frac{\e^{\i k_vv}}{\sqrt{2\pi a_+\cosh\tau}}\,\Psi_p^{(\lambda_0,\i k)}(\tau)
\enspace,
         \\   
E_p&=&\frac{\hbar^2}{2ma_+}\big(p^2+k_0^2\big)\enspace,
\end{eqnarray}
where $\lambda_0^2=k_0^2-2ma_eE/\hbar^2$ and the wave-functions for the
modified P\"oschl--Teller functions. Re-inserting $\cos u=\tanh\tau$
gives the solution in terms of the variable~$u$.

We also see from this example that the introduction of a third
variable $w$ \cite{GROat}, say, to a three-dimensional version of Darboux space
$\DIV$ allows separation of variables, where the additional quantum
number $k_0$ corresponds to the motion in $w$.

\begin{table}[t!]
\caption{\label{solutions} Solutions of the path integration for
  superintegrable potentials in   Darboux spaces%
}
\vspace{-0.5cm}
\begin{eqnarray}\begin{array}{l}\vbox{\small\offinterlineskip
\halign{&\vrule#&$\strut\ \hfil\hbox{#}\hfill\ $\cr
\noalign{\hrule}
height2pt&\omit&&\omit&\cr
&Space and Potential  &&Solution in terms of the wave-functions  &\cr
height2pt&\omit&&\omit&\cr
\noalign{\hrule}\noalign{\hrule}\noalign{\hrule}\noalign{\hrule}
height2pt&\omit&&\omit&\cr
&$\DI$                &&          &\cr
height2pt&\omit&&\omit&\cr
\noalign{\hrule}\noalign{\hrule}
height2pt&\omit&&\omit&\cr
&$V_1$:  $(u,v)$      &&Hermite polynomials $\times$ 
                        Parabolic cylinder functions      &\cr
&\qquad
Parabolic             &&No explicit solution              &\cr
height2pt&\omit&&\omit&\cr
\noalign{\hrule}
height2pt&\omit&&\omit&\cr
&$V_2$:  $(u,v)$      &&Hermite polynomials $\times$ 
                        Parabolic cylinder functions      &\cr
&\qquad
$(r,q)$               &&Hermite polynomials $\times$ 
                        Parabolic cylinder functions      &\cr
height2pt&\omit&&\omit&\cr
\noalign{\hrule}\noalign{\hrule}
height2pt&\omit&&\omit&\cr
&$\DII$               &&          &\cr
height2pt&\omit&&\omit&\cr
\noalign{\hrule}\noalign{\hrule}
height2pt&\omit&&\omit&\cr
&$V_1$:  $(u,v)$      &&Hermite polynomial $\times$
                        Whittaker functions$^*$           &\cr
&\qquad
Parabolic             &&No explicit solution              &\cr
height2pt&\omit&&\omit&\cr
\noalign{\hrule}
height2pt&\omit&&\omit&\cr
&$V_2$:  $(u,v)$      &&Laguerre polynomial $\times$
                        Whittaker functions$^*$           &\cr
&\qquad
Polar                 &&Gegenbauer polynomial $\times$
                        Whittaker functions$^*$           &\cr
&\qquad
Elliptic              &&No explicit solution              &\cr
height2pt&\omit&&\omit&\cr
\noalign{\hrule}
height2pt&\omit&&\omit&\cr
&$V_3$: Polar        &&Gegenbauer polynomials $\times$ 
                        Bessel functions      &\cr
&\qquad
Parabolic             &&Product of Whittaker functions$^*$ &\cr
&\qquad
Elliptic              &&No explicit solution               &\cr
height2pt&\omit&&\omit&\cr
\noalign{\hrule}\noalign{\hrule}
height2pt&\omit&&\omit&\cr
&$\DIII$               &&                                  &\cr
height2pt&\omit&&\omit&\cr
\noalign{\hrule}\noalign{\hrule}
height2pt&\omit&&\omit&\cr
&$V_1$:  Parabolic      
         &&Product of Hermite polynomials/Parabolic cylinder functions
                                                           &\cr
&\qquad
Translated parabolic    
         &&Product of Hermite polynomials/Parabolic cylinder functions
                                                           &\cr
height2pt&\omit&&\omit&\cr
\noalign{\hrule}
height2pt&\omit&&\omit&\cr
&$V_2$:  $(u,v)$ &&Gegenbauer polynomials $\times$ Whittaker
functions$^*$                                              &\cr
&\qquad
Polar            &&Gegenbauer polynomials $\times$ Whittaker
functions$^*$                                              &\cr
&\qquad
Parabolic             &&Product of Whittaker functions$^*$ &\cr
height2pt&\omit&&\omit&\cr
\noalign{\hrule}
height2pt&\omit&&\omit&\cr
&$V_3$:  Polar        &&Gegenbauer polynomials $\times$ Whittaker
functions$^*$                                             &\cr
&\qquad
Hyperbolic            &&No explicit solution              &\cr
height2pt&\omit&&\omit&\cr
\noalign{\hrule}
height2pt&\omit&&\omit&\cr
&$V_4$:  Hyperbolic   &&Product of Whittaker functions$^*$&\cr
&\qquad
Elliptic              &&No explicit solution              &\cr
height2pt&\omit&&\omit&\cr
\noalign{\hrule}\noalign{\hrule}
height2pt&\omit&&\omit&\cr
&$\DIV$               &&                                   &\cr
height2pt&\omit&&\omit&\cr
\noalign{\hrule}\noalign{\hrule}
height2pt&\omit&&\omit&\cr
&$V_1$:  $(u,v)$-system      
                    &&Product of hypergeometric functions  &\cr
&\qquad  Horospherical      
                    &&Product of Whittaker functions$^*$   &\cr
&\qquad
Elliptic            &&No explicut solution                 &\cr
height2pt&\omit&&\omit&\cr
\noalign{\hrule}
height2pt&\omit&&\omit&\cr
&$V_2$:  $(u,v)$    &&hypergeometric functions             &\cr
&\qquad
Degenerate Elliptic &&hypergeometric functions             &\cr
height2pt&\omit&&\omit&\cr
\noalign{\hrule}
height2pt&\omit&&\omit&\cr
&$V_3$:  Elliptic   &&hypergeometric functions             &\cr
&\qquad
Degenerate Elliptic &&hypergeometric functions             &\cr
height2pt&\omit&&\omit&\cr
\noalign{\hrule}}}
\nonumber\\
\hbox{($^*$ The notion Whittaker functions means for a disrete spectrum
Laguerre polynomials, and}
\nonumber\\
\hbox{for a continuous spectrum Whittaker
functions $W_{\mu,\nu}(z)$, respectievly $M_{\mu,\nu}(z)$.)}
\end{array}
\nonumber\end{eqnarray}
\end{table}

\setcounter{equation}{0}%
\section{Summary and Discussion}
\message{Summary and Discussion}
In this second paper we have finished the discussion of superintegrable
potentials on the Darboux spaces of non-constant curvature. The results are
very satisfactory. 
There are two potentials on $\DI$, four potentials on $\DII$, five potentials
on $\DIII$, and four potentials on $\DIV$, respectively.
We could solve many of the emerging quantum mechanical problems.
To give an overview, we summarize our results in Table \ref{solutions}. 
We list for each space the corresponding potentials
including the general form of the solution (if explicitly
possible). We omit the trivial potentials here, because they are
separable in all corresponding coordinate systems.

In the first Darboux space $\DI$ the superintegrable were related to the
Holt potential and a shifted isotropic harmonic oscillator in two-dimensional
Euclidean space. Whereas the solution in the coordinate $v$ can be expressed
in terms of the wave-functions for the radial harmonic oscillator
(Laguerre polynomials) and the shifted harmonic oscillator (Hermite
polynomials), the solution in the coordinate $u$ was determined by a boundary
condition for $u$. This gave wave-functions in terms of parabolic cylinder
functions and a transcendental equation for the bound state energy levels.
The corresponding solution in the rotated $(r,q)$-system was similar.
An explicit solution in parabolic coordinates could not be found.

In the second Darboux space there were three non-trivial superintegrable
potentials. The potentials were related to the Hold-potential, the isotropic
singular oscillator, and the Coulomb potential in two-dimensional
Euclidean space.
We found combinations of polynomial wave-functions for the
discrete states and combinations of polynomials and  Whittaker functions for
the scattering states. The discrete energy spectrum for the oscillator-related
potentials was usually given by a quadratic equation in the energy.
For the Coulomb-related potential we found an equation in eight order in the
energy, which could be studied in a special case.
Also, in the semiclassical limit, we found that the energy-spectra indeed had
the behavior of an harmonic oscillator and a Coulomb potential, respectively.

On $\DIII$ we had potentials related to a linear potential, a Coulomb
potential, and a shifted oscillator in two-dimensional flat space.
We found for the first potential an equation in fourth order in the energy
$E$, and quadratic equations in the energy $E$ for the second potential and
third potential. 
The coulomb-related potential showed again in the semiclassical limit the  
behavior of a Coulomb potential.
Of some special interest was the feature of the complex periodic Morse
potential for the separation of $V_3$ in polar coordinates. Such complex
potentials have attracted in the recent years some attention, because the
involved $\CP\CT$-symmetry in these potentials has the consequence that they
nevertheless have a real spectrum, e.g. 
\cite{BQZ,CJT,MOAH,Znojila,Znojilb}--\cite{Znojil}.
Such kind a potentials also appear as subsystems in the list of
superintegrable potentials on the complex Euclidean plane \cite{KMP7}.

A special feature in $\DIII$ was that there is already for the free motion a
positive continuous and a negative infinite discrete spectrum.
A similar feature also exists for the free quantum motion on the
$\SU(1,1)$ and $\SO(2,2)$ hyperboloid.

In the fourth Darboux space we found potentials which were related to the 
Morse and P\"oschl--Teller potential, and combined modified P\"oschl--Teller
potentials. The modified P\"oschl--Teller
potentials had, of course, solutions in terms of hypergeometric functions,
respectively Jacobi polynomials (discrete spectrum) and Jacobi functions
(scattering states).

We were able to solve the various path integral representations, because we
have now to our disposal not only the basic path integrals for the harmonic 
oscillator, the linear oscillator, the radial harmonic oscillator, and
the (modified) P\"oschl--Teller Potential, but also path integral
identities derived from path integration on harmonic spaces like the
elliptic and spheroidal path integral representations with its more 
complicated special functions. This includes also numerous transformation
techniques to find a particular solution based on one of the basic solutions.
Various Green  function analysis techniques can be applied to find not
only an expression for the Green  function but also for the
wave-functions and the energy spectrum. 
Usually, we stated in all cases the solution for the discrete spectrum
contribution, i.e.~the energy-spectrum and the bound-states wave-functions.
However, not in all cases we stated explictly the scattering states.
In the cases, where we omitted the explicit representation, this can be 
done in a straightforward way by inserting the corresponding solution by the
potential problem in question and inserting the various coupling constants and
scattering quantum numbers.

Let us also note that our solutions are often on a more or less formal level.
Neither have we specified an embedding space, nor have we specified 
boundary conditions on our spaces. For instance, in $\DI$ boundary 
conditions and the signature of the ambient space is very important, because
choosing a positive or a negative signature of the ambient space changes
the boundary conditions, and hence the quantization conditions 
\cite{GROPOe}. The same line of reasoning is, of course, valid in the
other three Darboux spaces. We have not discussed in detail special cases
of the parameters (say $a$ and $b$), including the limiting cases to flat
spaces or spaces with constant (negative) curvature. Such a discussion would
go far beyond the scope of this paper.

Let us finally mention an important observation due to \cite{KalninsKMWinter}.
At the end of their paper Kalnins et al. gave a list of superintegrable
potentials on the two-dimensional complex plane and complex sphere. As it
turns out\cite{KKPM}, all of the potentials on Darboux spaces can be generated
by taking a two-dimensional line element and dividing this line element by a
superintegrable potential belonging to a specific class.\footnote{%
The cases for two-dimensional flat space with two-dimensional superintegrable
potentials is discussed in \cite{GROau}. It turns out that the quantization
conditions for the bound energy states is always determined by an equation of 
eighth order in $E$.} 
Not every class generates a new potential on a Darboux space, some are simply
related by a coordinate transformation, and some potentials can be generated
from the Euclidean plane as well as the complex sphere. The appearance of the
complex sphere is especially obvious in the general elliptic coordinate
system on $\DIV$. Some of the various different potentials coming from the
complex plane and sphere are also related by the so-called ``coupling constant
metamorphosis''. Coupling constant metamorphosis always comes into play if the
energy $E$  of the quantum system appears in the form of 
$E\times\hbox{metric terms}$. This observation leads to the notion that every
nondegenerate superintegrable system in two dimensions is ``St\"ackel
equivalent'' to a super integrable system in a two-dimensional space of
constant curvature \cite{KKPM}.

In the language of path integrals
coupling constant metamorphosis comes from ``time-'' or ``space-time''
transformations (also called Duru--Kleinert transformations \cite{KLEo}).
Here the most important example is the Coulomb problem, where by means of a  
space-time transformation the Coulomb-coupling $\alpha$ just becomes a constant
and the emerging harmonic oscillator problem has the frequency
$\omega^2=-2E/m$, i.e.~the negative energy of the Coulomb problem appears as
an harmonic oscillator frequency.
As we have seen this kind of coupling constant metamorphosis or space-time
transformation, respectively, had been indispensable tools in the path
integral evaluations of the free motion and for the superintegrable
potentials, and we can use both notions as synonymous.

We did not go into details of three-dimensional generalization of the Darboux
spaces \cite{GROat}. Of course, it is possible to extend the notion of
superintegrability to three-dimensional Darboux spaces.
In particular, in three dimensions there are more of such potentials. 
In total, there are five maximally
superintegrable potentials \cite{GROPOa}, the first four of them also are
superintegrable on, including the singular harmonic oscillator,
the Holt potential and the Coulomb potential. 
New features will arise due to the fact that on three-dimensional
generalization of the more complicated
Darboux spaces $\DIII$ and $\DIV$, coordinate systems from the
three-dimensional complex sphere come into play \cite{KAMI}.
Studies along such lines will be performed in future investigations.

\subsection*{\bf Acknowledgments}
This work was supported by the Heisenberg--Landau program. 

The authors are grateful to Ernie Kalnins for fruitful and pleasant
discussions on superintegrability and separating coordinate systems. 
C.Grosche would like to thank the organizers of the ``XII. International
Conference on Symmetry Methods in Physics'',  July 3--8, Yerevan, Armenia, for
the warm hospitality during the stay in Yerevan. 

G.S.Pogosyan acknowledges the support of the Direcci\'on General de Asuntos
del Personal Acad\'emico, Universidad Nacional Aut\'onoma de M\'exico 
({\sc dgapa--unam}) by the grant 102603 {\it Optica Matem\'atica}, 
{\sc sep-conacyt} project 44845 and {\sl PROMEP} 103.5/05/1705.

 
\begin{appendix}
\setcounter{section}{1}
\def\theequation{\thesection.\arabic{equation}}
\section*
{A Path Integral for the Free Motion on $\DIV$ \\ in Degenerate Elliptic
  Coordinates ($\gamma=1$)} 
\addcontentsline{toc}{section}{A Path Integral for the Free Motion on $\DIV$
  in Degenerate Elliptic \\ Coordinates ($\gamma=1$)}
We start by considering the metric in elliptic coordinates
($\gamma=1$):
\begin{equation}
\d s^2=\Bigg[a_-\bigg(\frac{1}{\sinh^2\homega}+\frac{1}{\sin^2\hvphi}\bigg)
   -a_+\bigg(\frac{1}{\cosh^2\homega}-\frac{1}{\cos^2\hvphi}\bigg)\Bigg]
   (\d\homega^2+\d\hvphi^2)\enspace.
 \end{equation}
We formulate the path integral in the usual way. We perform the 
space-time transformation with the coordinate transformation 
$\cos\hvphi=\tanh\htau$ yielding
\begin{eqnarray}
&&\!\!\!\!\!\!\!\!\!\!
K(\homega'',\homega',\hvphi'',\hvphi';T)
=\pathint{\homega}\pathint{\hvphi}\sqrt{g}
\nonumber\\   &&\!\!\!\!\!\!\!\!\!\!\quad\times
\exp\Bigg[\frac{\i m}{2\hbar}\int_0^T\bigg(
\frac{a_-}{\sinh^2\homega}-\frac{a_+}{\cosh^2\homega}
+\frac{a_-}{\sin^2\hvphi}-\frac{a_+}{\cos^2\hvphi}\bigg)
(\dot\homega^2+\dot\hvphi^2)\dt\Bigg]
\nonumber\\   &&\!\!\!\!\!\!\!\!\!\!
  =\int_{-\infty}^\infty\frac{\d E}{2\pi\hbar}\,\e^{-\i ET/\hbar}
  \int_0^\infty\d s''\exp\bigg[\ih\bigg(a_-E-\frac{\hbar^2}{8m}\bigg)s''\bigg]
  K(\homega'',\homega',\htau'',\htau';s'')
         \\   &&\!\!\!\!\!\!\!\!\!\!
  \hbox{with the transformed path integral given by}
\nonumber\\   &&\!\!\!\!\!\!\!\!\!\!
  K(\homega'',\homega',\htau'',\htau';s'')
  =\pathints{\htau}\pathints{\homega}\cosh\htau
\nonumber\\   &&\!\!\!\!\!\!\!\!\!\!\ \times
  \exp\Bigg(\ih\ints \Bigg\{
  \frac{m}{2}(\dot\ttau^2+\cosh^2\ttau\dot\tomega^2)
  -\frac{\hbar^2}{2m}\Bigg[\frac{1}{\cosh^2\htau}
  \bigg(\frac{\lambda_-^2+\bviert}{\sinh^2\homega}
  -\frac{\lambda_+^2+\bviert}{\cosh^2\homega}+\viert\bigg)
  -\frac{\lambda_+^2+\bviert}{\sinh^2\htau}\Bigg]\Bigg\}\d s\Bigg),
\nonumber\\   &&\!\!\!\!\!\!\!\!\!\!
\end{eqnarray}
where $\lambda_{\pm}^2=\viert-2ma_{\pm}E/\hbar^2$.
The successive path integrations are of the modified P\"oschl--Teller type.
Therefore the solution can be written as follows:
\begin{equation}
K(\homega'',\homega',\hvphi'',\hvphi';T)
  =\int\d k\int p
  \Psi_k^{(\lambda_-,\lambda_+)}(\homega'')
  \Psi_k^{(\lambda_-,\lambda_+)\,*}(\homega')
  \Psi_p^{(\lambda_+,\i k)}(\htau'')
  \Psi_p^{(\lambda_+,\i k)\,*}(\tau')
  \,\e^{-\i\hbar Tp^2/2m}\enspace.
\end{equation}
with the energy spectrum
\begin{equation}
E_p=\frac{\hbar^2}{2ma_-}\bigg(p^2+\viert\bigg)\enspace
\end{equation}
and we can re-insert $\tanh\htau\to\cos\hvphi$.
The difference of the energy spectra in degenerate elliptic and elliptic
coordinates (interchanging of $a_+$ and $a_-$) can be removed by 
a shift of the coordinates $\tvphi$ and $\hvphi$ by $\pi/2$, respectively.
  
\setcounter{section}{2}
\def\theequation{\thesection.\arabic{equation}}
\section*
{B Path Integral for the Free Motion on $\DIV$ \\ 
in Degenerate Elliptic Coordinates ($\gamma=2$)}
\addcontentsline{toc}{section}{B Path Integral for the Free Motion on $\DIV$ 
in Degenerate Elliptic \\ Coordinates ($\gamma=2$)}%
\setcounter{equation}{0}%
We start by considering the metric in degenerate elliptic coordinates
($\gamma=2$):
\begin{equation}
\d s^2=\viert\left(\frac{a_+}{\sinh^22\tomega}+\frac{a_-}{\sin^22\tvphi}\right)
   (\d\tomega^2+\d\tvphi^2)\enspace.
\end{equation}
We formulate the path integral in the usual way. We scale both variables
by the factor $2$ and perform the space-time transformation with the
coordinate transformation $\cos\tvphi=\tanh\ttau$ yielding
($\lambda^2=\bviert-2ma_+E/\hbar^2$):
\begin{eqnarray}
&&K(\tomega'',\tomega',\tvphi'',\tvphi';T)
  =\half\pathint{\tomega}\pathint{\tvphi}
  \bigg(\frac{a_+}{\sinh^2\tomega}+\frac{a_-}{\sin^2\tvphi}\bigg)
\nonumber\\   && \qquad \times
  \exp\Bigg[\frac{\i m}{2\hbar}\int_0^T
  \bigg(\frac{a_+}{\sinh^2\tomega}+\frac{a_-}{\sin^2\tvphi}\bigg)
  (\dot\tomega^2+\dot\tvphi^2)\dt\Bigg]
\nonumber\\   &&
  =\int_{-\infty}^\infty\frac{\d E}{2\pi\hbar}\,\e^{-\i ET/\hbar}
  \int_0^\infty\d s''\exp\bigg[\ih\bigg(a_-E-\frac{\hbar^2}{8m}\bigg)s''\bigg]
  K(\tomega'',\tomega',\ttau'',\ttau';s'')
         \\   &&\hbox{with the transformed path integral given by}
\nonumber\\   &&
  K(\tomega'',\tomega',\ttau'',\ttau';s'')
  =\pathints{\ttau}\pathints{\tomega}\cosh\ttau
\nonumber\\   && \qquad \times
  \exp\left\{\ih\ints \left[
  \frac{m}{2}(\dot\ttau^2+\cosh^2\ttau\dot\tomega^2)
  -\frac{\hbar^2}{2m\cosh^2\ttau}\bigg(\frac{\lambda^2+\bviert}{\cosh^2\tomega}
  +\viert\bigg)\right]\d s\right\}
\nonumber\\   &&  
  =(\cosh\ttau'\cosh\ttau'')^{-1/2}
   \sum_{\pm}\int_{\bbbr}\frac{\d k\,k\sinh\pi k}
   {\cosh^2\pi\lambda+\sinh^2\pi k}
   P^{\i k}_{\i\lambda-1/2}(\pm\tanh\omega'')
   P^{-\i k}_{\i\lambda-1/2}(\pm\tanh\tomega')
\nonumber\\   && \qquad \times
   \sum_{\pm}\int_{\bbbr}\frac{\d p\,p\sinh\pi p}
   {\cosh^2\pi k+\sinh^2\pi p}
   P^{\i p}_{\i k-1/2}(\pm\tanh\ttau'')P^{-\i p}_{\i k-1/2}(\pm\tanh\ttau')
   \,\e^{-\i\hbar Tp^2/2m}\enspace.
\end{eqnarray}
Therefore we obtain the wave-functions and the energy-spectrum, respectively
\begin{eqnarray}
\Psi_{k,p}(\ttau,\tomega)&=&\frac{1}{\sqrt{2\cosh\ttau}}
   \Bigg(\frac{k\sinh\pi k}{\cosh^2\pi\lambda+\sinh^2\pi k}
   \frac{p\sinh\pi p}{\cosh^2\pi k+\sinh^2\pi p}\Bigg)^{1/2}
\nonumber\\   && \qquad \times
   P^{\i k}_{\i\lambda-1/2}(\pm\tanh\omega)
   P^{\i p}_{\i k-1/2}(\pm\tanh\ttau)
\end{eqnarray}
and $E_p=\frac{\hbar^2}{2ma_-}\big(p^2+\viert\big)$,
and we can re-insert $\tanh\ttau\to\cos\tvphi$.

\section{Superintegrable Potentials on $E(2,\bbbc)$.}
\setcounter{equation}{0}%
In this appendix we shortly discuss the path integral representation 
of superintegrable potentials on the two-dimensional complex Euclidean 
plane. A thorough path integral discussion on the real two-dimensional complex
Euclidean plane has been done in \cite{GROPOa}, and therefore these solutions
will not be repeated here, only some new due to the appearance of three more
potentials $V_5$--$V_7$. As Usual $P_1=-\i\hbar\partial_x$ and
$P_2=-\i\hbar\partial_y$ denote the momentum operators, and $M=y P_1-x P_2$ is
the angular momentum. The potentials now read as follows 
\cite{KKPM,KMP1,KMP2,KMP7}
\begin{equation} 
\left.
\begin{array}{ll}
 V_5=\frac{B}{2}(x-\i y)
&\underline{\hbox{Cartesian}}       \\
&\underline{\hbox{Semi-hyperbolic}} \\
&\hbox{Light Cone}                  \\[2mm]
\hline\\
 V_6=\dfrac{\alpha}{2\sqrt{x-\i y}}
&\underline{\hbox{Parabolic}}       \\
&\underline{\hbox{Semi-hyperbolic}} \\
&\hbox{Light Cone}                 \\[2mm]
\hline\\
 V_7=\half\bigg[\alpha\dfrac{x^2+y^2}{(x+\i y)^4}
          +\dfrac{\beta}{(x+\i y)^2}+\gamma(x^2+y^2) \bigg]\qquad
&\underline{\hbox{Polar}}           \\
&\hbox{Hyperbolic}                  \\
\end{array}\qquad\right\}
\end{equation} 
In the $\underline{\hbox{underlined cases}}$ we give a (formal) path integral
representation. 

\begin{table}[h!]
\caption{\label{cosytabE2C} 
Coordinate Systems on the Complex Plane $E(2,\bbbc)$}
\begin{eqnarray}\begin{array}{l}\vbox{\small\offinterlineskip
\halign{&\vrule#&$\strut\ \hfil\hbox{#}\hfill\ $\cr
\noalign{\hrule}
height2pt&\omit&&\omit&&\omit&\cr
&Coordinate System        &&Integrals of Motion           
                          &&Coordinates            &\cr
height2pt&\omit&&\omit&&\omit&\cr
\noalign{\hrule}\noalign{\hrule}
height2pt&\omit&&\omit&&\omit&\cr
&1. Cartesian, &&$I=p_1^2$  &&$x,y$   &\cr
&\quad ($x,y\in\bbbr$) &&  &&   &\cr
height2pt&\omit&&\omit&&\omit&\cr
\noalign{\hrule}
height2pt&\omit&&\omit&&\omit&\cr
&2. Polar      &&$I=m^2$           &&$x=\vrho\cos\vphi$   &\cr
&\quad ($\vrho>0,\vphi\in[0,\pi)$)       
               &&                  &&$x=\vrho\sin\vphi$   &\cr
height2pt&\omit&&\omit&&\omit&\cr
\noalign{\hrule}
height2pt&\omit&&\omit&&\omit&\cr
&3. Light Cone &&$I=(P_1+\i P_2)^2$           
                                   &&$\hat x=x-\i y$      &\cr
&\quad ($x,y\in\bbbr$) &&          &&$\hat y=x+\i y$,     &\cr
height2pt&\omit&&\omit&&\omit&\cr
\noalign{\hrule}
height2pt&\omit&&\omit&&\omit&\cr
&4. Elliptic   &&$I=M^2-a^2P_2^2$  &&$x=\cosh\omega\cos\alpha$  &\cr
&\quad ($\omega>0,\alpha\in[0,2\pi)$)   
               &&$a\not=0$         &&$y=\sinh\omega\sin\alpha$  &\cr
height2pt&\omit&&\omit&&\omit&\cr
\noalign{\hrule}
height2pt&\omit&&\omit&&\omit&\cr
&5. Parabolic        &&$I=\{M,P_2\}$           
                     &&$x=\half(\xi^2-\eta^2)$                  &\cr
&\quad ($\xi,\eta>0$)        &&  &&$y=\xi\eta$                  &\cr
height2pt&\omit&&\omit&&\omit&\cr
\noalign{\hrule}
height2pt&\omit&&\omit&&\omit&\cr
& 6. Hyperbolic   
         &&$I=M^2+(P_1+\i P_2)^2$           
         &&$x=\dfrac{u^2+u^2v^2+v^2}{2uv\vphantom{\Big[}}$      &\cr
&\quad ($u,v>0$)  
  &&     &&$y=\i\dfrac{u^2-u^2v^2+v^2}{2uv\vphantom{\Big[}}$    &\cr
height2pt&\omit&&\omit&&\omit&\cr
\noalign{\hrule}
height2pt&\omit&&\omit&&\omit&\cr
&7. Semi-hyperbolic        
         &&$I=\{M,P_1+\i P_2\}+(P_1-\i P_2)^2$           
         &&$x= \half(w-z)^2+\viert(w+z)$                        &\cr
&\quad ($w,z\in\bbbr$) 
  &&     &&$y=-\half(w-z)^2-\viert(w+z)$                        &\cr
height2pt&\omit&&\omit&&\omit&\cr
\noalign{\hrule}}}\end{array}\nonumber\end{eqnarray}
\end{table}

\subsection*{The Potential $V_5$.}
For the potential $V_5$ the corresponding Lagrangian has the form
\begin{equation}
\CL=\frac{m}{2}(\dot x^2+\dot y^2)-\frac{B}{2}(x-\i y)\enspace.
\end{equation}
Thus, we identify two linear potentials \cite{GROad,SCHUHd}
\begin{eqnarray}
&&\!\!\!\!\!\!\!\!
K^{(V_5)}(x'',x',y'',y';T)
\nonumber\\   &&\!\!\!\!\!\!\!\!
=\pathint{x}\pathint{y}\exp\left\{\ih\intt
\bigg[{m\over2}(\dot x^2+\dot y^2)-\frac{B}{2}(x-\i y)\bigg]\dt\right\}
\nonumber\\   &&\!\!\!\!\!\!\!\!
=\bigg({m\over2\pi\i\hbar T}\bigg)
\exp\bigg[\ih\bigg({m\over2}{(x''-x')^2+(x''-x')^2\over T}
 -{BT\over4}(x'+x''-\i y'-\i y'')\bigg)\bigg]\qquad\qquad\qquad\qquad
\\   &&\!\!\!\!\!\!\!\!
=\bigg({4m\over\hbar^2B}\bigg)^{4\over3}
\int_{\bbbr} dE\,\e^{-\i ET/\hbar}\int_{\bbbr} d\lambda
\nonumber\\   &&\!\!\!\!\!\!\!\!\qquad\times
\Ai\left[\bigg(x'-{2E+\lambda\over k}\bigg)
       \bigg({mB\over\hbar^2}\bigg)^{1\over3}\right]
\Ai\left[\bigg(x''-{2E+\lambda\over k}\bigg)
       \bigg({mB\over\hbar^2}\bigg)^{1\over3}\right]
\nonumber\\   &&\!\!\!\!\!\!\!\!\qquad\times
\Ai\left[\i\bigg(y'-{2E-\lambda\over k}\bigg)
       \bigg({mB\over\hbar^2}\bigg)^{1\over3}\right]
\Ai\left[\i\bigg(y''-{2E-\lambda\over k}\bigg)
       \bigg({mB\over\hbar^2}\bigg)^{1\over3}\right]\enspace,
\end{eqnarray}
with the continuous spectrum $E=\hbar^2p^2/2m$, and $\lambda$ is the second
separation constant.

For $V_5$ in the semi-hyperbolic coordinates we obtain for the corresponding
Lagrangian ($\dot w=\d w/\dt$)
\begin{equation} 
\CL_E=\frac{m}{2}(w-z)(\dot w^2-\dot z^2)-\frac{B}{2}(w+z)+E\enspace,
\end{equation}
which gives after a time transformation
($\dot w=\d w/\d s$, $\dot z=\d z/\d s$ and $\dt =(w-z)\d s$) a transformed
Lagrangian 
\begin{equation} 
\tilde\CL_E=\frac{m}{2}(\dot w^2-\dot z^2)-\frac{B}{2}(w^2-z^2)+E(w-z)\enspace.
\end{equation}
Therefore the potential $v_5$ has been transformed into the problem of a
shifted harmonic oscillator, whose solution is well-known. In order to
determine the path integral solution we consider the Green function of the
harmonic oscillator \cite{GRSh}, use the convolution formula for the kernel in
terms of a product of two Green functions
\begin{eqnarray}
K^{(V_5)}(w'',w',z'',z';T)
              &=&\int_{-\infty}^\infty\frac{\d E}{2\pi\hbar}\,\e^{-\i ET/\hbar}
   \int_0^\infty \d s'' K_w(w'',w';s'')\cdot K_z(z'',z';s'')
\nonumber\\   &=&
\int_{-\infty}^\infty\frac{\d E}{2\pi\hbar}\,\e^{-\i ET/\hbar}
\frac{\hbar}{2\pi\i}\int\d\CE G_w(E;w'',w';-\CE) G_z(E;z'',z';\CE)\,,
\qquad\quad
\end{eqnarray}
and obtain therefore
\begin{eqnarray}
&&\!\!\!\!\!\!\!\!\!\!\!
K^{(V_5)}(w'',w',z'',z';T)
 \nonumber\\   &&\!\!\!\!\!\!\!\!
=\pathint{w}\pathint{z}
\exp\left\{\ih\intt\bigg[\frac{m}{2}(w-z)(\dot w^2-\dot z^2)-\frac{B}{2}(w+z)
\bigg]\dt\right\}\qquad
 \nonumber\\   &&\!\!\!\!\!\!\!\!
  =\frac{1}{4\pi^2}\int_{-\infty}^\infty\d E\int\d\lambda
{m\over\pi\hbar^3}\sqrt{\frac{m}{B}}\,
\Gamma^2\bigg(\half-{E+\lambda\over\hbar\omega}\bigg)
 \nonumber\\   &&\!\!\!\!\!\!\!\!\qquad\times
  D_{-\half+{E+\lambda\over\hbar\omega}}\left[\sqrt{{2\over\hbar}\sqrt{mB}}\,
   \bigg(w_>-\frac{E}{b}\bigg)\right]
  D_{-\half+{E+\lambda\over\hbar\omega}}\left[-\sqrt{{2\over\hbar}\sqrt{mB}}\,
   \bigg(w_<-\frac{E}{b}\bigg)\right]
 \nonumber\\   &&\!\!\!\!\!\!\!\!\qquad\times
  D_{-\half+{E+\lambda\over\hbar\omega}}\left[\sqrt{{2\over\hbar}\sqrt{mB}}\,
   \bigg(z_>-\frac{E}{b}\bigg)\right]
  D_{-\half+{E+\lambda\over\hbar\omega}}\left[-\sqrt{{2\over\hbar}\sqrt{mB}}\,
   \bigg(z_<-\frac{E}{b}\bigg)\right]\,,
\end{eqnarray}
with the continuous spectrum $E=\hbar^2p^2/2m$, and $\lambda$ is the second
separation constant.
The Green function may be evaluated in  terms of even and odd parabolic
cylinder functions $E^{(0)}_\nu(z)$ and $E^{(1)}_\nu(z)$, e.g.
\cite{GROas,POGOa,GRSh,GROPOa}, which is omitted here.

\subsection*{The Potential $V_6$.}
Let us consider the two Lagrangians of the potential $V_6$ expressed in
parabolic and semi-hyperbolic coordinates, respectively
\begin{eqnarray}
\CL_E&=&\frac{m}{2}(\xi^2+\eta^2)(\dot\xi^2+\dot\eta^2)
         +\sqrt{2}\,\alpha\frac{\xi-\i\eta}{\xi^2+\eta^2}+E
\\   &=&\frac{m}{2}(w-z)(\dot w^2-\dot z^2)+\i\frac{\sqrt{2}\,\alpha}{w-z}+E
\enspace,
\end{eqnarray}
which gives after a time transformation
($\dot \xi=\d \xi/\d s$, $\dot \eta=\d \eta/\d s$ and $\dt =(\xi^2+\eta^2)\d s$
in parabolic coordinates; $\dot w=\d w/\d s$, $\dot z=\d z/\d s$ and $\dt
=(w-z)\d s$ in semi-hyperbolic coordinates) the transformed Lagrangians
\begin{eqnarray}
\tilde\CL_E&=&\frac{m}{2}(\dot\xi^2+\dot\eta^2)
         +\sqrt{2}\,\alpha(\xi-\i\eta)+(\xi^2+\eta^2)
\\   &=&\frac{m}{2}(\dot w^2-\dot z^2)+\i\sqrt{2}\,\alpha+E(w-z)\enspace.
\end{eqnarray}
In parabolic coordinates we have a shifted harmonic oscillator and
in semi-hyperbolic coordinates a linear potential plus a constant. The solution
is consequently almost identical to the corresponding solutions for the
potential $V_5$ with appropriate replacement of the coupling constants.
See also \cite{GROas,GRSh,GROPOa,POGOa} for more details

\subsection*{The Potential $V_7$.}
Let us consider the last potential $V_7$. In polar coordinates we have the
effective Lagrangian (note the additional $\hbar^2$-potential \cite{GRSh})
\begin{equation}
\CL=\frac{m}{2}(\dot\vrho^2+\vrho^2\dot\vphi^2-\omega^2)
    -\frac{\hbar^2}{2mr^2}\bigg(\alpha\,\e^{-4\i\vphi}-2\beta\,\e^{-2\i\vphi}
    -\viert\bigg)\enspace.
\end{equation}
In the variable $\vphi$ we have a complex periodic Morse potential, the same
kind of potentials we have encountered on $\DIII$ for $V_3$ in polar
coordinates. We identify $\alpha=4c_1^2$ and $\beta=c_2/c_1$. Furthermore we
see that the remaining path integral in the variable $\vrho$ is just a 
radial harmonic oscillator path integral. Putting everything together yields
\begin{eqnarray}
&&\!\!\!\!\!\!\!\!\!\!\!
K^{(V_7)}(\vrho'',\vrho',\vphi'',\vphi';T)
=\pathint{\vrho}\pathint{\vphi}
\nonumber\\ &&\!\!\!\!\!\!\!\!\!\!\!\qquad\times
\exp\Bigg\{\ih\int_0^T\Bigg[
\frac{m}{2}(\dot \vrho^2+\vrho^2\dot\vphi^2-\omega^2\vrho^2)
-\frac{\hbar^2}{2m\vrho^2}\Big(\alpha
\e^{-4\i\vphi}-2\beta\,\e^{-2\i\vphi}-\bviert\Big)
\Bigg]\dt\Bigg\}
\nonumber\\ &&\!\!\!\!\!\!\!\!\!\!\!
=\sum_{l=0}^\infty
\Phi_{[cMP],l}^{(c_1,c_2)}(\vphi'')\Phi_{[cMP],l}^{(c_1,c_2)}(\vphi')
\frac{m\omega}{\i\hbar\sin\omega T}
\exp\bigg[-\frac{m\omega}{2\i\hbar}({\vrho'}^2+{\vrho''}^2)\cot\omega T\bigg]
I_{l+2\hbox{$\frac{c_2}{c_1}$}+\half}
\bigg(\frac{m\omega \vrho'\vrho''}{\i\hbar\sin\omega T}\bigg),
\nonumber\\ &&\!\!\!\!\!\!\!\!\!\!\!
\end{eqnarray}
with the well-known expansion by means of the Hille--Hardy formula in terms of
Laguerre polynomials for $\vrho$. We leave the result as it stands.

\section{Superintegrable Potentials on $S(2,\bbbc)$.}
\setcounter{equation}{0}%
Let us shortly enumerate the superintegrable potentials on the complex sphere.
On the real two-dimensional sphere there are two superintegrable potentials,
a feature which has been already investigated, e.g.~\cite{GROPOb}.
On the complex two-dimensional sphere there are four more potentials which are
are listed in (\ref{Potentials-S2C}) \cite{KKPM,KAMI,KMP1}.
In the $\underline{\hbox{underlined cases}}$ we give a path integral
representation. These representations remain, however, on a formal level,
because the complex sphere is an abstract space and serves just as a tool to
find the relevant potentials. Going to the corresponding real spaces, i.e. the
sphere and the hyperboloid, respectively, requires the real representation of
the coordinate system in question, including the corresponding
path integral representation.

In Table \ref{cosytabS2C} we list the five coordinate systems on the
complex sphere $S(2,\bbbc)$ according to \cite{KKPM,KAMI,KMP1}.
Let us note that we can also use $v=\i\e^{-\i x}$ as a parameterization
in the horospherical system $(x,y\in\bbbr)$. As usual $J_1,J_2,J-3$ are the 
angular momentum operators in three dimensions.

\subsection*{The Potential $V_3$.}
Let us start superintegrable potential on the two-dimensional
complex sphere. It has the form
\begin{eqnarray}
V_3(\vec s)&=&\frac{\alpha}{s_3^2}
            +\frac{\beta}{(s_1-\i s_2)^2}
            +\gamma\frac{(s_1+\i s_2}{(s_1-\i s_2)^3}
\\       &=&
\frac{\alpha}{\cos^2\vtheta^2}+\beta\frac{\e^{-2\i\vphi}}{\sin^2\vtheta}
-\gamma\frac{\e^{-4\i\vphi}}{\sin^2\vtheta}\enspace,
\\       &=&
\e^{-2\i x}\bigg(\gamma y^2+\frac{\alpha}{y^2}+\beta\bigg)-\gamma\e^{-4\i x}
\enspace,
\end{eqnarray}
and we have inserted spherical and horospherical coordinates on the 
(complex) sphere, respectively.

\begin{equation} 
\left.
\begin{array}{ll}
 V_3(\vec s)=\dfrac{\alpha}{s_3^2}+\dfrac{\beta}{(s_1-\i s_2)^2}
     +\gamma\dfrac{s_1+\i s_2}{(s_1-\i s_2)^3} 
&\underline{\hbox{Spherical}}      \\
&\underline{\hbox{Horospherical}}  \\
&\hbox{Degenerate Elliptic I}      \\[2mm]
\hline\\
 V_4(\vec s)=\dfrac{\alpha}{(s_1-\i s_2)^2}
     +\dfrac{\beta s_3}{\sqrt{s_1^2+s_2^2}}
     +\dfrac{\i\gamma}{\sqrt{(s_1+\i s_2)(s_1-\i s_2)^2}} 
&\hbox{Spherical}                  \\
&\hbox{Degenerate Elliptic II}     \\[2mm]
\hline\\
 V_5(\vec s)=\dfrac{\alpha z_++c^2z_-}{\sqrt{(c^2z_--z_+)^2-4c^2z_3}}
&\hbox{Elliptic}                   \\
\qquad \qquad  
  +\dfrac{\beta(z_+-c^2z_-)(z_+z_-+z_3^2)}{z_3^2\sqrt{(c^2z_--z_+)^2-4c^2z_3}}
  +\gamma\dfrac{z_+z_-}{z_3^2}
&                                  \\
(z_\pm=s_1\pm\i s_2,z_3=\sqrt{1-s_1^2-s_2^2},c^2=\frac{1+r}{1-r})
&\hbox{Degenerate Elliptic I}      \\[2mm]
\hline\\
 V_6(\vec s)=\dfrac{\alpha}{(s_1-\i s_2)^2}
          +\dfrac{\beta s_3}{(x-\i y)^3}
          +\gamma\dfrac{1-4s_3^2}{(s_1-\i s_2)^4}
&\underline{\hbox{Horospherical}}  \\
&\hbox{Degenerate Elliptic II}     \\
\end{array}\qquad\right\}
\label{Potentials-S2C}
\end{equation} 

\begin{table}[h!]
\caption{\label{cosytabS2C}
Coordinate Systems on the Complex Sphere $S(2,\bbbc)$}
\begin{eqnarray}\begin{array}{l}\vbox{\small\offinterlineskip
\halign{&\vrule#&$\strut\ \hfil\hbox{#}\hfill\ $\cr
\noalign{\hrule}
height2pt&\omit&&\omit&&\omit&\cr
&Coordinate System        &&Integrals of Motion           
                          &&Coordinates            &\cr
height2pt&\omit&&\omit&&\omit&\cr
\noalign{\hrule}\noalign{\hrule}
height2pt&\omit&&\omit&&\omit&\cr
&1.Spherical              &&$L=J_3^2$   &&$s_1=\sin\vtheta\cos\vphi$      &\cr
&\quad$(\vtheta\in[0,\pi),\vphi\in[0,2\pi)$
                          && 
                          &&$s_2=\sin\vtheta\sin\vphi$, $s_3=\cos\vtheta$ &\cr
height2pt&\omit&&\omit&&\omit&\cr
\noalign{\hrule}
height2pt&\omit&&\omit&&\omit&\cr
&2.Elliptic
&&$L=J-1^2+rJ_2^2$        &&$s_1^2=\dfrac{(ru-1)(rv-1)}{1-r}$             &\cr
& &&                      &&$s_2^2=\dfrac{r(u-1)(v-1)}{1-r}$, $z^2=ruv$   &\cr
height2pt&\omit&&\omit&&\omit&\cr
\noalign{\hrule}
height2pt&\omit&&\omit&&\omit&\cr
&3. Horospherical
&&$L=(J_1+\i J_2)^2$ &&$s_1=\dfrac{\i}{2}\bigg(v+\dfrac{y^2-1}{v}\bigg)$  &\cr
&  &&                &&$s_2=\dfrac{\i}{2}\bigg(v+\dfrac{y^2-1}{v}\bigg)$,
                            $s_3=\i y/v$                                  &\cr
height2pt&\omit&&\omit&&\omit&\cr
\noalign{\hrule}
height2pt&\omit&&\omit&&\omit&\cr
&4. Degenerate
&&$L=(J_1+\i j_2)^2-c^2J_3^2$
                          &&$s_1+\i s_3=\dfrac{1}{\cosh\tau_1\cosh\tau_2}$&\cr
&\quad Elliptic 1            &&
   &&$s_2-\i s_3=\dfrac{\cosh\tau_2}{\cosh\tau_1}
                +\dfrac{\cosh\tau_1}{\cosh\tau_2}
               -\dfrac{1}{\cosh\tau_1\cosh\tau_2}$                        &\cr
&\quad$(\tau_{1,2}\in\bbbr)$ && &&$s_3=\tanh\tau_1\tanh\tau_2$             &\cr
height2pt&\omit&&\omit&&\omit&\cr
\noalign{\hrule}
height2pt&\omit&&\omit&&\omit&\cr
&5. Degenerate            &&$L=J_3(J_1-\i J_2)^2$
                          &&$s_1+\i s_2=\dfrac{1}{\xi\eta}$               &\cr
&\quad Elliptic 2         && 
   &&$s_1+\i s_2=-\viert\dfrac{(\xi^2-\eta^2)^2}{\xi\eta}$                &\cr
&\quad $(\xi,\eta>0)$ &&  &&$s_3=\half\dfrac{\xi^2+\eta^2}{\xi\eta}$      &\cr
height2pt&\omit&&\omit&&\omit&\cr
\noalign{\hrule}}}\end{array}\nonumber\end{eqnarray}
\end{table}

This potential has now in spherical coordinates in the $\vphi$-dependence  the
same structure as the potential $V_7$ on the complex plane, thus the solution
is the same ($c_{1,2}$ in the complex Morse potential appropriately).
In the $\vtheta$ dependence we obtain after the separation of $\vphi$ a
P\"oschl--Teller potential. In comparison to $V_7$ on the complex plane, we
must therefore replace the wave-functions in $\vrho$ in terms of Laguerre
polynomials by the P\"oschl--Teller wave-functions 
$\Phi_n^{(\tilde\alpha,l+2\hbox{$\frac{c_2}{c_1}$}+\half)}(\vtheta)$ 
($\tilde\alpha^2=2m\alpha/\hbar^2+\viert$) and we are done.
Summarizing we obtain
\begin{eqnarray}
&&\!\!\!\!\!\!\!\!\!\!\!
K^{(V_3)}(\vtheta'',\vtheta',\vphi'',\vphi';T)
=\pathint{\vtheta}\pathint{\vphi}\sin\vtheta
\nonumber\\ &&\!\!\!\!\!\!\!\!\!\!\!\qquad\times
\exp\Bigg\{\ih\intt\Bigg[
\frac{m}{2}(\dot\vtheta^2+\sin^2\vtheta\dot\vphi^2)
-\frac{\alpha}{\cos^2\vtheta}-\frac{1}{\sin^2\vtheta}
\bigg(\beta\e^{-2\i\vphi}-\gamma\e^{-4\i\vphi}-\viert\bigg)\Bigg]\dt\Bigg\}
\nonumber\\ &&\!\!\!\!\!\!\!\!\!\!\!
=(\sin\vtheta'\sin\vtheta'')^{-1/2}
\sum_{n=0}^\infty\sum_{l=0}^\infty
\Phi_{[cMP],l}^{(c_1,c_2)}(\vphi'')\Phi_{[cMP],l}^{(c_1,c_2)}(\vphi')
\Phi_n^{(l+2\hbox{$\frac{c_2}{c_1}$}+\half,\tilde\alpha)}(\vtheta'')
\Phi_n^{(l+2\hbox{$\frac{c_2}{c_1}$}+\half,\tilde\alpha)}(\vtheta')
\nonumber\\ &&\!\!\!\!\!\!\!\!\!\!\!\qquad\times
\exp\bigg[-\ih\frac{\hbar^2}{2m}
\Big(2n+l+2\hbox{$\frac{c_2}{c_1}$}+\hbox{$\frac{3}{2}$}\Big)^2T\bigg]\enspace.
\end{eqnarray}
In horospherical coordinates we have in the variable $y$ a radial harmonic
oscillator (set $\gamma=m\omega^2/2$, 
$\tilde\alpha^2=2m\alpha/\hbar^2+\viert$ ) and in the same way
($c_{1,2}$ in the complex Morse potential appropriately)
\begin{eqnarray}
&&\!\!\!\!\!\!\!\!\!\!\!
K^{(V_3)}(x'',x',y'',y';T)
=\pathint{x}\pathint{y}\,\e^{2\i x}
\nonumber\\ &&\!\!\!\!\!\!\!\!\!\!\!\qquad\times
\exp\Bigg\{\ih\intt\Bigg[\frac{m}{2}(\dot x+\e^{2\i x}\dot y^2)
-\e^{-2\i x}\bigg(\gamma y^2+\frac{\alpha}{y^2}+\beta\bigg)
-\gamma\e^{-4\i x}\Bigg]\dt\Bigg\}
\nonumber\\   &&\!\!\!\!\!\!\!\!\!\!\!
=\e^{-\i(x'+x'')}\sum_{n=0}^\infty\sum_{l=0}^\infty
\Psi_l^{(RHO,\tilde\alpha)}(y'')\Psi_l^{(RHO,\tilde\alpha)}(y')
\Phi_{[cMP],n}^{(c_1,c_2)}(\vphi'')\Phi_{[cMP],n}^{(c_1,c_2)}(\vphi')
\nonumber\\ &&\!\!\!\!\!\!\!\!\!\!\!\qquad\times
\exp\bigg[-\ih\frac{\hbar^2}{2m}
\Big(n+2\hbox{$\frac{c_2}{c_1}$}+1\Big)^2T\bigg]\enspace,
\end{eqnarray}
and the $\Psi_l^{(RHO,\tilde\alpha)}(y)$ are the wave-functions of the radial
harmonic oscillator \cite{GRSh}.

\subsection*{The Potential $V_6$.}
As the last potential we consider $V_6$. We have
(set $\gamma=-m\omega^2/8$)
\begin{eqnarray}
 V_6(\vec s)&=&\dfrac{\alpha}{(s_1-\i s_2)^2}
          +\dfrac{\beta s_3}{(x-\i y)^3}
          +\gamma\dfrac{1-4s_3^2}{(s_1-\i s_2)^4}
       \\   &=&
\e^{-2\i x}\frac{m}{2}\omega^2\bigg(y+\frac{\i\beta}{m\omega^2}\bigg)^2
          -\e^{-2\i x}\bigg(\alpha+\frac{\beta^2}{2m\omega^2}\bigg)
          -\gamma\e^{-4\i x}\enspace,
\end{eqnarray}
and we have inserted horospherical coordinates. This potential is in the
variable $y$ a shifted harmonic oscillator, however, the shift is a complex
one. In the variable $x$ we have the complex periodic Morse potential.
Again we encounter a complex potential, this time a $\CP\CT$-symmetric harmonic
oscillator with spectrum $E_l=\hbar\omega(l+\half)$, e.g.~\cite{Znojilb}.
Consequently we have in a similar way as before 
($c_{1,2}$ in the complex Morse potential appropriately, set 
$\kappa=\i\beta/m\omega^2$):
\begin{eqnarray}
&&\!\!\!\!\!\!\!\!\!\!\!
K^{(V_6)}(x'',x',y'',y';T)
=\pathint{x}\pathint{y}\,\e^{2\i x}
\nonumber\\ &&\!\!\!\!\!\!\!\!\!\!\!\qquad\times
\exp\Bigg\{\ih\intt\Bigg[\frac{m}{2}(\dot x^2+\e^{2\i x}\dot y^2)
-\Bigg(\frac{m}{2}\omega^2\bigg(y+\frac{\i\beta}{m\omega^2}\bigg)^2
          +\bigg(\alpha+\frac{\beta^2}{2m\omega^2}\bigg)\Bigg)\e^{-2\i x}
          -\gamma\e^{-4\i x}\Bigg]\dt\Bigg\}
\nonumber\\   &&\!\!\!\!\!\!\!\!\!\!\!
=\e^{-\i(x'+x'')}\sum_{l=0}^\infty
\Psi_l^{(HO)}(y'')\Psi_l^{(HO)}(y')
\sum_{n=0}^\infty
\Phi_{[cMP],n}^{(c_1,c_2)}(\vphi'')\Phi_{[cMP],n}^{(c_1,c_2)}(\vphi')
\nonumber\\ &&\!\!\!\!\!\!\!\!\!\!\!\qquad\times
\exp\bigg[-\ih\frac{\hbar^2}{2m}
\Big(n+2\hbox{$\frac{c_2}{c_1}$}+1\Big)^2T\bigg]\enspace,
\end{eqnarray}
and the $\Psi_l^{(HO,\kappa)}(y)$ are the wave-functions of the shifted
harmonic oscillator \cite{GRSh}. 
The representations of the potentials $V_4$ and $V_5$ in the separating
coordinate systems lead to intractable powers in the various
coordinates, respectively powers of $\cosh\tau_{1,2}$, i.e.~highly anharmonic
terms which cannot be treated. The same holds for $V_3$ and $V_6$ in the
remaining separating coordinate systems. This concludes the discussion.

\end{appendix}


\input cyracc.def
\font\tencyr=wncyr10
\font\tenitcyr=wncyi10
\font\tencpcyr=wncysc10
\def\cyrrm{\tencyr\cyracc}
\def\cyrit{\tenitcyr\cyracc}
\def\cyrcp{\tencpcyr\cyracc}
\addcontentsline{toc}{section}{References}%
\renewcommand{\baselinestretch}{0.925}%
\noindent%

\bigskip
\vbox{\centerline{\ }
\centerline{\quad\epsfig{file=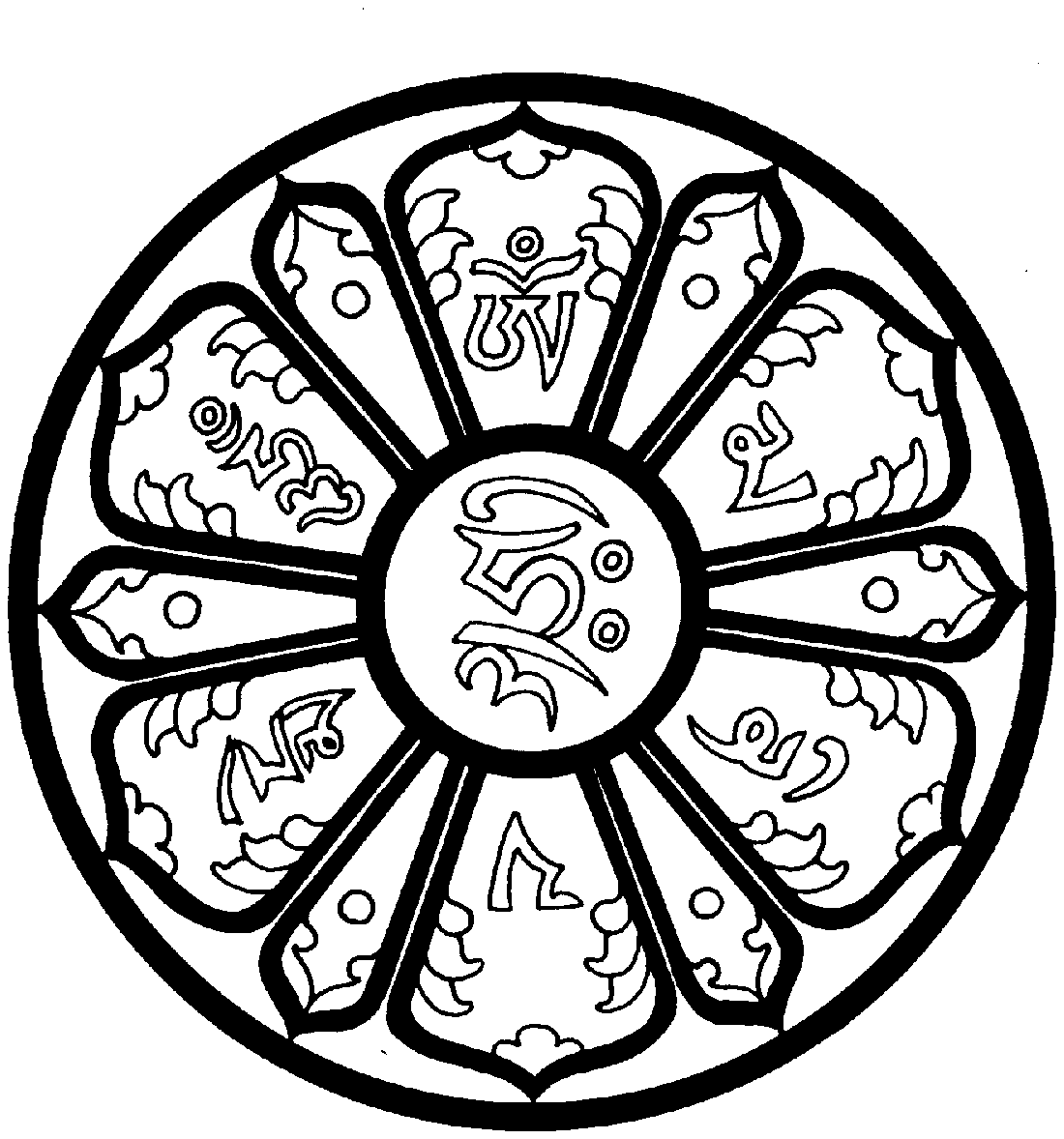,width=5cm}}}


\begin{thebibliography}{99}
\message{Bibliography}
\small
\bibitem{BBRR}
Bender, C.M., Brod, J.; Refig, A., Reuter, M.E.:
The $\CC$ Operator in $\CP\CT$-Symmetric Quantum Theories.
{\it J.\,Phys.\,A: Math.\,Gen.}\ {\bf 37} (2004) 10139--10165.
\bibitem{BJb}
B\"ohm, M., Junker, G.: Path Integration Over Compact and Noncompact
Rotation Groups. {\it J.\,Math.\,Phys.}\ {\bf 28} (1987) 1978--1994.
\bibitem{BQZ}
Bagchi, B., Quesne, C., Znojil, M.:
Generalized Continuity Equation and Modified Normalization in PT-Symmetric
Quantum Mechanics.
{\it Mod.Phys.Lett.}\ {\bf A 16} (2001) 2047--2057.
\bibitem{CJT}   
Cannata, F,. Junker, J., Trost, J.:
Schr\"odinger Operators with Complex Potential but Real Spectrum.
{\it Phys.\,Lett.}\ {\bf A 246} (1998) 219--226.
\bibitem{DASYPS}
Daskaloyannis, C., Ypsilantis, K.:
Unified Treatment and Classification of Superintegrable Systems with 
Integrals Quadratic in Momenta on a Two Dimensional Manifold.
{\it J.\,Math.\,Phys.}\ {\bf 45} (2006) 042904. 
\bibitem{DORW}
Del Olmo, M.A., Rodr\'\ii guez, M.A., Winternitz, P.: The Conformal Group 
$\SU(2,2)$ and Integrable Systems on a Lorentzian Hyperboloid. 
{\it Fortschr.\,Phys.}\ {\bf 44} (1996) 199--233.
\newline
Integrable Systems Based on $\SU(p,q)$ Homogeneous Manifolds.
{\it J.\,Math.\,Phys.}\ {\bf 34} (1993) 5118--5139.
\bibitem{EVA}
Evans, N.W.: Superintegrability in Classical Mechanics.
{\it Phys.\,Rev.}\ {\bf A 41} (1990) 5666--5676.
\newline
Group Theory of the Smorodinsky-Winternitz System.
{\it J.\,Math.\,Phys.}\ {\bf 32} (1991) 3369--3375.
\newline
Super-Integrability of the Winternitz System.
{\it Phys.\,Lett.}\ {\bf A 147} (1990) 483--486.
\bibitem{FH}
Feynman, R.P., Hibbs, A.: {\it Quantum Mechanics and Path Integrals}.
McGraw Hill, New York, 1965.
\bibitem{FMSUW}
Fri\v s, J., Mandrosov, V., Smorodinsky, Ya.A., Uhlir, M., Winternitz, P.:
On Higher Symmetries in Quantum Mechanics.
{\it Phys.\,Lett.}\ {\bf 16} (1965) 354--356.
\newline
Fri\v s, J., Smorodinski\u\ii, Ya.A., Uhl\'\ii\v r, M., Winternitz, P.:
Symmetry Groups in Classical and Quantum Mechanics.
{\it Sov.J.Nucl.\,Phys.}\ {\bf 4} (1967) 444--450.
\bibitem{GRA}
Gradshteyn, I.S., Ryzhik, I.M.: {\it Table of Integrals, Series, and
Products}. Academic Press, New York, 1980.
\bibitem{GROb}
Grosche, C.: The Path Integral on the Poincar\'e Upper Half-Plane With
a Magnetic Field and for the Morse Potential.
{\it Ann.\,Phys.\,$($N.Y.$)$} {\bf 187} (1988) 110--134.
\bibitem{GROf}  
Grosche, C.: The Path Integral on the Poincar\'e Disc, the Poincar\'e Upper 
Half-Plane and on the Hyperbolic Strip.
{\it Fortschr.\,Phys.}\ {\bf 38} (1990) 531--569.
\bibitem{GROad}
Grosche, C.: {\it Path Integrals, Hyperbolic Spaces, and Selberg Trace
Formul\ae}. World Scientific, Singapore, 1996.
\bibitem{GROas}
Grosche, C.: Path Integration on Darboux Spaces.
{\it Phys.\,Part.\,Nucl.}\ {\bf 37} (2006) 368--389.
\bibitem{GROat}
Grosche, C.: Path Integral Approach for Spaces of Non-constant Curvature in
Three Dimensions.
DESY Report, DESY 05--221, November 2005, \texttt{quant-ph/0511135}. 
To appear in {\it Proceedings of the ``II. International Workshop on
  Superintegrable Systems in Classical and Quantum Mechanics'', Dubna, Russia,
  June 27 - July 1, 2005}, {\it Physics Atomic Nuclei} {\bf 69} (2006).
\bibitem{GROau}
Grosche, C.: Path Integral Approach for for Quantum Motion 
on Spaces of Non-constant Curvature According to Koenigs.
DESY Report,  DESY 06--140, August 2006, \texttt{quant-ph/0608231}.
To appear in {\it Proceedings of the ``XII. International Conference on
  Symmetry Methods  in Physics'', July 3--8, Yerevan, Armenia, 2006}, 
{\it Physics Atomic Nuclei} {\bf 70} (2007).
\bibitem{GKPSa}
Grosche, C., Karayan, Kh., Pogosyan, G.S., Sissakian, A.N.: Quantum
Motion on the Three-Dimensional Sphere: The Ellipso-Cylindrical Bases.
{\it J.\,Phys.\,A: Math.\,Gen.}\ {\bf 30} (1997) 1629--1657.
\bibitem{GROPOa}
Grosche, C., Pogosyan, G.S., Sissakian, A.N.: Path Integral Discussion for 
Smoro\-dinsky-Winternitz Potentials: I.~Two- and Three-Dimensional Euclidean 
Space. {\it Fortschr.\,Phys.}\ {\bf 43} (1995) 453--521.
\bibitem{GROPOb}
Grosche, C., Pogosyan, G.S., Sissakian, A.N.: Path Integral Discussion for 
Smoro\-dinsky-Winternitz Potentials: II.~The Two- and Three-Dimensional Sphere.
{\it Fort\-schr.Phys.}\ {\bf 43} (1995) 523--563.
\bibitem{GROPOc}
Grosche, C., Pogosyan, G.S., Sissakian, A.N.: Path-Integral Approach to 
Superintegrable Potentials on the Two-Dimensional Hyperboloid.
{\it Phys.\,Part.\,Nucl.}\ {\bf 27} (1996) 244--278.
\bibitem{GROPOd}
Grosche, C., Pogosyan, G.S., Sissakian, A.N.: Path Integral Approach for 
Superintegrable Potentials on the Three-Dimensional Hyperboloid. 
{\it Phys.\,Part.\,Nucl.}\ {\bf 28} (1997) 486--519.
\bibitem{GROPOe}
Grosche, C., Pogosyan, G.S., Sissakian, A.N.: Path Integral Approach for 
Superintegrable Potentials on Spaces of Non-constant Curvature: I.
Darboux Spaces $\DI$ and $\DII$.
{\it DESY preprint} DESY 06-113, July 2006, \texttt{quant-ph/0608083}.
{\it Phys.\,Part.\,Nucl.}, to appear.
\bibitem{GRSh} 
Grosche, C., Steiner, F.: {\it Handbook of Feynman Path Integrals}.
{\it Springer Tracts in Modern Physics} {\bf 145}.
Springer, Berlin, Heidelberg, 1998.
\bibitem{KALc}
Kalnins, E.G.: 
On the Separation of Variables for the Laplace Equation 
$\Delta\psi+K^2\psi=0$ in Two- and Three-Dimensional Minkowski Space.
{\it SIAM J.\,Math.\,Anal.}\ {\bf 6} (1975) 340--374.
\bibitem{KAL}  
Kalnins, E.G.: {\it Separation of Variables for Riemannian Spaces of
Constant Curvature}. Longman Scientific \&\ Technical, Essex, 1986.
\bibitem{KKM}
Kalnins, E.G., Kress, J.M., Miller, W.Jr.:
Jacobi, Ellisoidal Coordinates, and Superintegrable Systems.
{\it J.Nonlinear Math.Phys.}\ {\bf 12} (2005). 209-229.
\bibitem{KalninsKMWinter}
Kalnins, E.G., Kress, J.M., Miller, W.Jr., Winternitz, P.:
Superintegrable Systems in Darboux Spaces.
{\it J.\,Math.\,Phys.}\ {\bf 44} (2003) 5811--5848.
\bibitem{KKPM}
Kalnins, E.G., Kress, J.M., Pogosyan, G.S., Miller, W.Jr.:
Completeness of Superintegrability in Two-Dimensional Constant-Curvature
Spaces. 
{\it J.\,Phys.\,A: Math.\,Gen.}\ {\bf 34} (2001) 4705--4720.
\bibitem{KalninsKWinter}
Kalnins, E.G., Kress, J.M., Winternitz, P.:
Superintegrability in a Two-Dimensional Space of Non-constant Curvature.
{\it J.\,Math.\,Phys.}\ {\bf 43} (2002) 970--983.
\bibitem{KAMIb}
Kalnins, E.G.,  Miller Jr., W.: Lie Theory and Separation of Variables.~4. The 
Groups $\SO(2,1)$ and $\SO(3)$. {\it J.\,Math.\,Phys.}\ {\bf 15} (1974)
1263--1274. 
\bibitem{KAMI}
Kalnins, E.G., Miller Jr., W.: The Wave Equation, $\OO(2,2)$, and Separation 
of Variables on Hyperboloids.
{\it Proc.\,Roy.\,Soc.Edinburgh} {\bf A 79} (1977) 227--256.
\newline
Kalnins, E.G., Miller Jr., W.: 
Lie Theory and the Wave Equation in Space-Time. 2. The Group $\SO(4,\bbbc)$.
{\it SIAM J.\,Math.\,Anal.}\ {\bf 9} (1978) 12--33.
\bibitem{KMP5}  
Kalnins, E.G., Miller Jr., W., Hakobyan, Ye.M., Pogosyan, G.S.:
Superintegrability on the Two-Dimensional Hyperboloid II.
{\it J.\,Math.\,Phys.}\ {\bf 40} (1999) 2291--2306.
\bibitem{KMP3}
Kalnins, E.G., Miller Jr., W., Pogosyan, G.S.:
Superintegrability and Associated Polynomial Solutions.
Euclidean Space and the Sphere in Two Dimensions.
{\it J.\,Math.\,Phys.}\ {\bf 37} (1996) 6439--6467.
\bibitem{KMP4}
Kalnins, E.G., Miller Jr., W., Pogosyan, G.S.:
Superintegrability on the Two-Dimensional Hyperboloid.
{\it J.\,Math.\,Phys.}\ {\bf 38} (1997) 5416--5433.
\bibitem{KMP1}
Kalnins, E.G., Miller Jr., W., Pogosyan, G.S.:
Completeness of Multiseparable Superintegrability on the Complex
2-Sphere. {\it J.\,Phys.\,A: Math.\,Gen.}\ {\bf 33} (2000) 6791--6806.
\bibitem{KMP2}
Kalnins, E.G., Miller Jr., W., Pogosyan, G.S.:
Completeness of Multiseparable Superintegrability in $E_{2,C}$
{\it J.\,Phys.\,A: Math.\,Gen.}\ {\bf 33} (2000) 4105--4120.
\bibitem{KMP7}
Kalnins, E.G., Miller Jr., W., Pogosyan, G.S.:
Superintegrability on Two-Dimensional Complex Euclidean Space.
In {\it Algebraic Methods in Physics. A Symposium for the 60th
Birthdays of Ji\v r\'\ii Patera and Pavel Winternitz}, pp.95--103.
{\it CRM Series in Mathematical Physics, Eds.: 
Yvan Saint-Aubin, Luc Vinet}. Springer, Berlin, Heidelberg, 2001.
\bibitem{KMP2a}
Kalnins, E.G., Miller Jr., W., Pogosyan, G.S.:
Completeness of Multiseparable Superintegrability in Two Dimensions.
{\it Phys.\,Atomic.\,Nucl.}\ {\bf 65} (2002) 1033--1035
\bibitem{KMP6}
Kalnins, E.G., Miller Jr., W., Pogosyan, G.S.:
Exact and quasi-exact solvability of two-dimensional superintegrable
quantum systems. I. Euclidean space. \texttt{math-ph/0412035}.
\bibitem{KLEo}
Kleinert, H.: {\it Path Integrals in Quantum Mechanics, Statistics and
Polymer Phys\-ics}. World Scientific, Singapore, 1990.
\bibitem{KOENIGS}
Koenigs, G.: Sur les g\'eod\'esiques a int\'egrales quadratiques.  A note
appearing in ``Lecons sur la th\'eorie g\'en\'erale des
surface''. Darboux, G., Vol.4, 368--404, Chelsea Publishing, 1972.
\bibitem{POGOa}
Lutsenko, I.V., Pogosyan, G.S., Sisakyan, A.N.,TerAntonyan, V.M.:
Hydrogen Atom as Indicator of Hidden Symmetry of a Ring-Shaped
Potential;
{\it Theor.\,Math.\,Phys.}\ {\bf 83} (1990) 633--639.
\bibitem{MOAH}
Mostafazadeh, A., Batal, A.:
Physical Aspects of Pseudo-Hermitian and $PT$-Symmetric Quantum Mechanics.
{\it J.\,Phys.\,A: Math.\,Gen.}\ {\bf A 37} (2004) 11645--11680. 
\bibitem{OLE}
{\cyrrm Olevs\cydot ki\u i, M.N.: Triortogonal\cprime nye sistemy v
prostranstvakh postoyanno\u i krivizny, v kotorykh uravnenie} $\Delta_2u+
\lambda u=0$ {\cyrrm dopus\cydot kaet polnoe razdelenie peremennyh}.
{\cyrit Mat.\,Sb.}\ {\bf 27} (1950) 379--426.
\newline
[Olevski\u\ii, M.N.: Triorthogonal Systems in Spaces of Constant Curvature in 
which the Equation $\Delta_2u+\lambda u=0$ Allows the Complete Separation of 
Variables. {\it Math.\,Sb.}\ {\bf 27} (1950) 379--426 (in Russian)].
\bibitem{PI}
Peak, D., Inomata, A.: Summation Over Feynman Histories in Polar
Coordinates. {\it J.\,Math.\,Phys.}\ {\bf 10} (1969) 1422--1428.
\bibitem{SCHUHd}
Schulman, L.S.: {\it Techniques and Applications of Path Integration}.
John Wiley \&\ Sons, New York, 1981.
\bibitem{USH}
Ushveridze, A.: {\it Quasi-exactly Solvable Models in Quantum
Mechanics}. Bristol, Institute of Physics Publishing, 1994.
\bibitem{WSUF}
Winternitz, P., Smorodinski\u\ii, Ya.A., Uhlir, M., Fris, I.:
Symmetry Groups in Classical and Quantum Mechanics.
{\it Sov.\,J.\,Nucl.\,Phys.}\ {\bf 4} (1967) 444--450.
\bibitem{WOJ}
Wojciechowski, S.:
Superintegrability of the Calogero-Moser System.
{\it Phys.Lett.}\ {\bf A 95} (1983) 279--281.
\bibitem{Znojilb}
Znojil, M.:
$\CP\CT$-Symmetric Harmonic Oscillators.
{\it Phys.Lett.}\  {\bf A 259} (1999) 220--223.
\bibitem{Znojila}
Znojil, M.:
Exact solution for Morse oscillator in PT-Symmetric Quantum Mechanics.
{\it Phys.Lett.}\  {\bf A 264} (1999) 108--111.
\bibitem{Znojil}
Znojil, M.:
Workshop on  Pseudo-Hermitian Hamiltonians in Quantum Physics. I--III. 
{\it Czech.J.Phys.}\ {\bf 54} (2004) 1--156.
{\it Czech.J.Phys.}\ {\bf 54} (2004) 1005--1148.
{\it Czech.J.Phys.}\ {\bf 55}  (2005) 1045--1192.
\newline
And \texttt{http://gemma.ujf.cas.cz/\%7Eznojil/conf/index.html}.
\end{thebibliography}
\end{document}